\documentclass[preprint,superscriptaddress,amsmath,amssymb,floatfix,nofootinbib,letterpaper]{revtex4}
\usepackage{epsfig,subfigure}
\usepackage{epstopdf}
\usepackage[utf8]{inputenc}
\usepackage{graphicx}
\usepackage{amssymb}
\usepackage{amsxtra}
\usepackage{amsmath}
\usepackage{booktabs,multirow,tabularx}
\usepackage{slashed}
\usepackage{float}
\usepackage{placeins}
\usepackage{rotating}
\usepackage{lscape}
\usepackage{color}
\usepackage{hyperref}
\usepackage{soul}

\newcommand{\bea}{\begin{eqnarray}}
\newcommand{\eea}{\end{eqnarray}}

\def\beq{\begin{equation}}
\def\eeq{\end{equation}}

\DeclareUnicodeCharacter{2212}{-}

\begin{document}

\title{Beyond the Standard Model Effective Field Theory: \\The Singlet Extended Standard Model}

\author{Shekhar Adhikari}
\email{s869a465@ku.edu}
\affiliation{Department of Physics and Astronomy, University of Kansas, Lawrence, Kansas, 66045~ U.S.A.}

\author{Ian M. Lewis}
\email{ian.lewis@ku.edu}
\affiliation{Department of Physics and Astronomy, University of Kansas, Lawrence, Kansas, 66045~ U.S.A.}

\author{Matthew Sullivan}
\email{msullivan1@bnl.gov}
\affiliation{Department of Physics and Astronomy, University of Kansas, Lawrence, Kansas, 66045~ U.S.A.}
\affiliation{High Energy Theory Group, Physics Department, Brookhaven National Laboratory,
Upton, New York, 11973~ U.S.A.}


\begin{abstract}
One of the assumptions of simplified models is that there are a few new particles and interactions accessible at the LHC and all other new particles are heavy and decoupled.  The effective field theory (EFT) method provides a consistent method to test this assumption.  Simplified models can be augmented with higher order operators involving the new particles accessible at the LHC.  Any UV completion of the simplified model will be able to match onto these beyond the Standard Model EFTs (BSM-EFT).  In this paper we study the simplest simplified model: the Standard Model extended by a real gauge singlet scalar.  In addition to the usual renormalizable interactions, we include dimension-5 interactions of the singlet scalar with Standard Model particles.  As we will show, even when the cutoff scale is $3$~TeV, these new effective interactions can drastically change the interpretation of Higgs precision measurements and scalar searches.  In addition, we discuss how power counting in a BSM-EFT depends strongly on the processes and parameter space under consideration.  Finally, we propose a $\chi^2$ method to consistently combine the limits from new particle searches with measurements of the Standard Model.  Unlike imposing a hard cutoff on heavy resonance rates, our method allows fluctuations in individual channels that are consistent with global fits.
\end{abstract}

\maketitle

\section{Introduction}
\label{sec:intro}
The Large Hadron Collider (LHC) has had two very successful runs.  While no new physics beyond the Standard Model (BSM) has been discovered, we may yet expect it to show up in currently unanalyzed data or in future runs at the LHC.  In the absence of discoveries of more complete models such as Supersymmetry, extra dimensions, or composite Higgs models, it is useful to study simplified models~\cite{Alves:2011wf}.  A frequent assumption of simplified models is that there are at most a handful of new particles accessible at LHC energies, while all additional new particles are too heavy to be produced.  However, this raises the question: can the effects of the inaccessible new particles be truly neglected?   For example, consider a simplified model with a new up-type vector like quark (VLQ).  If there is a new scalar in the theory,  even if the scalar cannot be directly produced, it can mediate new loop level decays of the VLQ into photons and gluons~\cite{Kim:2018mks}.  Indeed, in certain regions of parameter space, these decay modes can be dominant~\cite{Kim:2018mks,Alhazmi:2018whk,Criado:2019mvu}, fundamentally changing the phenomenology of the simplified VLQ model.

The most ``model independent'' method to determine the effects of new, heavy particles is an effective field theory (EFT).  An EFT is a power expansion in inverse powers of some new physics scale $\Lambda$:
\begin{eqnarray}
\mathcal{L}=\mathcal{L}_{\rm ren}+\sum_{n=5}^{\infty}\sum_k \frac{f_{k,n}}{\Lambda^{n-4}}\mathcal{O}_{k,n},
\end{eqnarray}
where $f_{k,n}$ are Wilson coefficients, $\mathcal{L}_{\rm ren}$ is the renormalizable Lagrangian, and $\mathcal{O}_{k,n}$ are dimension-$n$ operators.  In the Standard Model EFT (SMEFT)~\cite{Buchmuller:1985jz,Grzadkowski:2010es,Brivio:2017vri}, $\mathcal{L}_{\rm ren}$ and $\mathcal{O}_{k,n}$ consist of SM fields and are invariant under SM symmetries.  To test the stability of simplified models against heavy new physics, this framework needs to be extended to the Beyond the Standard Model EFT (BSM-EFT)~\cite{Kim:2018mks,Alhazmi:2018whk,Criado:2019mvu,Dawson:2016ugw,Bauer:2016hcu,Carmona:2016qgo,Anisha:2019nzx,Karmakar:2019vnq,Crivellin:2016ihg,DiazCruz:2001tn,Bar-Shalom:2018ure,Chala:2017sjk,Alanne:2017oqj,Goertz:2019vht,Nagai:2014cua,Nagai:2019tgi}.  In the BSM-EFT, $\mathcal{L}_{\rm ren}$ and $\mathcal{O}_{k,n}$ consist of SM and simplified model fields and are invariant under the symmetries of the simplified model.  This approach is agnostic about the high scale new physics since any UV completion of a simplified model will match onto the BSM-EFT.

In this paper we study the BSM-EFT of the simplest possible extension of the SM, the addition of a real scalar singlet $S$~\cite{OConnell:2006rsp,Barger:2007im,Bowen:2007ia}.  Beyond being the simplest extension of the SM, the singlet model can help provide a strong first order electroweak phase transition necessary of electroweak baryogenesis~\cite{Choi:1993cv,Profumo:2007wc,Espinosa:2011ax,Curtin:2014jma,Chen:2017qcz}.  At the renormalizable level, the new singlet only enters the scalar potential, and its interactions with fermions and gauge bosons are inherited by its mixing with the SM Higgs boson.  However, it is highly unlikely that a singlet scalar would appear without any new physics.  For example, even if it can give rise to a strong first order electroweak phase transition, in order to successfully have electroweak baryogenesis, new sources of CP violation are needed~\cite{Xiao:2015tja,Cline:2017jvp,Chao:2017oux,Bell:2019mbn}.  In fact, it has been shown~\cite{Espinosa:2011eu,Cline:2012hg,Huang:2018aja} that the BSM-EFT for the real scalar singlet can provide the CP violation necessary for electroweak baryogenesis.

Our analysis will consist of two major portions: reinterpreting Higgs precision measurements in the singlet extended SM and reinterpreting searches for new heavy scalars.    After electroweak symmetry breaking (EWSB), the new singlet scalar and Higgs boson will mix.  Without the new EFT interactions, this mixing results in a universal suppression of Higgs boson production rates.  Hence, Higgs precision measurements have a very simple interpretation~\cite{Robens:2015gla,Buttazzo:2015bka,Robens:2016xkb,Lewis:2017dme,Ilnicka:2018def,Martin-Lozano:2015dja}.  However, the BSM-EFT will introduce new interactions between the Higgs boson and fermions/gauge bosons.  As we will show, these can significantly alter the interpretation of Higgs measurements.  A similar argument can be made for constraints coming from heavy scalar searches.  At the renormalizable level, the new scalar inherits all of its interactions with fermions and gauge bosons from the SM Higgs boson.  Hence, its production rates are the same as a heavy Higgs boson but suppressed by a mixing angle.  Similarly, its decay rates are the same as a heavy Higgs boson suppressed by a mixing angle, except when a di-Higgs resonance is kinematically available.  That is, at the renormalizable model, the phenomenology is well defined.  As we will show, with the introduction of new interactions between the scalar and fermions/gauge bosons the phenomenology can significantly change.  Even though it is typically assumed that heavy new physics can be neglected, we will show that even in the simplest of all simplified models this assumption must be called into question.

This paper is an extension of work in Ref.~\cite{Dawson:2016ugw}, where only effective interactions between the scalar singlet and gauge bosons were considered.  We should note that the full BSM-EFT was considered in Ref.~\cite{Bauer:2016hcu}.  However, they also considered dimension-6 SMEFT operators.  While these effects can be important, we are interested in the question of how the EFT including new particles can change the phenomenology of the simplified models.  Hence, we will focus on dimension-5 operators involving SM gauge bosons, SM fermions, and the new scalar singlet.  In addition, we will include the most up-to-date Higgs precision data and searches for scalar singlets.  Also, we give a robust discussion of power counting the BSM-EFT and propose a new $\chi^2$ analysis to combine heavy resonance searches with precision measurements.

In Section~\ref{sec:Model}, we develop the BSM-EFT for the real scalar singlet.  The EFT power counting in a BSM-EFT can change from the usual SMEFT power counting, as we will discuss in Sec.~\ref{sec:PC}.  The effects of the new operators on Higgs production and decay are shown in Sec.~\ref{sec:ProdDec}, and results from fitting to Higgs signal strengths are given in Sec.~\ref{sec:Higgs}.  In Sec.~\ref{sec:chi2}, we propose a $\chi^2$ analysis for heavy resonance search limits, and in Sec.~\ref{sec:search}, the final results of heavy scalar resonances and their combination with Higgs signal strengths are given.  We conclude in Sec.~\ref{sec:conc}.  A discussion about assumptions in Higgs signal rate calculations is given in App.~\ref{app:WidthXSApprox}, the Feynman rules are given in Appendix~\ref{app:FeynRules}, the experimental results we fit to are given in App.~\ref{app:data}, and various parameter space limits are given in App.~\ref{app:limits}.

\section{Model}
\label{sec:Model}
We consider the SM extended by a real gauge singlet scalar, $S$, and will not impose an additional $Z_2$ upon $S$.  In order to focus on the effects of new physics on the scalar singlet properties, we will consider only dimension-5 EFT operators.  For simplicity, we will also only focus on CP even operators.  These are the lowest order effective operators that include a scalar singlet~\cite{Dawson:2016ugw,Bauer:2016hcu}.  At dimension-5, the only SMEFT operators are those that contribute to Majorana neutrino masses~\cite{Weinberg:1979sa,Buchmuller:1985jz,Grzadkowski:2010es}, which are not relevant for LHC analyses.  Hence, these will be neglected and the BSM-EFT will only consist of operators including the new singlet scalar.

Adapting the notation of Refs.~\cite{Chen:2014ask}, to order $\Lambda^{-1}$ the scalar potential is:  
\begin{eqnarray}
V(\Phi,S)&=&-\mu^2\Phi^\dagger\Phi+\lambda(\Phi^\dagger\Phi)^2+\frac{a_1}{2}\Phi^\dagger\Phi S+\frac{a_2}{2}\Phi^\dagger\Phi S^2+\frac{a_3}{2\Lambda}\Phi^\dagger\Phi S^3+\frac{a_4}{2\Lambda}(\Phi^\dagger \Phi)^2 S\nonumber\\
&&+b_1 S+\frac{b_2}{2}S^2+\frac{b_3}{3}S^3+\frac{b_4}{4}S^4+\frac{b_5}{5\Lambda}S^5\label{eq:Pot}
\end{eqnarray}
where $\Phi=(0,\phi_0/\sqrt{2})^T$ is the SM Higgs doublet in the unitary gauge, $\phi_0=h+v$ is the neutral scalar component of $\Phi$, $h$ is the Higgs boson, and $\langle\phi_0\rangle=v$ is the SM Higgs vacuum expectation value (vev).  Since $S$ is not charged under any symmetry, its vev does not break any symmetry and results in an unphysical redefinition of parameters~\cite{Chen:2014ask,Lewis:2017dme}.  Hence, without loss of generality we can impose $\langle S\rangle = 0$.

After EWSB, the Higgs boson $h$ and scalar singlet have the same quantum numbers and can mix:  
\begin{eqnarray}
\begin{pmatrix}h_1\\h_2\end{pmatrix} = \begin{pmatrix}\cos\theta & \sin\theta\\ -\sin\theta &\cos\theta \end{pmatrix}\begin{pmatrix} h\\S\end{pmatrix},
\end{eqnarray}
where $h_{1,2}$ are mass eigenstates with masses $m_{1,2}$.  We will assume $m_1=125~{\rm GeV}<m_2$, since the other mass hierarchy is strongly constrained by LEP~\cite{Robens:2015gla}.  
With the masses, mixing, and vevs, we can now solve for five parameters in the potential
\begin{eqnarray}
\mu^2&=&\frac{1}{2}\left(\cos^2\theta\,m^2_1+\sin^2\theta\,m_2^2\right)\,,\\
\lambda&=&\frac{\cos^2\theta\,m_1^2+\sin^2\theta\,m_2^2}{2\,v^2}\,,\nonumber\\
a_1&=&\sin2\theta\,\frac{m_1^2-m_2^2}{v}-a_4\frac{v^2}{\Lambda}\,,\nonumber\\
b_1&=&\frac{1}{4}\sin2\theta\,v\,\left(m_2^2-m_1^2\right)+a_4\frac{v^4}{8\,\Lambda}\,,\nonumber\\
b_2&=&\cos^2\theta\,m_2^2+\sin^2\theta\,m_1^2-\frac{1}{2}a_2\,v^2.\nonumber
\end{eqnarray}
These are $\mathcal{O}(v/\Lambda)$ corrections on the relationships founds in Refs.~\cite{Chen:2014ask,Lewis:2017dme}.
The free parameters of the scalar potential are then
\begin{eqnarray}
m_1=125~{\rm GeV},\,m_2,\,v=246~{\rm GeV},\,\langle S\rangle=0,\,\theta,\,a_2,\,a_3,\,a_4,\,b_3,\,b_4,\,b_5.
\end{eqnarray}

The scalar potential gives rise to important trilinear scalar couplings after EWSB:
\begin{eqnarray}
V(h_1,h_2)\supset \frac{1}{3!}\lambda_{111}h_1^3+\frac{1}{2}\lambda_{211}h_1^2h_2,
\end{eqnarray}
where
\begin{eqnarray}
\lambda_{111}&=&\frac{3\,m_1^2}{v}\cos^3\theta+2\,b_3\sin^3\theta+3\,a_2v\cos\theta\sin^2\theta+\frac{3\,a_3v^2}{2\,\Lambda}\sin^3\theta+\frac{3\,a_4v^2}{\Lambda}\cos^2\theta\sin\theta,\\
\lambda_{211}&=&-\frac{m_2^2+2\,m_1^2}{v}\cos^2\theta\sin\theta+2\,b_3\cos\theta\sin^2\theta+a_2v\sin\theta\left(2\cos^2\theta-\sin^2\theta\right)\nonumber\\
&&+\frac{3\,a_3v^2}{2\,\Lambda}\cos\theta\sin^2\theta+\frac{a_4v^2}{\Lambda}\cos\theta\left(\cos^2\theta-2\sin^2\theta\right).\label{eq:l211}
\end{eqnarray}
When kinematically allowed, the coupling $\lambda_{211}$ gives rise to resonant double Higgs production via the decay $h_2\rightarrow h_1h_1$.  The Higgs trilinear coupling $\lambda_{111}$ can alter the nonresonant di-Higgs rate away from SM predictions.

There are important theoretical constraints on the scalar potential, Eq.~(\ref{eq:Pot}), to consider.  Limits from the potential affect the allowed values of $\lambda_{211}$ and can have a significant impact on the $h_2\rightarrow h_1h_1$ branching ratio~\cite{Chen:2014ask}.  First, there are quintic terms $S(\Phi^\dagger\Phi)^2$,\,$S^3\Phi^\dagger\Phi$ and $S^5$ that dominate at large field values and can be negative, indicating an unstable potential.  We only consider parameter space where the global minimum is inside the field value region $|S|<\Lambda$ and $|\phi_0|<\Lambda$ and not along the boundaries.  Above the cutoff scale, it assumed new physics comes in and stabilizes the potential.  Second, the potential is much more complicated than the SM and has many different minimum even inside the allowed field value regions.  The singlet vev cannot contribute to the $W$ and $Z$ masses.  Hence, the Higgs vev must give the correct masses and we only consider parameter space where the global minimum is $\langle \phi_0\rangle=v=246$~GeV and $\langle S\rangle = 0$.  Finally, in the scalar potential we require all dimensionless parameters to be bounded by $4\pi$ and all dimensionful parameters to be bounded by $\Lambda$.

In addition to the scalar potential, the scalar singlet obtains new interactions with SM fermions and gauge bosons~\cite{Dawson:2016ugw,Bauer:2016hcu,Carmona:2016qgo}.   Current measurements of the observed Higgs boson are only sensitive to third generation quarks, and second and third generation leptons.  Hence, we will only consider those interactions in addition to the gauge bosons.  The relevant effective operators in the fermion mass eigenbasis are then:
\begin{eqnarray}
\mathcal{L}_{EFT}&\supset& \frac{g_s^2}{16\pi^2}\frac{f_{GG}}{\Lambda}S\,G_{\mu\nu}^AG^{A,\mu\nu}+\frac{g^2}{16\pi^2}\frac{f_{WW}}{\Lambda}S\,W_{\mu\nu}^aW^{a,\mu\nu}+\frac{{g'}^2}{16\pi^2}\frac{f_{BB}}{\Lambda}S\,B_{\mu\nu}B^{\mu\nu}\label{eq:EFT}\\
&&-\sqrt{2}\left(\frac{f_\mu}{\Lambda}\frac{m_\mu}{v}S\,\overline{L}_2 \Phi\,\mu_R+\frac{f_\tau}{\Lambda}\frac{m_\tau}{v}S\,\overline{L}_3 \Phi\,\tau_R+\frac{f_b}{\Lambda}\frac{m_b}{v}S\,\overline{Q}_3 \Phi\, b_R+\frac{f_t}{\Lambda}\frac{m_t}{v}S\,\overline{Q}_3 \widetilde{\Phi}\,t_R+{\rm h.c.}\right),\nonumber
\end{eqnarray}
where $L_{2,3}$ are second and third generation lepton $SU(2)_L$ doublets, $Q_3$ is the third generation quark $SU(2)_L$ doublet, $\mu_R,\,\tau_R,\,b_R,\,t_R$ are $SU(2)_L$ singlets, and $m_\mu,m_\tau,m_b,m_t$ are the masses of the relevant fermions.  All Wilson coefficients are assumed to be real.  The Feynman rules from Eqs.~(\ref{eq:Pot},\ref{eq:EFT}) can be found in Appendix~\ref{app:FeynRules}.

\section{Power Counting}
\label{sec:PC}
In traditional SMEFT counting, the amplitude squared terms should be truncated to the same order as the Lagrangian.  As an example, consider a baryon and lepton number conserving SMEFT amplitude to dimension-8\footnote{In SMEFT, dimension-5 and dimension-7 operators violate lepton and/or baryon number~\cite{Weinberg:1979sa,Degrande:2012wf,Lehman:2014jma,Henning:2015alf,Kobach:2016ami}.}
\begin{eqnarray}
\mathcal{A}_{\rm SMEFT}\sim \mathcal{A}_{\rm ren}+\frac{1}{\Lambda^2}\mathcal{A}_{6,{\rm SMEFT}}+\frac{1}{\Lambda^4}\mathcal{A}_{8,{\rm SMEFT}}+\mathcal{O}(\Lambda^{-6}),
\end{eqnarray} 
where $\mathcal{A}_{\rm ren}$ is the dimension-4 renormalizable amplitude, and $\mathcal{A}_{n,SMEFT}$ are SMEFT amplitudes originating from operators at dimension-$n$.  The amplitude squared is then
\begin{eqnarray}
|\mathcal{A}_{\rm SMEFT}|^2\sim |\mathcal{A}_{\rm ren}|^2+\frac{1}{\Lambda^2}\mathcal{A}_{\rm ren}\mathcal{A}_{6,{\rm SMEFT}}+\frac{1}{\Lambda^4}|\mathcal{A}_{6,{\rm SMEFT}}|^2+\frac{1}{\Lambda^4}\mathcal{A}_{\rm ren}\mathcal{A}_{8,{\rm SMEFT}}+\mathcal{O}(\Lambda^{-6}).
\end{eqnarray}
As can be clearly seen, at the amplitude squared level, the dimension-8 term is of the same order as the dimension-6 squared term.  Hence, for self-consistency, if only the dimension-6 term is included in the amplitude, then the amplitude squared should also be truncated at $\mathcal{O}(\Lambda^{-2})$.

According to this argument, since the interactions in Eqs.~(\ref{eq:Pot},\ref{eq:EFT}) are truncated at dimension-5 the squared amplitudes should be truncated at $\mathcal{O}(\Lambda^{-1})$.  Here we note that while this is the SMEFT procedure, in the model presented the counting is more complicated due to the unknown scalar mixing angle.  First, consider $h_1$ single production and decay.  The relevant singlet scalar interactions are all dimension-5 or higher.  Hence, to order $\Lambda^{-2}$, amplitudes for $h_1$ production and decay are schematically
\begin{eqnarray}
\mathcal{A}_{h_1}\sim \cos\theta\mathcal{A}_{\rm ren}+\cos\theta\,\frac{\mathcal{A}_{6,{\rm SMEFT}}}{\Lambda^2}+\sin\theta\left(\frac{\mathcal{A}_{5,S}}{\Lambda}+\frac{\mathcal{A}_{6,S}}{\Lambda^2}\right)+\mathcal{O}(\Lambda^{-3}),\label{eq:amph1}
\end{eqnarray}
where $\mathcal{A}_{5,S}$ and $\mathcal{A}_{6,S}$ are, respectively, dimension-5 and dimension-6 operators involving the scalar singlet $S$.  Note that due to mixing among the scalars after EWSB, in the production and decay of the mass eigenstate $h_1$, the SMEFT and renormalizable terms are proportional to $\cos\theta$ and the singlet scalar EFT terms are proportional to $\sin\theta$.  The amplitude squared is then
\begin{eqnarray}
|\mathcal{A}_{h_1}|^2&\sim& \cos^2\theta|\mathcal{A}_{\rm ren}|^2+\sin\theta\,\cos\theta\,\frac{\mathcal{A}_{\rm ren}\mathcal{A}_{5,S}}{\Lambda}\label{eq:Ah1sq}\\
&&+\frac{1}{\Lambda^2}\left(\sin^2\theta |\mathcal{A}_{5,S}|^2+\sin\theta\cos\theta\mathcal{A}_{\rm ren}\mathcal{A}_{6,S}+\cos^2\theta\mathcal{A}_{\rm ren}\mathcal{A}_{6,{\rm SMEFT}}\right)+\mathcal{O}(\Lambda^{-3}).\nonumber
\end{eqnarray}
In the small mixing angle limit, the SM and SMEFT contributions dominate, and the usual power counting is valid.  In the large mixing angle limit, $\sin\theta\rightarrow\pm1$, the $\cos\theta$ terms go to zero and the amplitude squared is
\begin{eqnarray}
|\mathcal{A}_{h_1}|^2\xrightarrow[|\sin\theta|\rightarrow 1]{} \frac{|\mathcal{A}_{5,S}|^2}{\Lambda^2}+\frac{\mathcal{A}_{6,S}\mathcal{A}_{5,S}}{\Lambda^3}+\mathcal{O}(\Lambda^{-4}).\label{eq:amp2h1}
\end{eqnarray}
Hence, the dimension-5 squared piece dominates the dimension-6 terms.  That is, in the large mixing angle limit, we can take the full dimension-5 amplitude squared and not violate power counting rules.

For $h_2$ single production and decay the relevant singlet scalar renormalizable interaction comes from the potential and induces $h_2\rightarrow h_1h_1$ when kinematically allowed.  This process depends on $\lambda_{211}$.  From Eq.~(\ref{eq:l211}), it is clear that the renormalizable piece of $\lambda_{211}$ is proportional to $\sin\theta$.  Hence, all renormalizable contributions to $h_2$ single production and decay are proportional to $\sin\theta$ and the amplitude is schematically
\begin{eqnarray}
\mathcal{A}_{h_2}\sim \sin\theta\mathcal{A}_{\rm ren}+\sin\theta\,\frac{\mathcal{A}_{6,{\rm SMEFT}}}{\Lambda^2}+\cos\theta\left(\frac{\mathcal{A}_{5,S}}{\Lambda}+\frac{\mathcal{A}_{6,S}}{\Lambda^2}\right)+\mathcal{O}(\Lambda^{-3}).
\end{eqnarray}
Now, in the large mixing angle limit $\sin\theta\rightarrow \pm1$, the amplitude becomes SM-like and the SMEFT power counting is correct.  While in the small mixing angle limit the dimension-5 term is the leading term, similar to Eq.~(\ref{eq:amph1}).  Hence, the leading term in the amplitude squared is the dimension-5 squared piece and the full dimension-5 amplitude squared does not violate power counting.  Note that $h_2h_2$ production depends on $\lambda_{221}$ which is not $\sin\theta$ mixing angle suppressed as shown in Eq.~(\ref{eq:l221}) and studied in Ref.~\cite{Chen:2017qcz}.  That is $h_2h_2$ production the power counting changes again.

As this discussion makes clear, the power counting in BSM-EFT depends intimately on the what parameter space is being considered and exactly what processes are under consideration.  We expect that LHC limits will force this model into the small mixing angle limit.  Hence, to test the validity of the EFT, for Higgs precision measurements we will compare $\mathcal{O}(\Lambda^{-1})$ rates to $\mathcal{O}(\Lambda^{-2})$ rates.  For scalar singlet searches we will always keep rates at $\mathcal{O}(\Lambda^{-2})$.

\section{$h_1$ Production and Decay}
\label{sec:ProdDec}
After mixing with the singlet scalar, the observed Higgs boson $h_1$ obtains additional, BSM-EFT couplings to gauge bosons and fermions via Eq.~(\ref{eq:EFT}).  These additional couplings will change the partial widths of $h_1$. In this section, we show the numerical dependence of the relevant branching ratios on the various Wilson coefficients.  

The total width of $h_1$ is
\begin{eqnarray}
\Gamma_1&=&\Gamma(h_1\rightarrow b\overline{b})+\Gamma(h_1\rightarrow c\overline{c})+\Gamma(h_1\rightarrow gg)\label{eq:h1Width}\\
&&+\Gamma(h_1\rightarrow \gamma\gamma)+\Gamma(h_1\rightarrow W^\pm W^{\mp,*})+\Gamma(h_1\rightarrow ZZ^*)\nonumber\\
&&+\Gamma(h_1\rightarrow \tau^+\tau^-)+\Gamma(h_1\rightarrow\mu^+\mu^-).\nonumber
\end{eqnarray}
Higher order QCD corrections are included in the numerical studies.  The partial widths $\Gamma(h_1\rightarrow b\overline{b})$ and $\Gamma(h_1\rightarrow c\overline{c})$ are calculated to next-to-next-to-leading order (NNLO) in QCD~\cite{Gorishnii:1990zu,Chetyrkin:1996sr,Chetyrkin:1996ke,Chetyrkin:1997vj,Djouadi:2005gi}; $\Gamma(h_1\rightarrow\gamma\gamma)$~\cite{Zheng:1990qa,Djouadi:1990aj,Dawson:1992cy,Djouadi:1993ji,Melnikov:1993tj,Inoue:1994jq,Spira:1995rr,Fleischer:2004vb,Harlander:2005rq,Anastasiou:2006hc,Aglietti:2006tp} is calculated at NLO with the exact top mass effects\cite{Spira:1997dg}; and $\Gamma(h_1\rightarrow gg)$~\cite{Djouadi:1991tka,Dawson:1990zj,Spira:1995rr} and $\Gamma(h_1\rightarrow Z\gamma)$~\cite{Spira:1991tj,Bonciani:2015eua,Kara:2015oes,Gehrmann:2015dua} are calculated to NLO in QCD~\cite{Djouadi:2005gi}  by reweighting the exact LO quark loop amplitudes, including all quark mass effects, by the NLO top loop amplitudes calculated in the infinite top quark mass limit:
\begin{eqnarray}
\mathcal{A}_{q,NLO}=\mathcal{A}_{q,LO}\frac{\mathcal{A}_{NLO,m_t\rightarrow \infty}}{\mathcal{A}_{LO,m_t\rightarrow \infty}},
\end{eqnarray}
where $\mathcal{A}_{q,NLO}$ are the quark loop amplitudes we use in our fits, $\mathcal{A}_{q,LO}$ are the exact LO quark loop amplitudes including all quark mass effects, and $\mathcal{A}_{LO,m_t\rightarrow \infty},\mathcal{A}_{NLO,m_t\rightarrow\infty}$  are the LO and NLO top loop amplitudes, respectively, including the top quark contributions calculated in the infinite top quark mass limit.  For $h_1\rightarrow gg$ there is also a three point contribution, $g-g-h_1$, from the BSM-EFT.  The NLO correction differs from the $m_t\rightarrow\infty$ limit of the SM by $(1+11\alpha_s/4\pi)$~\cite{Djouadi:1991tka,Dawson:1990zj,Spira:1995rr,deBlas:2018tjm}, which we take into account.  Finally, for loop level decays $\Gamma(h_1\rightarrow Z\gamma)$ and $\Gamma(h_1\rightarrow\gamma\gamma)$ we include contributions from $t,b,c,\tau,\mu$ and $W$, while for $\Gamma(h_1\rightarrow gg)$ we include $t,b$ and $c$.

In addition to the Higgs boson mass, our input parameters are the same as the LHC Higgs Cross Section Working Group~\cite{deFlorian:2016spz}:
\begin{gather}
G_F=1.16637\times10^{-5}~{\rm GeV}^{-2},\, M_W=80.35797~{\rm GeV},\,M_Z=91.15348~{\rm GeV},\nonumber\\
m_t^{OS}=173~{\rm GeV},\,\overline{m}_b(\overline{m}_b)=4.18~{\rm GeV},\,\overline{m}_c(3~{\rm GeV})=0.986~{\rm GeV},\nonumber\\
m_\tau=1.77682~{\rm GeV},\,m_\mu=0.1056583715~{\rm GeV},\nonumber\\
\alpha_s(M_Z)=0.118,\nonumber
\end{gather}
where bars indicate $\overline{\rm MS}$ parameters, the superscript $OS$ indicates parameters evaluated in the on-shell scheme, and masses inside parentheses indicate the renormalization scale at which the parameters are set.  

For calculating $h_1$ decay rates, we use one renormalization scale for all parameters, including running quark masses. We set the renormalization scale to the Higgs mass for tree level processes, while for loop level processes we set the renormalization scale to half of the Higgs mass.  For partial widths to quarks, light quark masses are evaluated in the $\overline{\rm MS}$ scheme. For the partial widths $\Gamma(h_1\rightarrow Z\gamma)$ and $\Gamma(h_1\rightarrow gg)$, pole masses for all quarks are used, while for the loop level $\Gamma(h_1\rightarrow \gamma\gamma)$, the running $\overline{\rm MS}$ masses normalized to the pole masses are used\cite{Spira:1997dg}. For this purpose, we use the relationship between the the on-shell and $\overline{\rm MS}$ masses in Ref.~\cite{Djouadi:2005gi}.

\begin{figure}[tb]%
\begin{center}
\subfigure[]{\includegraphics[width=0.45\textwidth,clip]{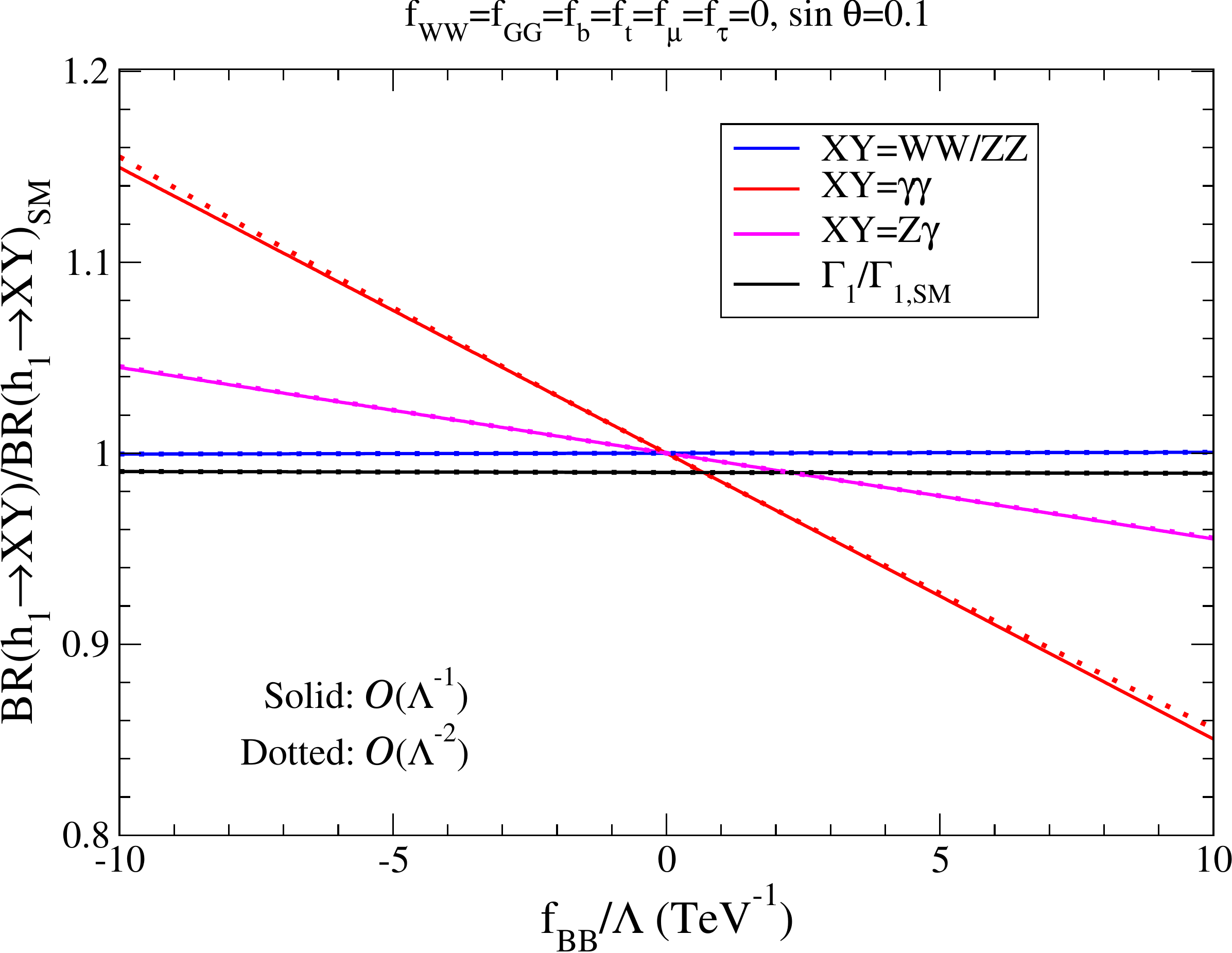}}
\subfigure[]{\includegraphics[width=0.45\textwidth,clip]{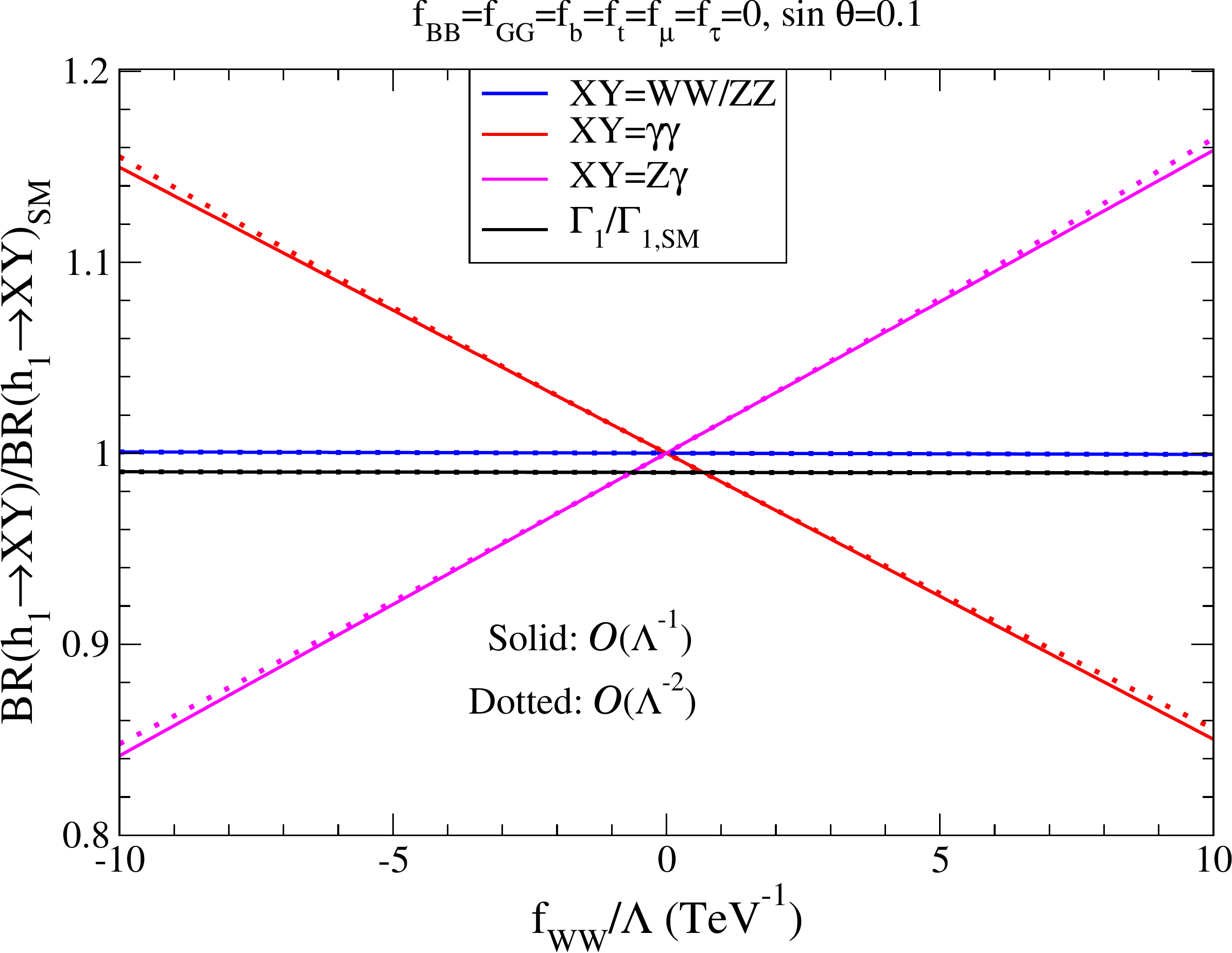}}
\subfigure[]{\includegraphics[width=0.45\textwidth,clip]{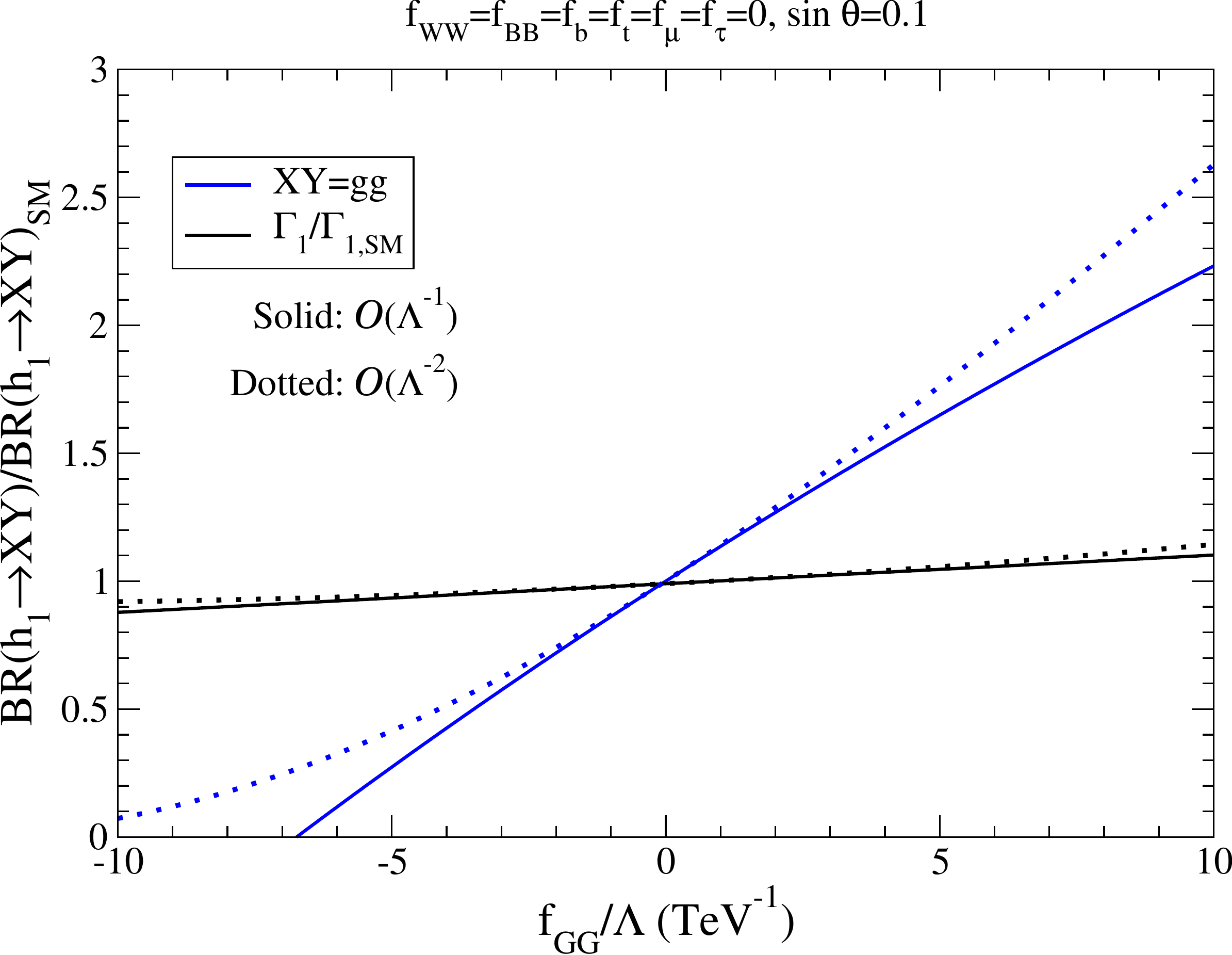}}
\end{center}
\caption{\label{fig:DecGauge} Dependence of various branching ratios of $h_1$ and $h_1$ total width normalized to their SM values as a function of the gauge boson Wilson coefficients (a) $f_{BB}$, (b) $f_{WW}$, and (c) $f_{GG}$.  All other Wilson coefficients are set to zero.  The partial widths are calculated at both (solid) $\mathcal{O}(\Lambda^{-1})$ and (dotted) $\mathcal{O}(\Lambda^{-2})$.  Different colors indicate different final states.  }
\end{figure}

\begin{figure}[tb]%
\begin{center}
\subfigure[]{\includegraphics[width=0.45\textwidth,clip]{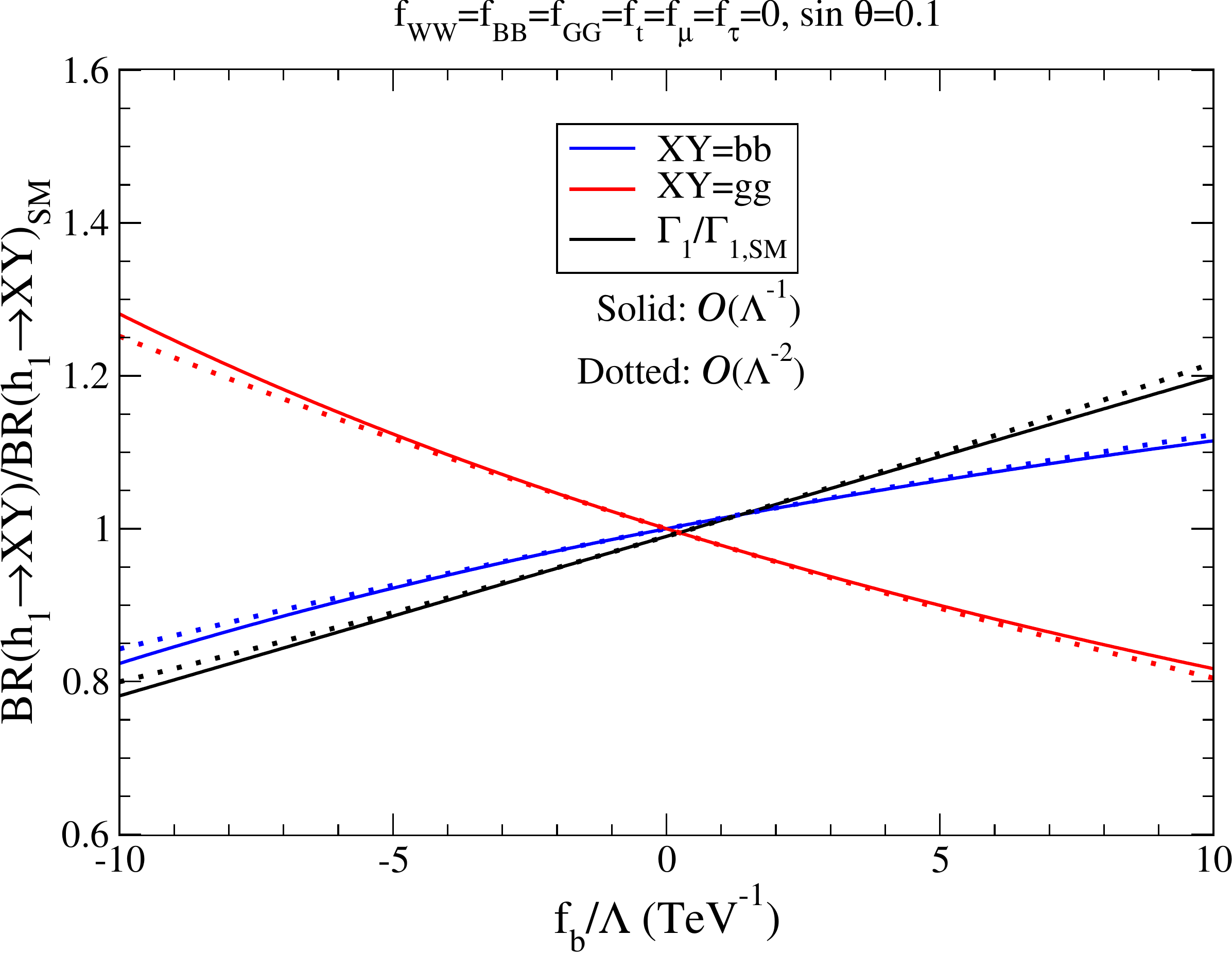}}
\subfigure[]{\includegraphics[width=0.45\textwidth,clip]{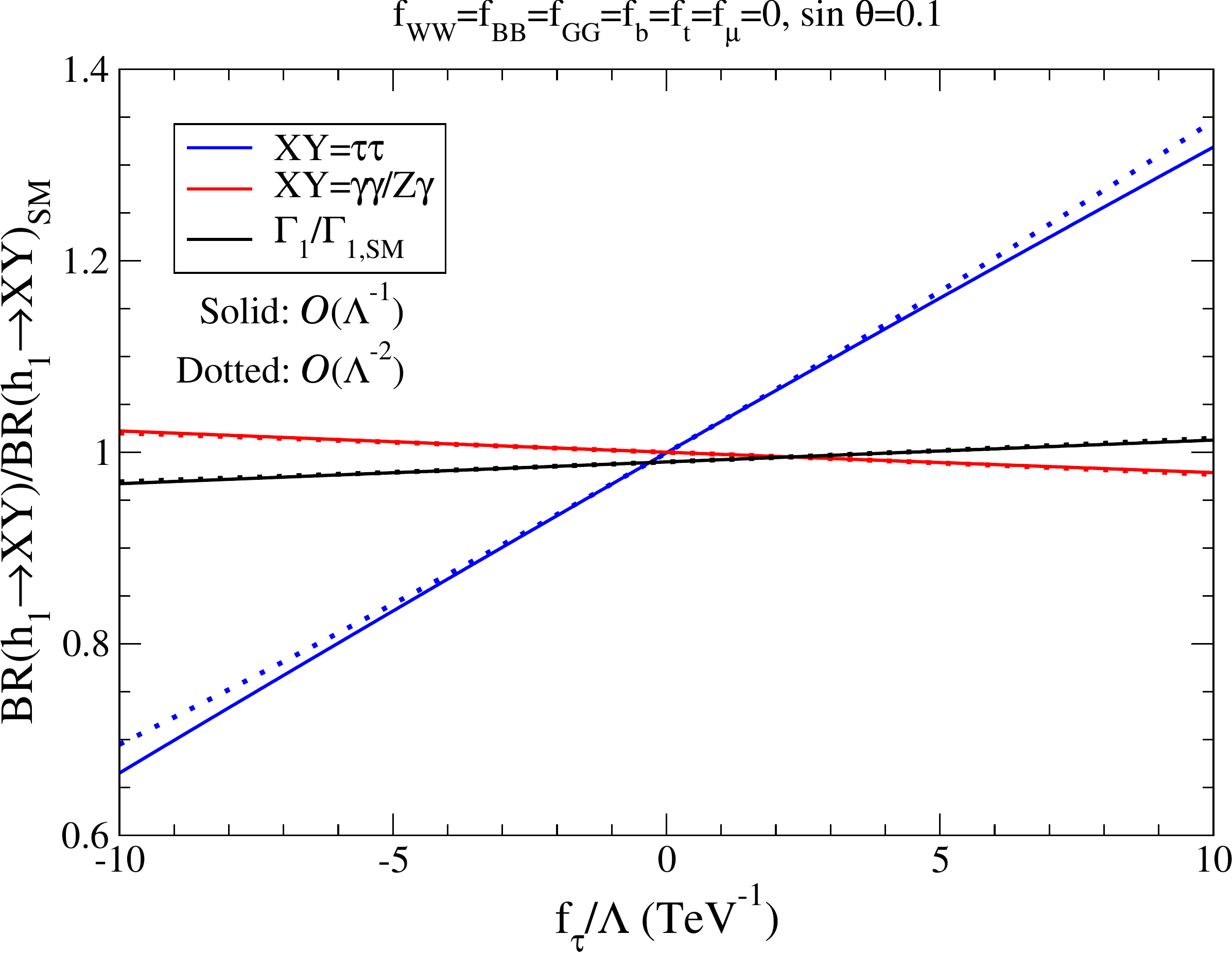}}
\subfigure[]{\includegraphics[width=0.45\textwidth,clip]{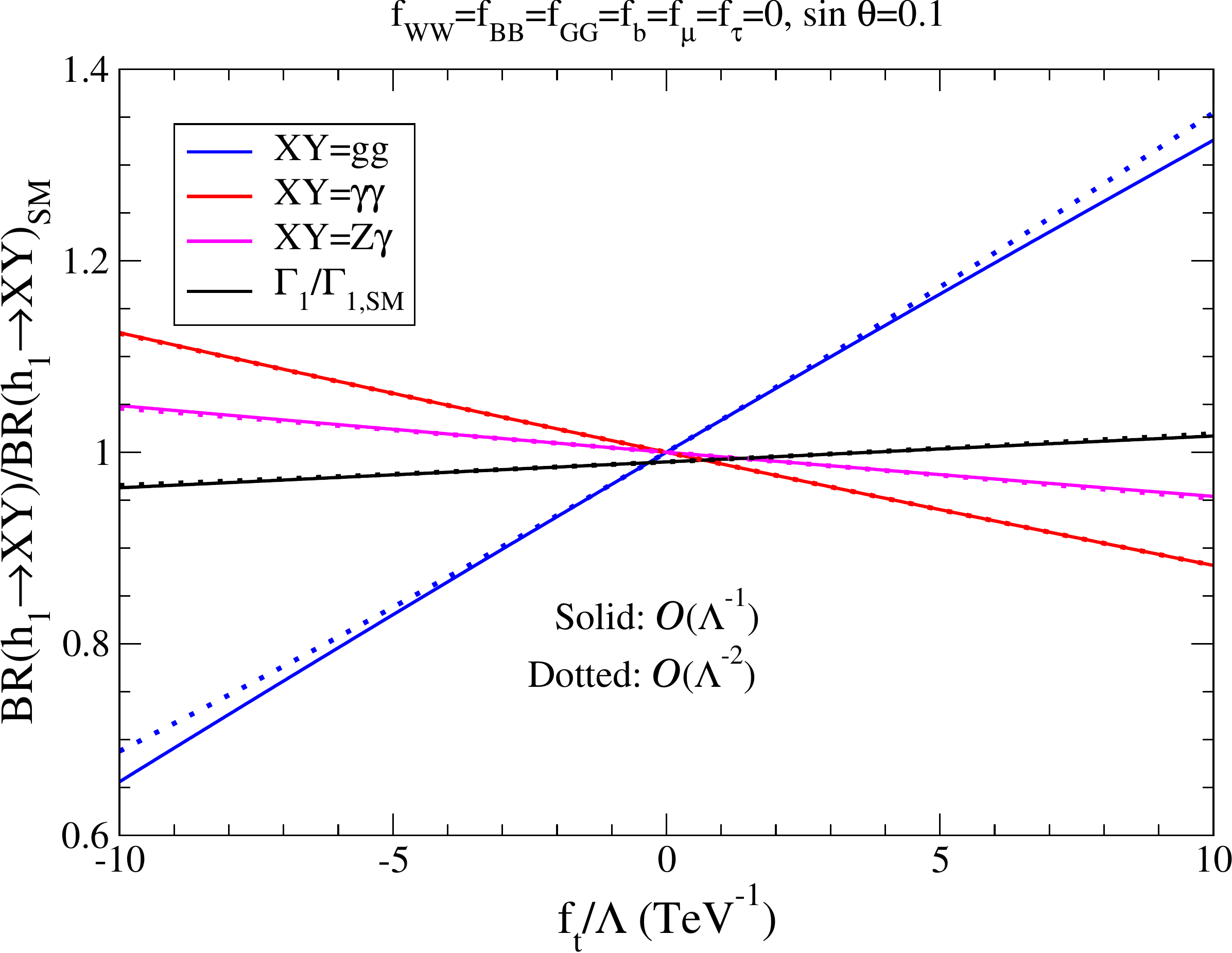}}
\end{center}
\caption{\label{fig:DecFerm} Same as Fig.~\ref{fig:DecGauge} for the fermion Wilson coefficients (a) $f_{b}$, (b) $f_{\tau}$, and (c) $f_{t}$.}
\end{figure}

In Figs.~\ref{fig:DecGauge} and~\ref{fig:DecFerm}, we show the dependence of various Higgs branching ratios and the total Higgs width on the gauge boson and fermion Wilson coefficients with a scalar mixing angle of $\sin\theta=0.1$.  We consider one Wilson coefficient at a time and show branching ratios for which the Wilson coefficients make a direct contribution to the partial widths.  Additionally, deviations in the total width are important in fits to Higgs precision data since $\Gamma_1$ enters all Higgs branching ratios.  Hence, we also show the dependence of the total width on the Wilson coefficients.

From Fig.~\ref{fig:DecGauge}, it is clear that the $WW$ and $ZZ$ partial widths have very little dependence on the Wilson coefficients $f_{BB}$ and $f_{WW}$.  This can be understood by noting that these decays are tree level in the SM, while the EFT contributions are suppressed by a loop factor, a small mixing angle, and a heavy scale.  The situation changes for SM loop level decays.  Both $h_1\rightarrow\gamma\gamma$ and $h_1\rightarrow Z\gamma$ depend strongly on $f_{BB}$ and $f_{WW}$, with deviations from SM predictions up to $15\%$.  Similarly, $h_1\rightarrow gg$ strongly depends on $f_{GG}$, with order one deviations from the SM.  Finally, the total width has little dependence on $f_{BB}$ and $f_{WW}$ since $\gamma\gamma$ and $Z\gamma$ have negligible contributions to $\Gamma_1$.  However, $\Gamma_1$ depends more strongly on $f_{GG}$ due to the larger $h_1\rightarrow gg$ partial width.  These results are consistent with Ref.~\cite{Dawson:2016ugw}.

The branching ratios into fermionic final states depend strongly on the fermion Wilson coefficients, as evidenced in Figs.~\ref{fig:DecFerm}(a,b).  The decay to $\tau\tau$ varies as much as $\sim30\%$ from SM predictions, while the bottom quark final state varies by around $\sim 20\%$.  It is striking that $h_1\rightarrow bb$ depends less on $f_b$ than $h_1\rightarrow\tau\tau$ depends on $f_\tau$.  This can be understood by noting that the total width of $h_1$ depends strongly on $f_b$, but very little on $f_\tau$.  Hence, the variation in the $h_1\rightarrow bb$ partial width is somewhat compensated by the variation in $\Gamma_1$.  Similarly, while the partial width of $h_1\rightarrow gg$ has little dependence on $f_b$, the ${\rm BR}(h_1\rightarrow bb)$ varies up to $\sim 20-30\%$ due to the variation in the total width.  Finally, all the loop level processes depend relatively strongly on the top quark Wilson coefficient, as seen in Fig.~\ref{fig:DecFerm}(c).  In particular, ${\rm BR}(h_1\rightarrow\gamma\gamma)$ and ${\rm BR}(h_1\rightarrow gg)$ vary upward of $\sim15\%$ and $\sim30$\%, respectively.  

Finally, the branching ratios and widths calculated to (solid) $\mathcal{O}(\Lambda^{-1})$ and (dotted) $\mathcal{O}(\Lambda^{-2})$ are shown in Figs.~\ref{fig:DecGauge} and~\ref{fig:DecFerm}.  For most final states and the total width, both the $\mathcal{O}(\Lambda^{-1})$ and $\mathcal{O}(\Lambda^{-2})$ results agree well.  This indicates that the BSM-EFT is valid in these regions of parameter space.  The only exception is the dependence of $h_1\rightarrow gg$ on $f_{GG}$.  However, as we will show in the next section, the fits to the Higgs precision data also indicate the BSM-EFT is valid in the allowed parameter regions.

While we do not explicitly show the variation of the Higgs production cross section, it should be noted that gluon fusion is the main production mode.  For on-shell $h_1$ decay, the LO gluon fusion production rate is
\begin{eqnarray}
\sigma_{ggF}(pp\rightarrow h_1)=\frac{\pi^2}{8\,m_1\,S}\,L\,\Gamma(h_1\rightarrow gg),\label{eq:ggF}
\end{eqnarray}
where the parton luminosity is
\begin{eqnarray}
L=\int_{\ln(\sqrt{\tau_0})}^{-\ln(\sqrt{\tau_0})}dy\,g(\sqrt{\tau_0}e^y)\,g(\sqrt{\tau_0}e^{-y}),
\end{eqnarray}
where $\sqrt{S}$ is the hadronic center-of-momentum energy and $\tau_0=m_1^2/S$.  As shown in Figs.~\ref{fig:DecGauge}(c) and~\ref{fig:DecFerm}(c), this production rate will have a strong dependence on $f_{GG}$ and $f_t$.  Other subdominant but important production modes are Higgs production in association with $W/Z$ ($Wh_1/Zh_1$) and vector boson fusion (VBF).  The relevant Wilson coefficients for these production modes are $f_{BB}$ and $f_{WW}$.  However, as evidenced in Figs.~\ref{fig:DecGauge}(a,b) $Wh_1$, $Zh_1$, and VBF production will have little dependence on $f_{BB}$ and $f_{WW}$.

\section{Higgs Signal Strengths}
\label{sec:Higgs}
Now we perform a fit to the Higgs precision data.  The effects of the additional interactions on the Higgs measurements are parameterized using Higgs signal strengths:
\begin{eqnarray}
\mu_i^f=\frac{\sigma_i(pp\rightarrow h_1)}{\sigma_{i,SM}(pp\rightarrow h_1)} \frac{{\rm BR}(h_1\rightarrow f)}{{\rm BR}_{SM}(h_1\rightarrow f)},
\end{eqnarray}
where $i$ is the initial state, $f$ is the final state, and the subscript $SM$ indicates SM values.  We combine the signal strengths into a chi-square:
\begin{eqnarray}
\chi^2_{h_1}=\sum_{i,j}\frac{(\mu_i^f-\hat{\mu}_i^f)^2}{(\delta_i^f)^2},\label{eq:Chi2h1}
\end{eqnarray}
where $\mu_i^f$ is a calculated signal strength, $\hat{\mu}_i^f$ is a signal strength measured at the LHC, and $\delta_i^f$ is the one standard deviation uncertainty on $\hat{\mu}_i^f$.  We combine measurements from both ATLAS and CMS at the 13 TeV LHC.  The set of signal strengths we use can be found in Tables~\ref{tab:ATLAS} and~\ref{tab:CMS} in Appendix~\ref{app:data}.  

\begin{figure}[htb]
\begin{center}
\subfigure[]{\includegraphics[width=0.4\textwidth,clip]{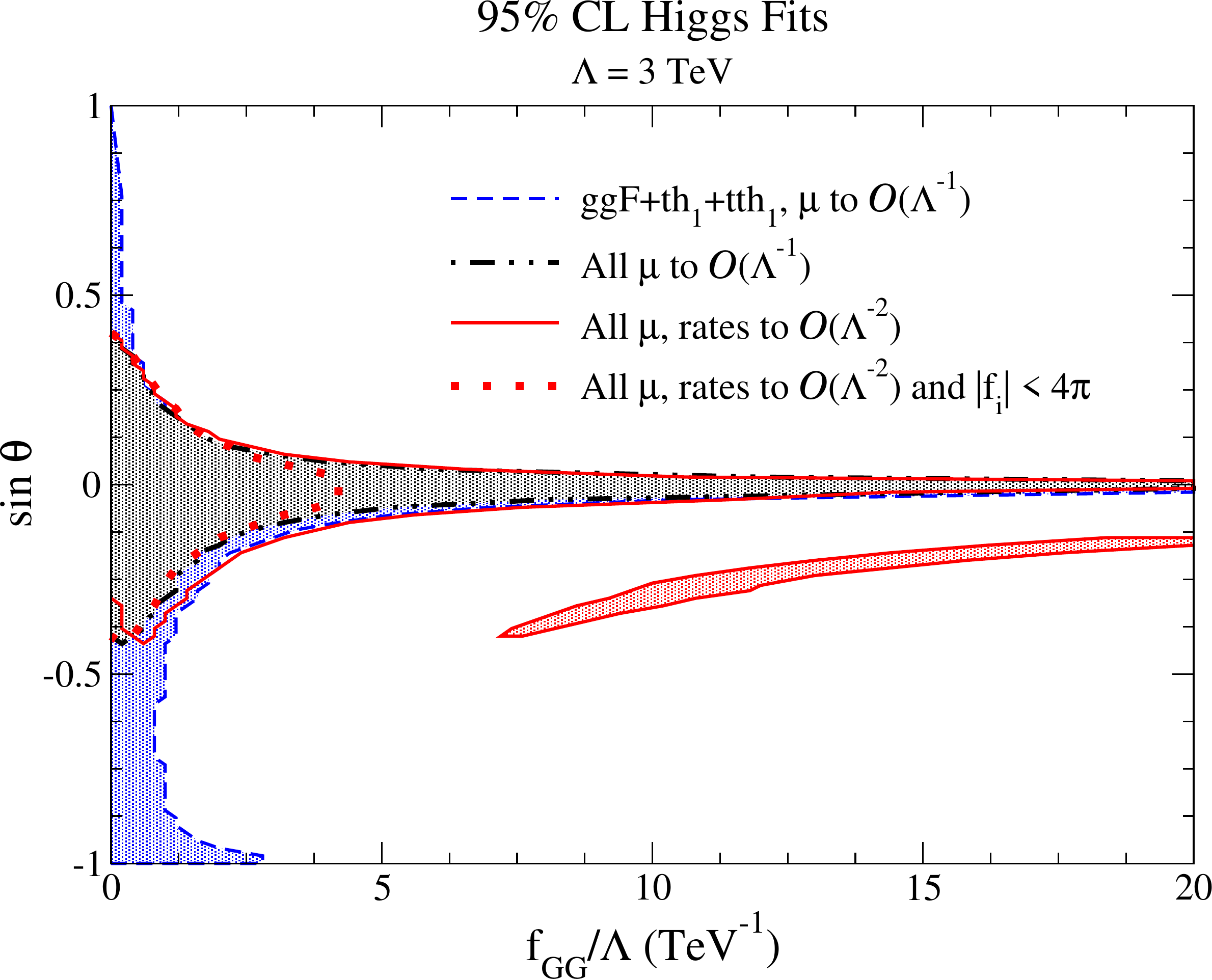}}
\subfigure[]{\includegraphics[width=0.4\textwidth,clip]{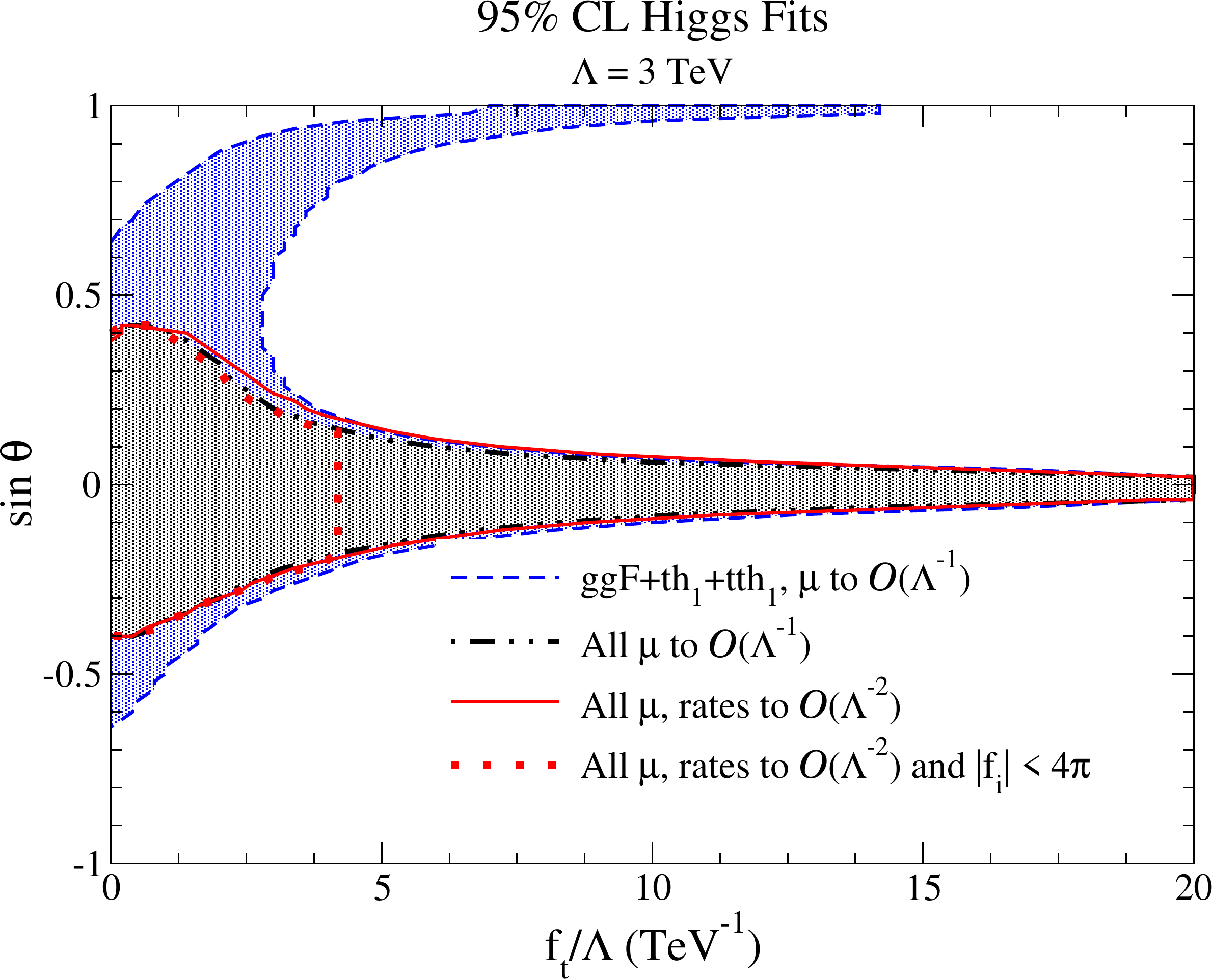}}\\\vspace{-0.1in}
\subfigure[]{\includegraphics[width=0.4\textwidth,clip]{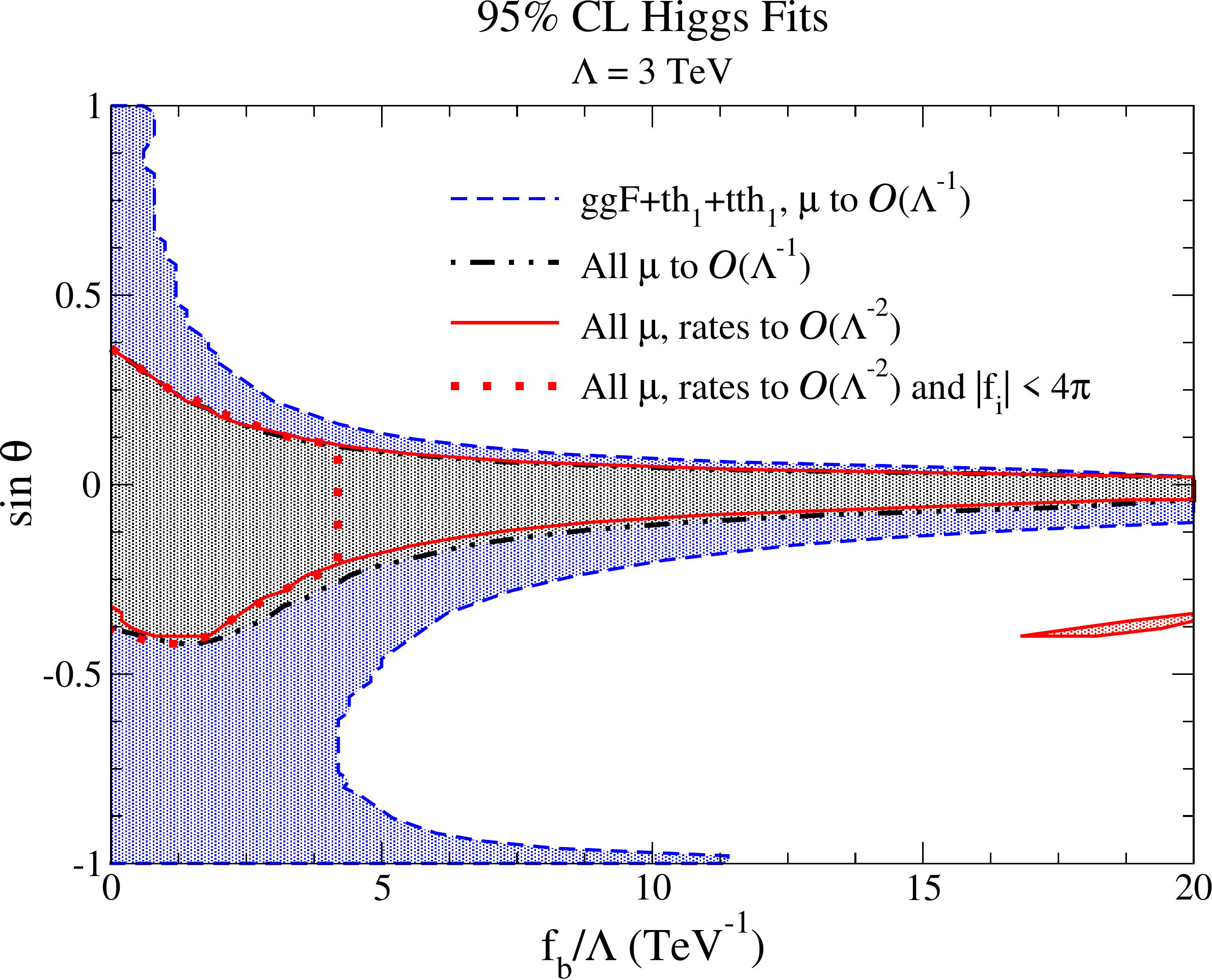}}
\subfigure[]{\includegraphics[width=0.4\textwidth,clip]{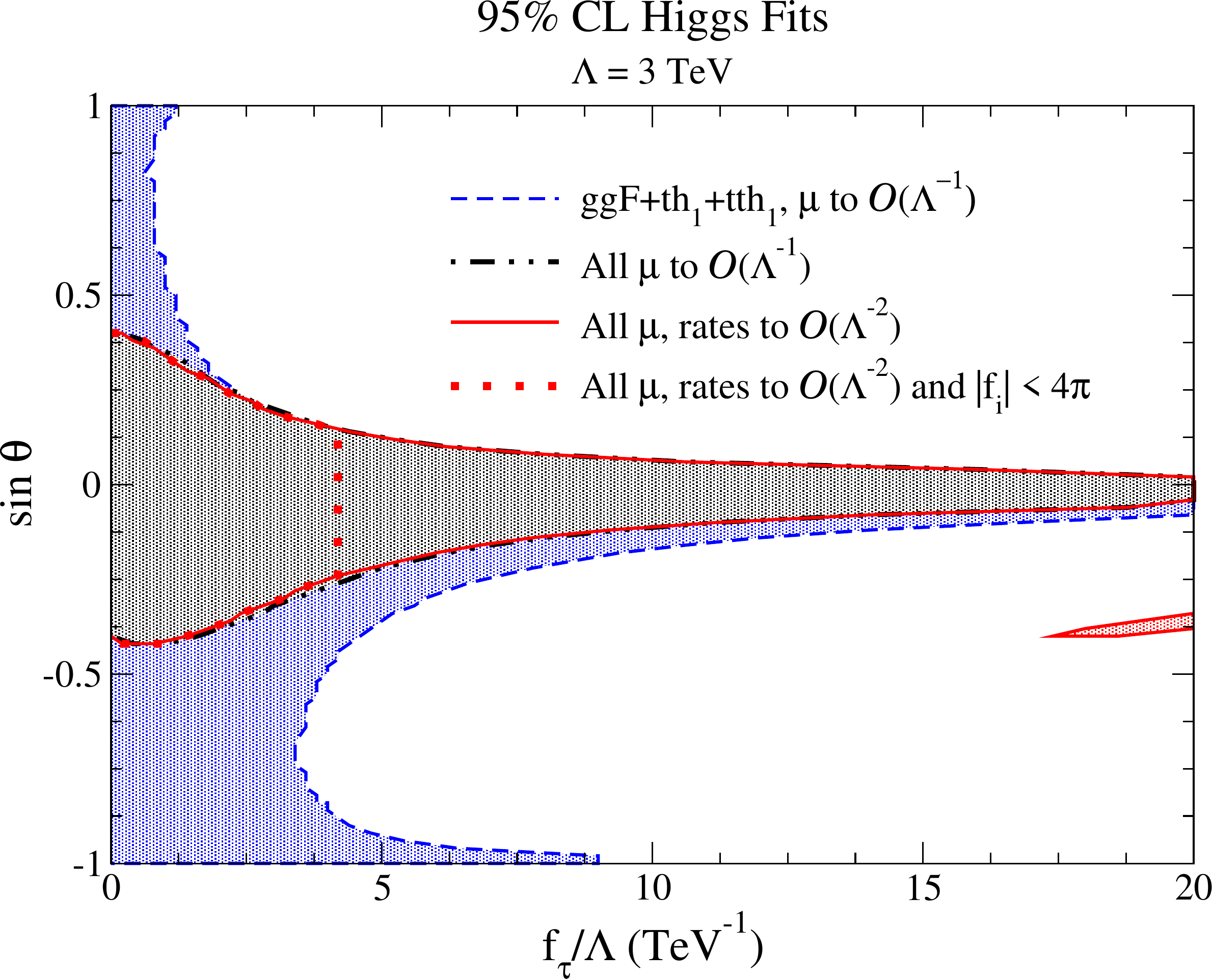}}\\\vspace{-0.1in}
\subfigure[]{\includegraphics[width=0.4\textwidth,clip]{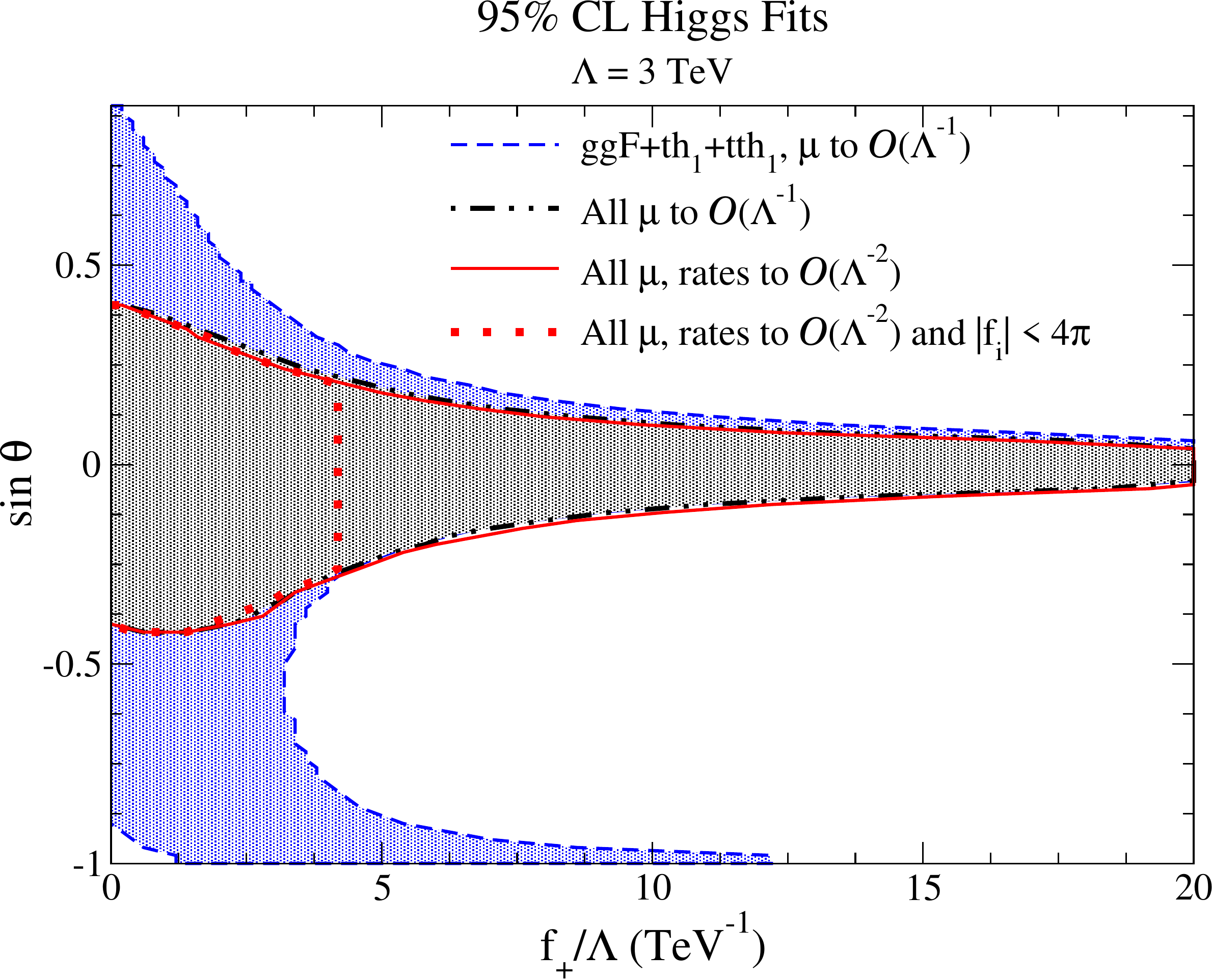}}
\subfigure[]{\includegraphics[width=0.4\textwidth,clip]{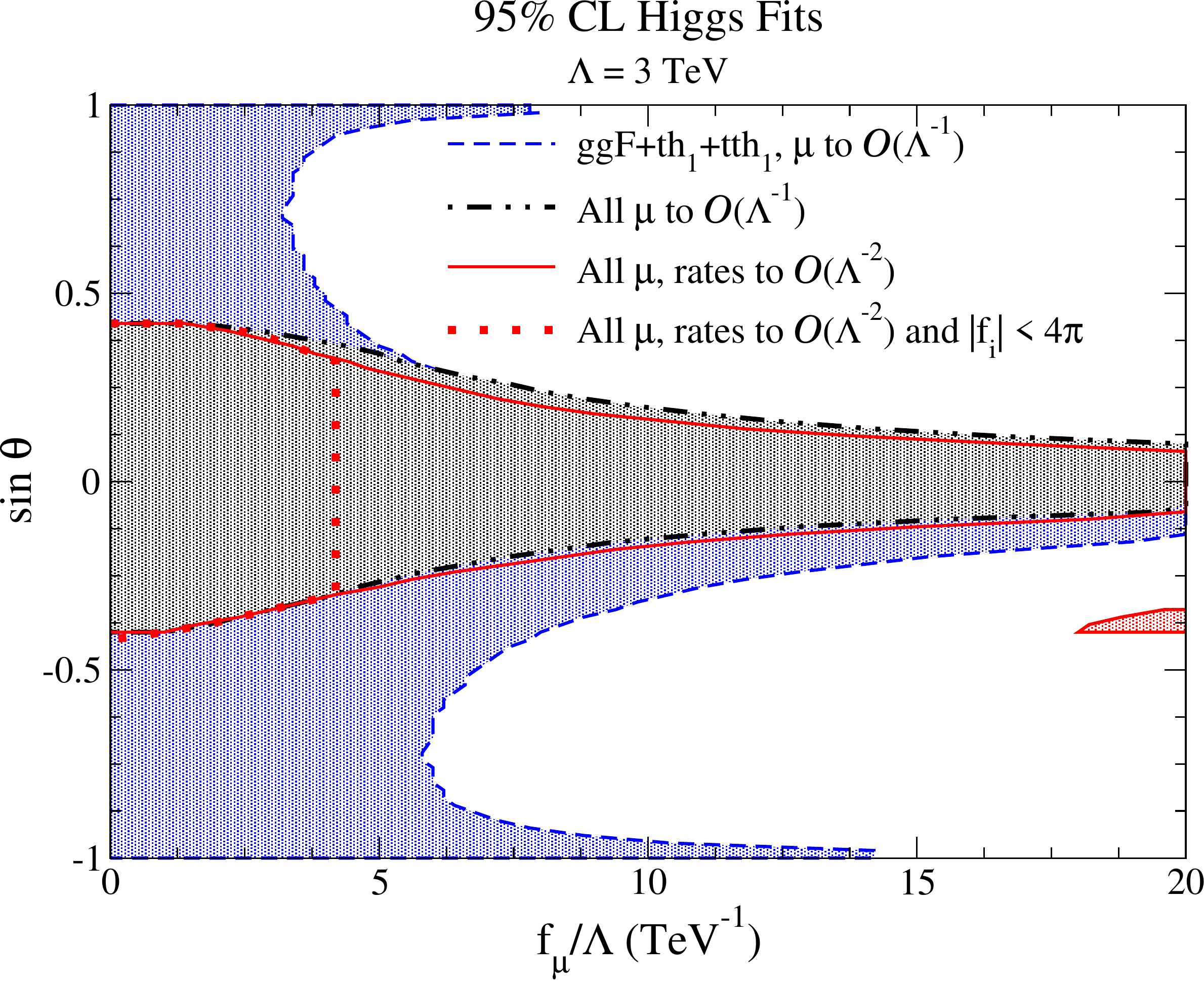}}\\\vspace{-0.3in}
\end{center}
\caption{\label{fig:Higgs} 95\% CL regions from Higgs precision data for $\sin\theta$ vs (a) $f_{GG}$, (b) $f_t$, (c) $f_b$, (d) $f_\tau$, (e) $f_+$, and (f) $f_\mu$.  Parameters not shown in a plot are profiled over. For $\mu^i_f$ expanded to $\mathcal{O}(\Lambda^{-1})$: (blue dashed) $tth_1+th_1+ggF$ initial states only, and   (black solid) all initial and final states.  For widths and cross sections kept to $\mathcal{O}(\Lambda^{-2})$: (red solid) all initial and final states and (red dotted) keeping $|f_i|<4\pi$  at a new physics scale of $\Lambda=3$~TeV.  Regions inside contours are allowed.  
}
\end{figure}

For the gluon fusion (ggF) production rate, the NLO cross section results are only known for the effective gluon and top quark couplings~\cite{Deutschmann:2017qum}.  To be consistent across our fits, we also include the bottom quark EFT couplings.  Since these cross sections are not available, we use the approximation
\begin{eqnarray}
\frac{\sigma_{ggF}(pp\rightarrow h_1)}{\sigma_{ggF,SM}(pp\rightarrow h_1)}\approx\frac{\Gamma(h_1\rightarrow gg)}{\Gamma_{SM}(h_1\rightarrow gg)},\label{eq:ggF1}
\end{eqnarray}
to calculate the contributions from $f_b$.  The validity of this approximation for top quark-Higgs and gluon-Higgs effective operators is discussed in Appendix~\ref{app:WidthXSApprox}.  We note that this is a standard approximation~\cite{Dawson:2016ugw,deBlas:2018tjm}, and is indeed good to $~\sim3-5\%$ for most top quark-Higgs and gluon-Higgs EFT contributions. In our fits we use the NLO cross section results of Ref.~\cite{Deutschmann:2017qum} to calculate the $f_t$ and $f_{GG}$ contributions to the signal strength, and the approximation of Eq.~(\ref{eq:ggF1}) for the $f_b$ contributions:
\begin{eqnarray}
\frac{\sigma_{ggF}(pp\rightarrow h_1)}{\sigma_{ggF,SM}(pp\rightarrow h_1)}&=&\cos^2\theta+\frac{\cos\theta\,\sin\theta}{(\Lambda/{\rm TeV})}\left(0.517\,f_t+1.45\,f_{GG}-0.0281\,f_b\right)\nonumber\\
&&+\frac{\sin^2\theta}{(\Lambda/{\rm TeV})^2}\left(0.0626\,f_t^2+0.492\,f_{GG}^2+8.53\times10^{-4}\,f_b^2\right.\nonumber\\
&&\left.+0.351\,f_t\,f_{GG}-8.62\times10^{-3}\,f_b\,f_t-0.0226\,f_b\,f_{GG}\right).
\end{eqnarray}
Corrections up to N$^3$LO in QCD are known for gluon fusion~\cite{Harlander:2016hcx,Anastasiou:2016hlm,Brooijmans:2016vro} and Higgs decays to gluons~\cite{Contino:2014aaa,Djouadi:2018xqq}.
We also include $Wh_1$, $Zh_1$, Higgs production in association with a $t\overline{t}$ pair ($t\overline{t}h_1$), Higgs production in association with a top plus jet or top plus W (collectively $th_1$), and VBF.  For these production modes the model is implemented in \texttt{MadGraph5\_aMC@NLO}~\cite{Alwall:2014hca} via \texttt{FeynRules}~\cite{Alloul:2013bka}.  The default \texttt{NNPDF2.3LO} pdf sets~\cite{Ball:2012cx} are used and for $h_1$ production modes the renormalization and factorization scales are set to the sum of the final state particle masses.  For the VBF mode, we apply the cuts~\cite{Azzi:2019yne}
\begin{eqnarray}
p_T^j>20~{\rm GeV},\, |\eta^j|<5,\,|\Delta\eta^{jj}|>3,\,{\rm and}\, m_{jj}>130~{\rm GeV},
\end{eqnarray}
where $p_T^j$ are jet transverse momenta, $\eta^j$ are jet pseudorapidity, $\Delta\eta^{jj}$ is the difference in the jet pseudorapidity, and $m_{jj}$ is the di-jet invariant mass.  These production and decay modes are calculated at LO in QCD, and it is hoped that most of the QCD corrections cancel in the ratio of the cross sections used for the signal strengths.  However, it should be pointed out that in the SMEFT, for some observables the QCD corrections can be strongly dependent upon the EFT operators~\cite{Baglio:2017bfe,Baglio:2018bkm,Baglio:2019uty,Baglio:2020oqu}.

Once all the signal strengths are known, we perform a fit to the Wilson coefficients and scalar mixing angle.  As shown in Appendix~\ref{app:FeynRules}, the $h_1\rightarrow\gamma\gamma$ decay depends on the combination $f_{BB}+f_{WW}$ but not $f_{BB}-f_{WW}$.  Also, Fig.~\ref{fig:DecGauge} shows that processes with external $W$ and $Z$ bosons do not depend strongly on $f_{WW}$ and $f_{BB}$.  Hence, we define 
\begin{eqnarray}
f_{\pm}=\frac{1}{\sqrt{2}}\left(f_{BB}\pm f_{WW}\right).
\end{eqnarray}
Now $h_1\rightarrow \gamma\gamma$ will constrain {\it only} $f_+$, and $Wh_1,\,Zh_1,\,$VBF$,\,h_1\rightarrow WW^*,$ and $h_1\rightarrow Z Z^*$ have negligible dependence on both $f_{\pm}$. Hence, we set $f_-=0$.  Hence, the following parameters are fit using just Higgs signal strengths:
\begin{eqnarray}
\sin\theta,\,f_{GG},\,f_+,\,f_b,\,f_t,\,f_\mu,\,{\rm and}\,f_\tau.
\end{eqnarray}

In Fig.~\ref{fig:Higgs} we show the results of the $\chi^2$ fits to Higgs data at 95\% CL.  As can be seen from Eq.~(\ref{eq:Ah1sq}), the squared amplitudes are invariant under the simultaneous parity transformations: $\sin\theta\rightarrow -\sin\theta$ and all Wilson coefficients $f_i\rightarrow -f_i$.  Hence, only $f_i>0$ results are shown and contain all the information.  The results are shown for  (blue-dashed) just $tth_1+th_1$+ggF production modes and (black/red) including all signal strengths.  As can be seen, if only gluon fusion and Higgs production in association with top quarks are used, all values of $\sin\theta$ are allowed.  This is because ggF,$\,tth_1,$ and $th_1$ depend relatively strongly on $f_t$ and $f_{GG}$.  Hence, deviations in $\sin\theta$ can be compensated for by changes in $f_t$ and $f_{GG}$.  The major effect of vector boson fusion and Higgs production in association with $W^\pm$ or $Z$ is to eliminate the largest $\sin\theta$ regions.  As discussed above in Sec.~\ref{sec:ProdDec}, VBF,$\,Wh_1,$ and $Zh_1$ do not depend strongly on the Wilson coefficients\footnote{This is also true in SMEFT~\cite{Corbett:2015ksa}.}.  Hence, the production rates for these modes are approximately the SM rate suppressed by the mixing angle $\cos^2\theta$:
\begin{eqnarray}
\sigma_{VBF/Vh_1}(pp\rightarrow h_1)\approx \cos^2\theta\, \sigma_{VBF/Vh_1,SM}(pp\rightarrow h_1),
\end{eqnarray}
where $Vh_1=Zh_1,Wh_1$.  Limits on these production rates then essentially place limits on the scalar mixing angle.  Additionally, at large Wilson coefficient values, fits including only $ggF,tth_1,$ and $th_1$ agree well with the full fit, except for $f_b$, $f_\mu$, and $f\tau$.  This is can be understood by noting that the strong constraints on $h_1\rightarrow \tau^+\tau^-$, $h_1\rightarrow\mu^+\mu^-$, and $h_1\rightarrow bb$ come from VBF,$\,Wh_1,$ and $Zh_1$.

Figure~\ref{fig:Higgs} also compares various calculations of the full $\chi^2$ to determine the validity of the limits on the BSM-EFT.  As discussed in Sec.~\ref{sec:PC}, for small mixing angles, the power counting of $h_1$ production and decay is expected to follow the usual power counting of the SMEFT.  
  To check the validity of the power counting in our fits, Fig.~\ref{fig:Higgs} shows the (black dash-dot-dot) signal strengths expanded to $\mathcal{O}(\Lambda^{-1})$ and  (red solid) signal strengths calculated by keeping cross sections and widths to $\mathcal{O}(\Lambda^{-2})$.  For the bulk of the distributions, these two scenarios largely agree with each other showing that the the limits are compatible with the BSM-EFT power counting.  However, if the full $\Lambda^{-2}$ dependence is kept, new 95\% CL regions open up at large Wilson coefficients.

From general perturbativity arguments, it is expected that the Wilson coefficients are bounded $|f_i|<4\pi$.  The red dotted contours in Fig.~\ref{fig:Higgs} show the results of requiring $|f_i|<4\pi$ for a new physics scale of $\Lambda=3$~TeV.  Comparing to the red solid lines, it can be seen that in the relevant regions, the fits are consistent with perturbative Wilson coefficients.  Comparing the power counting and perturbativity constraints, it is clear that the new allowed parameter regions that appear at $\mathcal{O}(\Lambda^{-2})$ but not $\mathcal{O}(\Lambda^{-1})$ are not consistent with perturbativity.  Hence, imposing the perturbativity constraints automatically guarantees Higgs precision measurements are fully compatible with BSM-EFT power counting.

\begin{figure}[tb]
\begin{center}
\includegraphics[width=0.45\textwidth,clip]{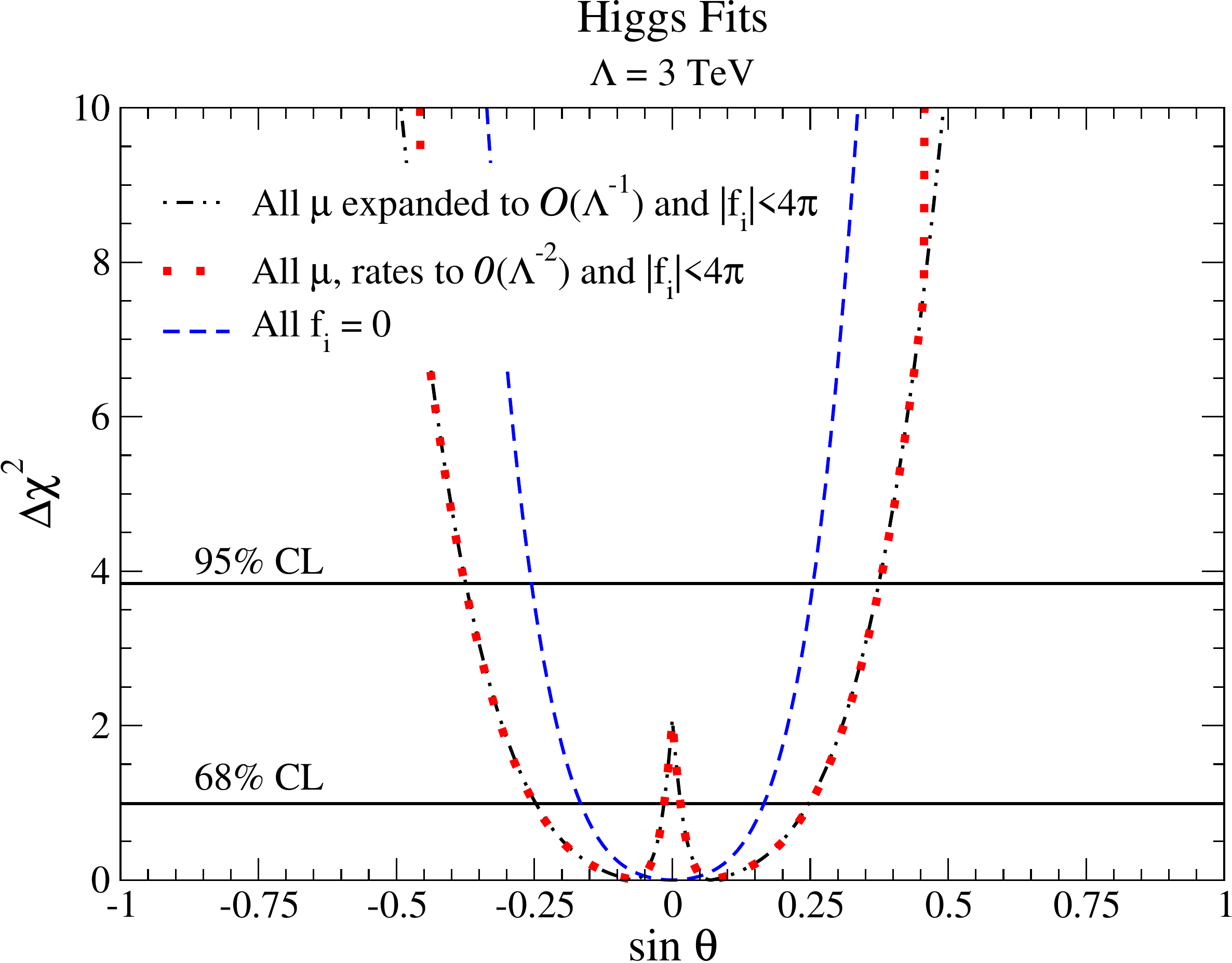}
\end{center}
\caption{\label{fig:sth} One-dimensional fits to $\sin\theta$ using Higgs signal strengths with all other parameters profiled over for (black dash-dot-dot) signal strengths expanded to $\mathcal{O}(\Lambda^{-1})$, (red dotted) cross sections and widths kept at $\mathcal{O}(\Lambda^{-2})$, and (blue dashed) all dimension-5 operators set to zero.  The new physics scale is $\Lambda=3$~TeV and Wilson coefficients are required to be $|f_i|<\pi$.}
\end{figure}

Finally, in Fig.~\ref{fig:sth}, we show the one-dimensional fits to $\sin\theta$ using Higgs data with dimension-5 operators and in the renormalizable singlet extend SM without dimension-5 operators.  As can be seen, even with a new physics scale of $\Lambda=3$~TeV, the dimension-5 operators make a substantial impact on the interpretation of Higgs data.  Also, consistently expanding signal strengths to $\mathcal{O}(\Lambda^{-1})$ gives the same result as keeping all cross sections and widths to $\mathcal{O}(\Lambda^{-2})$.  The conclusion is that the BSM-EFT is valid for Higgs measurements and, even if we assume a new physics scale beyond the current reach of the LHC, the effects of this new physics on the singlet extended SM cannot be ignored.

\section{Including Heavy Resonance Searches}
\label{sec:Heavy}
Heavy scalars are searched for regularly at the LHC. The EFT couplings of $h_1,h_2$ are inherited by the mixing of the SM Higgs with the scalar $S$.  Hence, in production and decay of $h_1$, the Wilson coefficients and mixing angle always appear in the combination $\sin\theta f_i$ whereas in the production and decay of $h_2$, they appear in the combination $\cos\theta\,f_i$.  As a result, heavy resonance searches are expected to give complementary information to Higgs signal strengths.

First, we describe how heavy resonance searches are incorporated into our $\chi^2$ fits, then we give the results.

\subsection{$\chi^2$ for Heavy Resonance Searches}
\label{sec:chi2}
Similar to the Higgs signal strengths, it is assumed that scalar resonance searches are Gaussian and a $\chi^2$ fit is performed:
\begin{eqnarray}
\left(\chi^{f}_{i,h_2}\right)^2=\left(\frac{\sigma_i^f-\hat{\sigma}_i^f}{\delta \sigma_i^f}\right)^2,
\end{eqnarray}
where $(\chi^{f}_{i,h_2})^2$ is a chi-square of a single $h_2$ process, $\sigma_i^f$ is the calculated cross section for initial state $i$ into final state $f$, $\hat{\sigma}_i^f$ is the measured cross section at the LHC, and $\delta\sigma_i^f$ is the one standard deviation uncertainty on $\hat{\sigma}_i^f$.  To calculate the cross section, both the SM rate as well as the new physics contribution must be included.  Using the narrow width approximation, we have
\begin{eqnarray}
\sigma_i^f=\sigma_{i,SM}^f+\sigma(i\rightarrow h_2){\rm BR}(h_2\rightarrow f).
\end{eqnarray}
While applying experimental bounds using the narrow width approximation is standard~\cite{Bechtle:2011sb,Bechtle:2013wla,Bechtle:2020pkv}, depending on the region of parameter space, the interference between the SM and heavy scalar resonances can be $\mathcal{O}(10\%)$ for $gg\rightarrow ZZ$, $gg\rightarrow W^+W^-$, and $gg\rightarrow h_1h_1$~\cite{Kauer:2015hia,Dawson:2015haa,Greiner:2015ixr,Carena:2018vpt} even in the on-shell region.  The $gg\rightarrow ZZ$ and $gg\rightarrow WW$ processes contribute $\sim 5\%$ to the total $ZZ$ and $WW$ rates~\cite{Dicus:1987dj,Glover:1988fe,Binoth:2005ua,Dawson:2013lya,Binoth:2006mf,Caola:2015ila,vonManteuffel:2015msa,Caola:2015psa,Caola:2015rqy}.  As we will show, after Higgs measurements and resonant searches are combined, the width is indeed narrow.

Typically, the observed and expected 95\% CL upper limits on resonance production are reported.  Assuming that there are no large fluctuations away from the SM predictions, a SM cross section is measured in all new physics searches.  The allowed fluctuations away from the SM cross section at 95\% CL are then the expected 95\% CL upper limits on new resonance cross sections.  That is, the uncertainty on the cross section is approximated as
\begin{eqnarray}
\delta \sigma_i^f\approx \hat{\sigma}_{i,Exp}^f/1.96,
\end{eqnarray}
where $\hat{\sigma}_{i,Exp}^f$ is the expected 95\% CL upper limit on the resonance cross section.  Again assuming there are no large excesses, the measured cross section is mostly SM-like with a small deviation given by the difference in the observed and expected bounds:
\begin{eqnarray}
\hat{\sigma}_i^f\approx \sigma_{i,SM}^f+\hat{\sigma}_{i,Obs}^f-\hat{\sigma}_{i,Exp}^f,
\end{eqnarray}
where $\hat{\sigma}_{i,Obs}^f$ is the observed 95\% CL upper limit on the resonance cross section.  With these approximations, we finally have 
\begin{eqnarray}
\left(\chi^{f}_{i,h_2}\right)^2=\left(\frac{\sigma(i\rightarrow h_2){\rm BR}(h_2\rightarrow f)+\hat{\sigma}_{i,Exp}^f-\hat{\sigma}_{i,Obs}^f}{\hat{\sigma}_{i,Exp}^f/1.96}\right)^2.\label{eq:Chi2Scalar_init}
\end{eqnarray}

One final complication is if $\hat{\sigma}_{i,Obs}^f<\hat{\sigma}_{i,Exp}^f$ then according to Eq.~(\ref{eq:Chi2Scalar_init}) the best fit signal cross section $\sigma(i\rightarrow h_2){\rm BR}(h_2\rightarrow f)$ will be negative, which is nonsensical.  We propose to alter the definition in Eq.~(\ref{eq:Chi2Scalar_init}) to 
\begin{eqnarray}
\left(\chi^{f}_{i,h_2}\right)^2=\begin{cases}
\displaystyle\left(\frac{\sigma(i\rightarrow h_2){\rm BR}(h_2\rightarrow f)+\hat{\sigma}_{i,Exp}^f-\hat{\sigma}_{i,Obs}^f}{\hat{\sigma}_{i,Exp}^f/1.96}\right)^2 & {\rm if~} \hat{\sigma}_{i,Obs}^f\geq \hat{\sigma}_{i,Exp}^f\\
\displaystyle\left(\frac{\sigma(i\rightarrow h_2){\rm BR}(h_2\rightarrow f)}{\hat{\sigma}_{i,Obs}^f/1.96}\right)^2&{\rm if~} \hat{\sigma}_{i,Obs}^f< \hat{\sigma}_{i,Exp}^f.
\end{cases}~\label{eq:Chi2Scalar}
\end{eqnarray}
The second line forces the best fit value of $\sigma(i\rightarrow h_2){\rm BR}(h_2\rightarrow f)$ to be bounded from below by zero.  Also in the second line, the uncertainty has been changed from the expected to observed signal rate.  If the best fit value of the signal cross section is at zero, then $\hat{\sigma}_{i,Obs}^f$ is how far away it can fluctuate from zero at 95\% CL.  Hence, this form of the $\chi^2$ allows for upward fluctuations with a best fit value of the signal cross section away from zero as well as bounding the best fit value of the cross section to be positive.  

The usual use of the reported 95\% CL upper bounds is to put a strict upper bound on resonance cross sections: $\sigma(i\rightarrow h_2){\rm BR}(h_2\rightarrow f)<\hat{\sigma}_{i,Obs}^f$~\cite{Bechtle:2011sb,Bechtle:2013wla,Bechtle:2020pkv}.  To check that our proposal is consistent, it must be checked that this interpretation can be derived from Eq.~(\ref{eq:Chi2Scalar}).  Assuming a one-parameter fit, the value of the resonance cross section at the minimum $\chi^2$ is 
\begin{eqnarray}
\left[\sigma(i\rightarrow h_2){\rm BR}(h_2\rightarrow f)\right]_{\chi^2_{\rm min}}=\begin{cases} 
\hat{\sigma}_{i,Obs}^f-\hat{\sigma}_{i,Exp}^f & {\rm if~} \hat{\sigma}_{i,Obs}^f\geq \hat{\sigma}_{i,Exp}^f\\
0 &{\rm if~} \hat{\sigma}_{i,Obs}^f< \hat{\sigma}_{i,Exp}^f\end{cases}
\end{eqnarray}
Then, the one-parameter fit limit is found by requiring $\Delta\chi^2=\chi^2-\chi^2_{\rm min}<3.84$, where $\chi_{\rm min}^2$ is the minimum $\chi^2$.  It can then be shown that Eq.~(\ref{eq:Chi2Scalar}) gives the limit
\begin{eqnarray}
\sigma(i\rightarrow h_2){\rm BR}(h_2\rightarrow f) < \hat{\sigma}_{i,Obs}^f,
\end{eqnarray}
which is consistent with the usual interpretation of these bounds.  

With these results, all heavy scalar searches can be combined into one $\chi^2$:
\begin{eqnarray}
\chi^2_{h_2}=\sum_{i,f}\left(\chi^{f}_{i,h_2}\right)^2.\label{eq:Chi2h2}
\end{eqnarray}
Unlike imposing a hard cutoff on the heavy resonance rates, this method will allow for fluctuations in some channels that are consistent with a global fit at 95\% CL.  The combined limits from scalar searches and Higgs measurements are found by combing the $\chi^2$ in Eqs.~(\ref{eq:Chi2h1}) and (\ref{eq:Chi2h2}):
\begin{eqnarray}
\chi^2_{\rm Tot}=\chi^2_{h_1}+\chi^2_{h_2}.
\end{eqnarray}

\subsection{Results for Heavy Resonance Searches and Higgs Precision}
\label{sec:search}
As with the Higgs boson, the main production channel of $h_2$ is gluon fusion due to the large gluon parton luminosities.  Hence, we will only consider the ggF initial state.  To calculate this we reweight partial widths with the NNLO+NNLL SM-like Higgs predictions provided by the LHC Higgs Cross Section Working Group~\cite{deFlorian:2016spz}:
\begin{eqnarray}
\sigma_{ggF}(pp\rightarrow h_2)=\sigma_{ggF,SM}^{NNLO+NNLL}(pp\rightarrow h_2)\frac{\Gamma(h_2\rightarrow gg)}{\Gamma_{SM}(h_2\rightarrow gg)},
\end{eqnarray}
where the subscript $SM$ indicates the prediction for a SM-like Higgs boson at a mass $m_2$.  The same higher order corrections used for $h_1$ decays, as discussed in Sec.~\ref{sec:ProdDec}, are incorporated into $h_2$ decays.  We also correctly account for the $(1+11\alpha_s/4\pi)$ difference in the NLO contributions to $h_2\rightarrow gg$ from quark loops and $g-g-h_2$ contact interactions~\cite{Djouadi:1991tka,Dawson:1990zj,Spira:1995rr,deBlas:2018tjm}, as discussed previously.  For $h_2$ decay rates the renormalization scale is set to $m_2$ for tree level decays and $m_2/2$ for loop level decays.  The values of $\hat{\sigma}_{ggF,Obs}^f$ and $\hat{\sigma}_{ggF,Exp}^f$ for the final states under consideration are given in Tables~\ref{tab:ATLsearch} and~\ref{tab:CMSsearch} in App.~\ref{app:data}.   

For all results presented in this section we keep cross sections and widths to order $\Lambda^{-2}$.  As discussed in Sec.~\ref{sec:PC}, this is valid power counting for $h_2$ processes in the small mixing limit.  Finally, we always require that Wilson coefficients are bounded by $|f_i|<4\pi$ and fit to all relevant Wilson coefficients and scalar trilinear couplings:
\begin{eqnarray}
\sin\theta,\,f_{GG},\,f_+,f_-,\,f_b,\,f_t,\,f_\tau,\,f_\mu,\,{\rm and}\,\lambda_{211}.
\end{eqnarray}
As with Higgs rates the $h_2$ rates are invariant under the simultaneous parity transformation $\sin\theta\rightarrow -\sin\theta$ and $f_i\rightarrow -f_i$.  Hence, we only show results for $f_i>0$ and results for $f_i<0$ can be found by performing the transformation.

\begin{figure}[tb]
\begin{center}
\subfigure[]{\includegraphics[width=0.45\textwidth,clip]{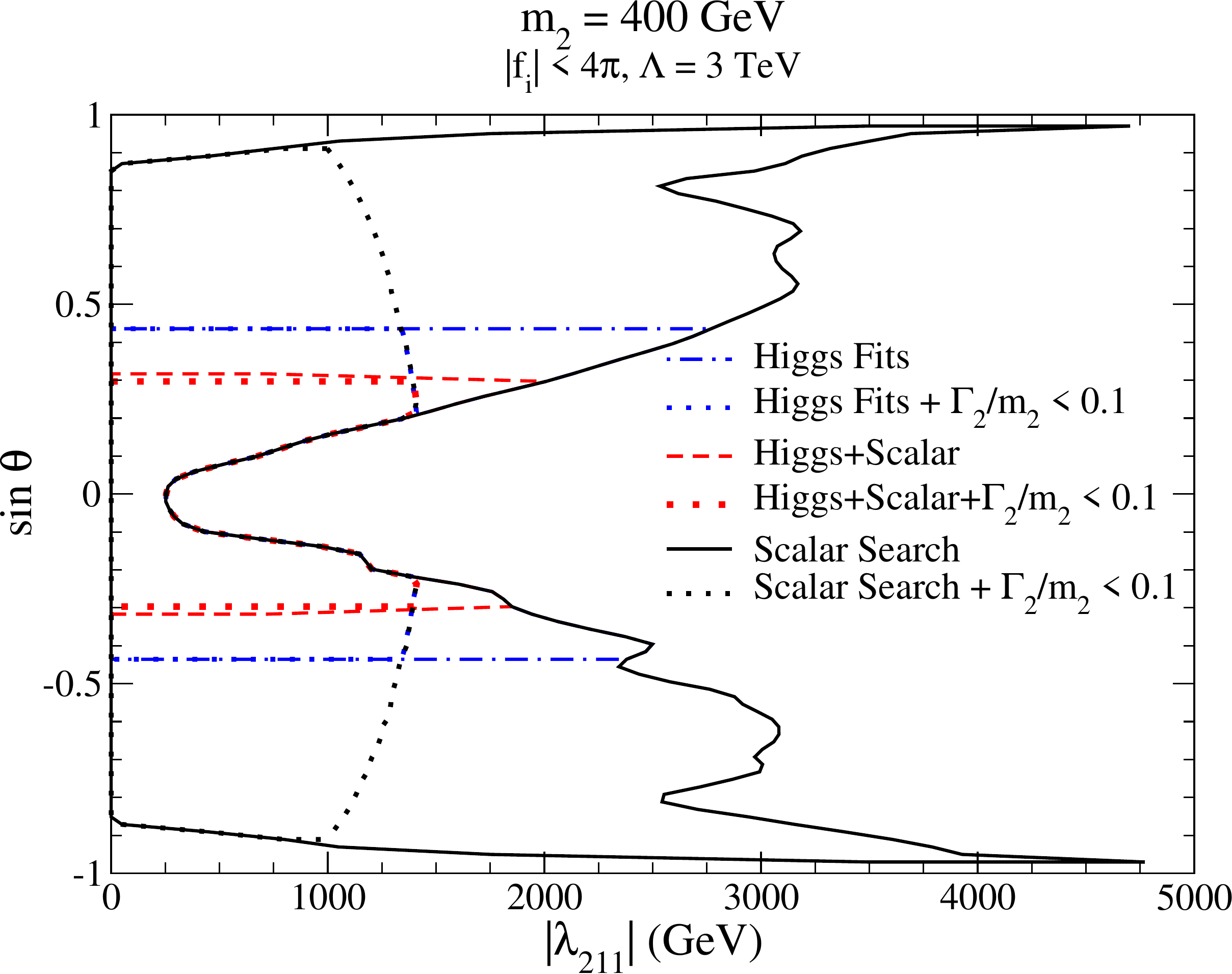}}
\subfigure[]{\includegraphics[width=0.45\textwidth,clip]{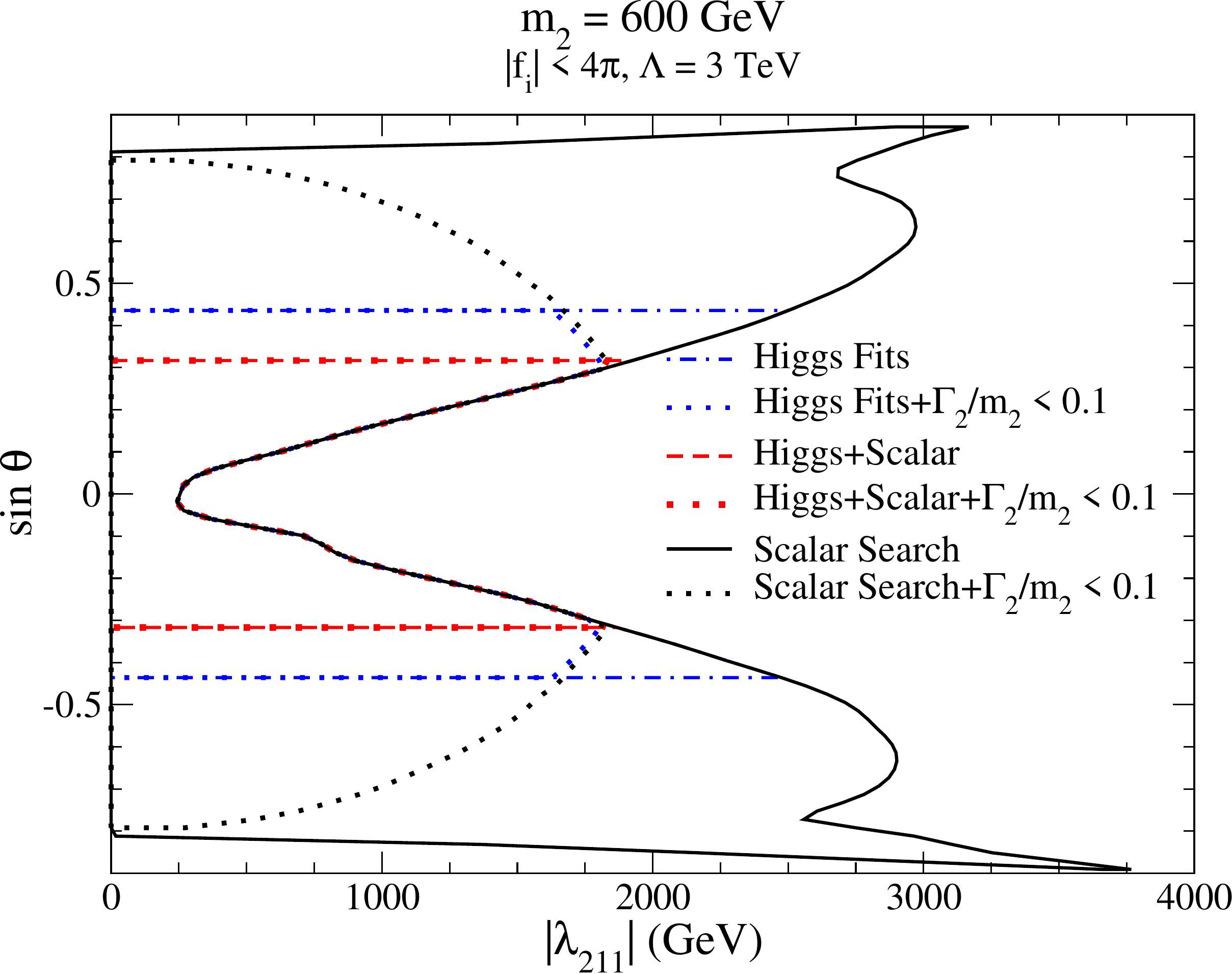}}
\end{center}
\caption{\label{fig:Scalar1} 95\% CL allowed regions from  (black solid) heavy scalar searches, (black dotted) heavy scalar searches with $\Gamma_2<0.1\,m_2$, (blue dot-dashed) Higgs measurements, (blue dotted) Higgs measurements with $\Gamma_2<0.1\,m_2$, (red dashed) combined heavy scalar searches and Higgs measurements, and (red dotted) combined heavy scalar searches and Higgs measurements with $\Gamma_2<0.1\,m_2$.  These are shown for $\lambda_{211}$ vs $\sin\theta$ with all other parameters profiled over.  Two scalar masses are considered: (a) $m_2=400$~GeV and (b) $m_2=600$~GeV.  The new physics scale is $\Lambda=3$~TeV.  The regions within the contours are allowed.}
\end{figure}

\begin{figure}[htb]
\begin{center}
\subfigure[]{\includegraphics[width=0.42\textwidth,clip]{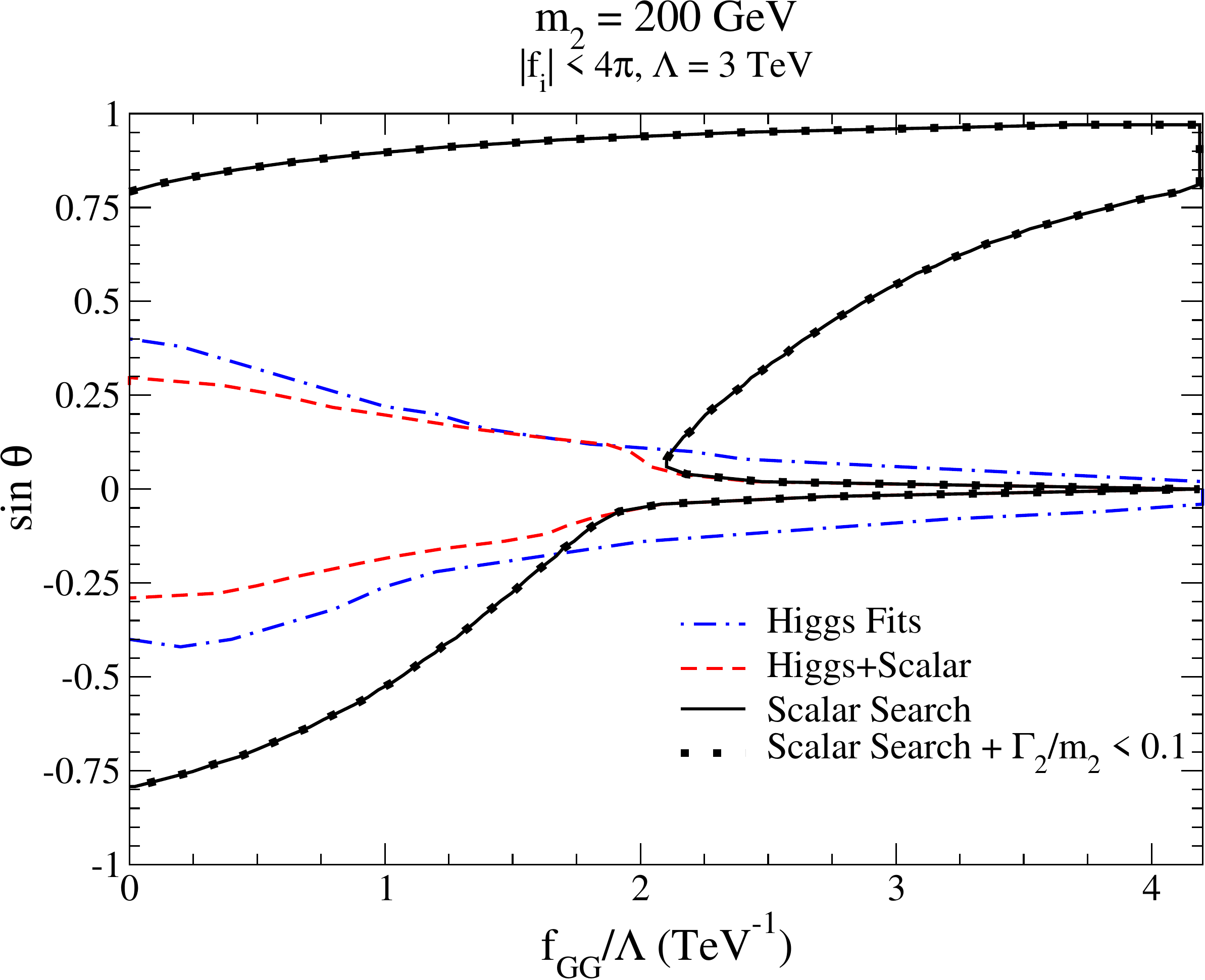}}
\subfigure[]{\includegraphics[width=0.42\textwidth,clip]{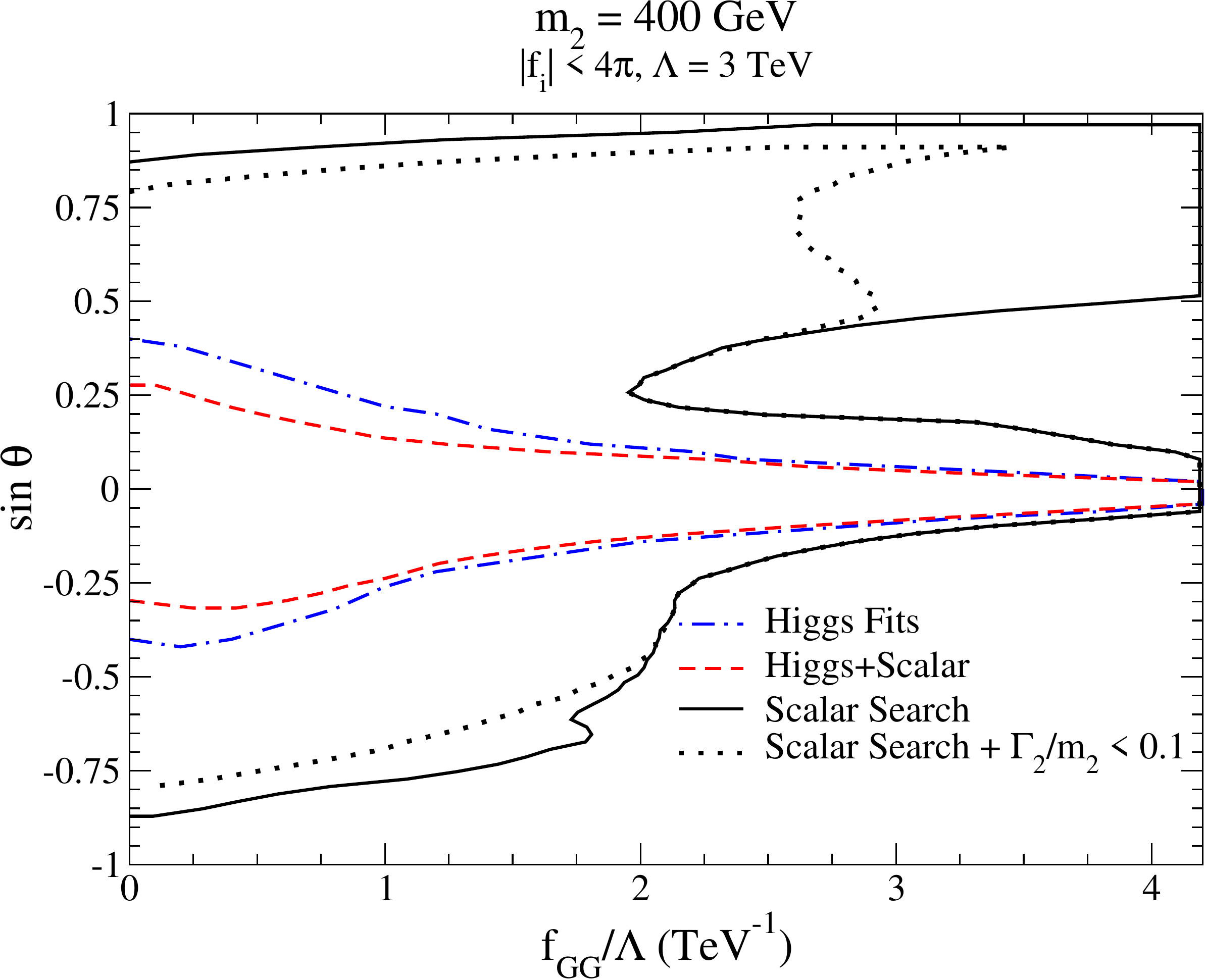}}\\\vspace{-0.1in}
\subfigure[]{\includegraphics[width=0.42\textwidth,clip]{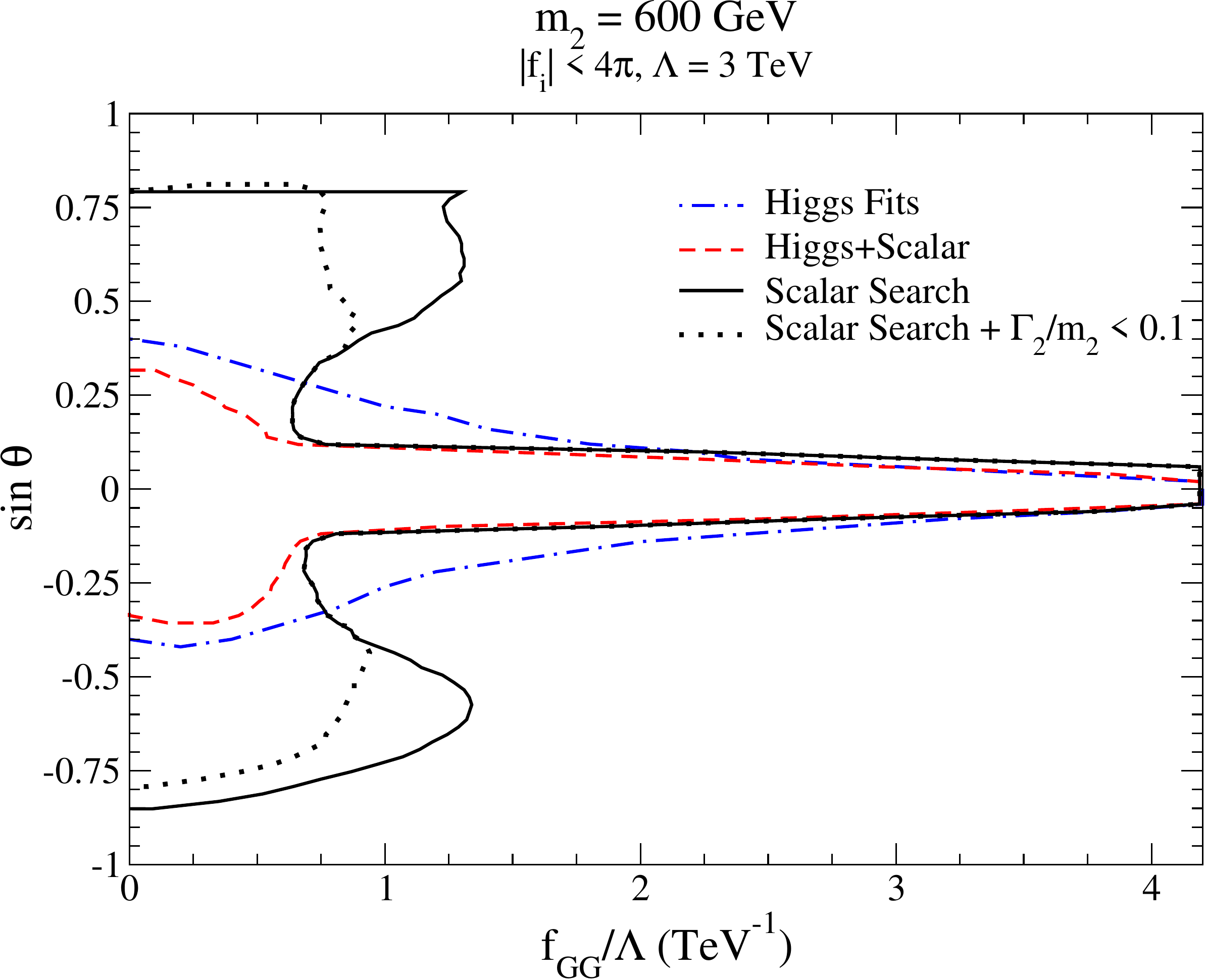}}
\subfigure[]{\includegraphics[width=0.42\textwidth,clip]{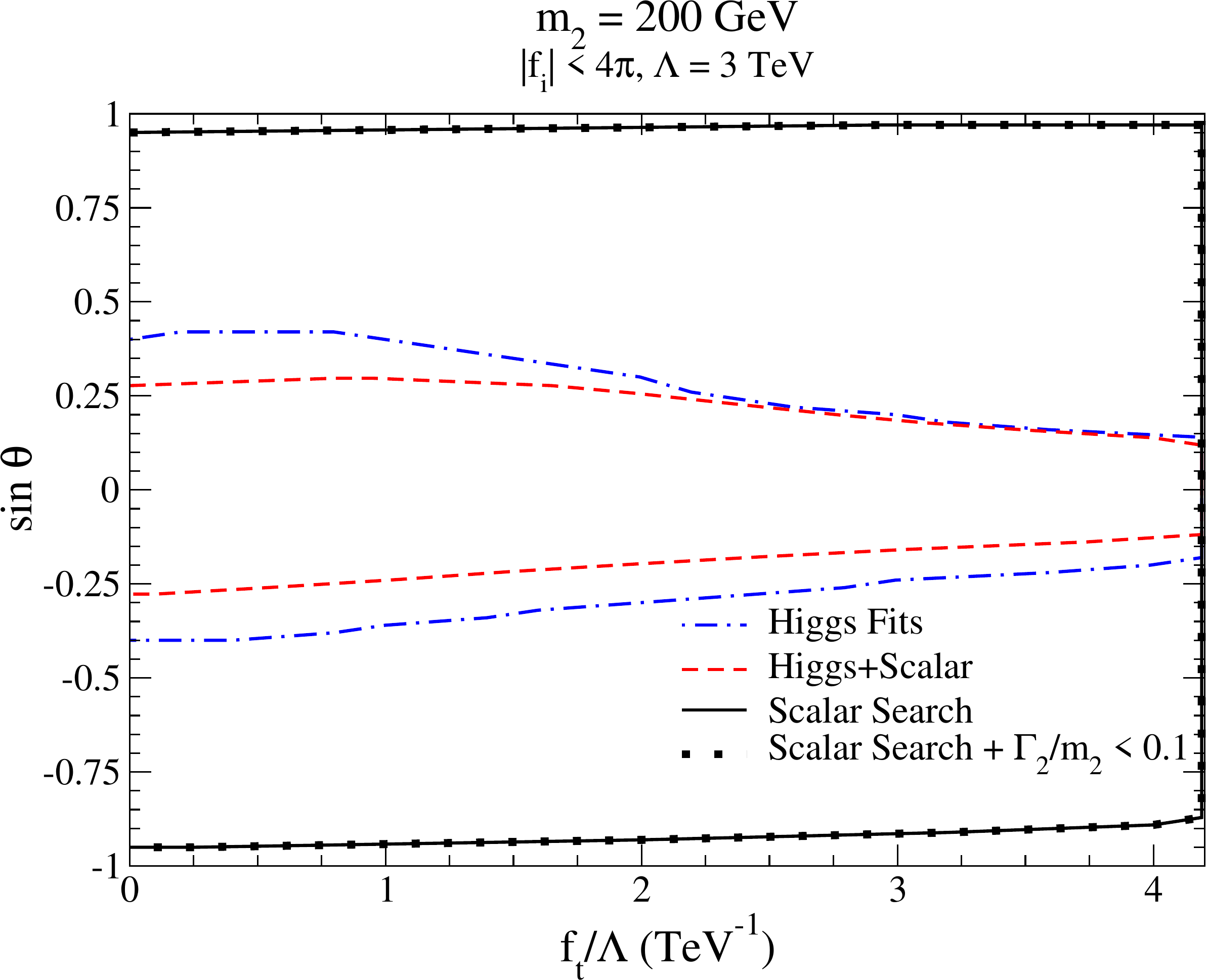}}\\\vspace{-0.1in}
\subfigure[]{\includegraphics[width=0.42\textwidth,clip]{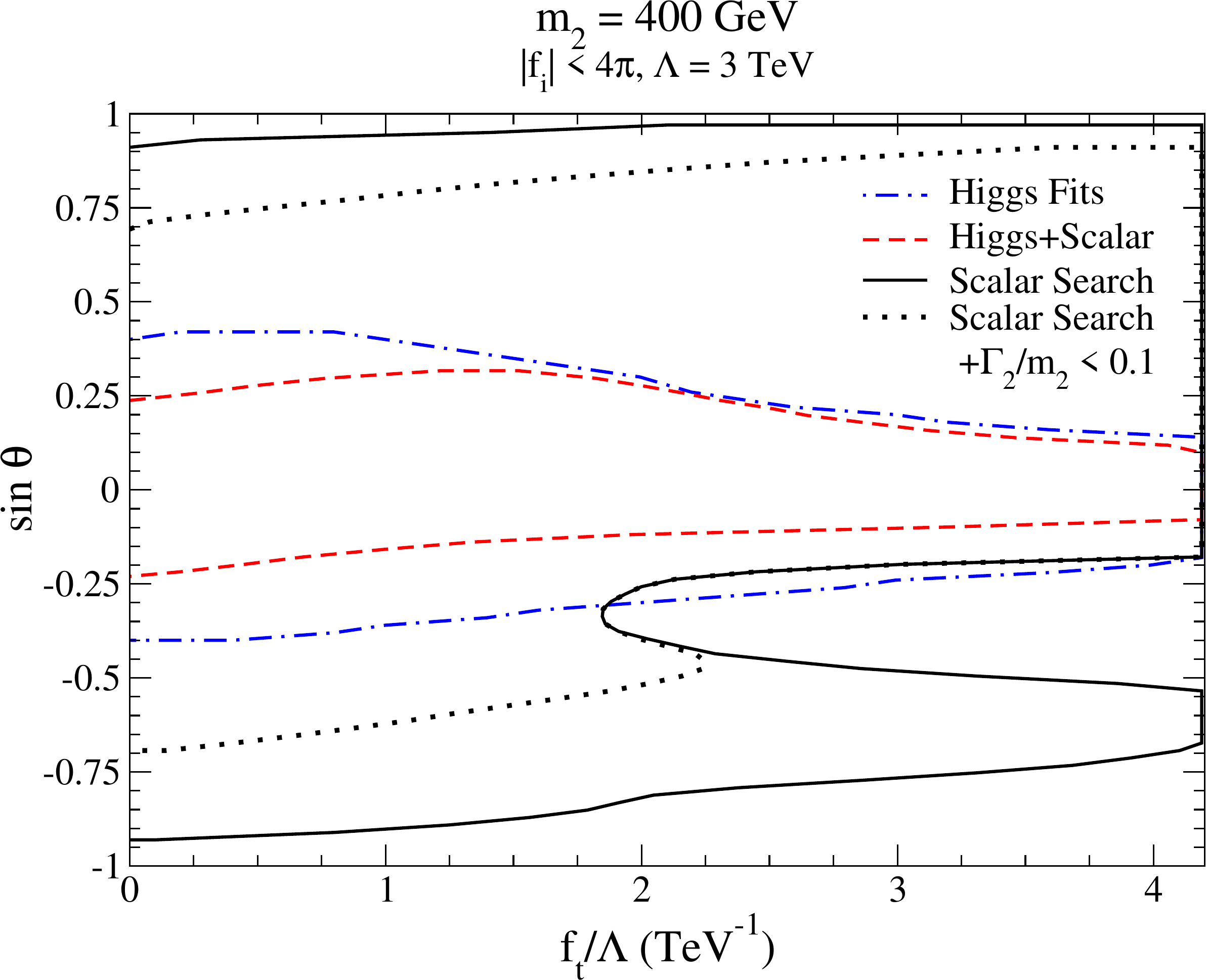}}
\subfigure[]{\includegraphics[width=0.42\textwidth,clip]{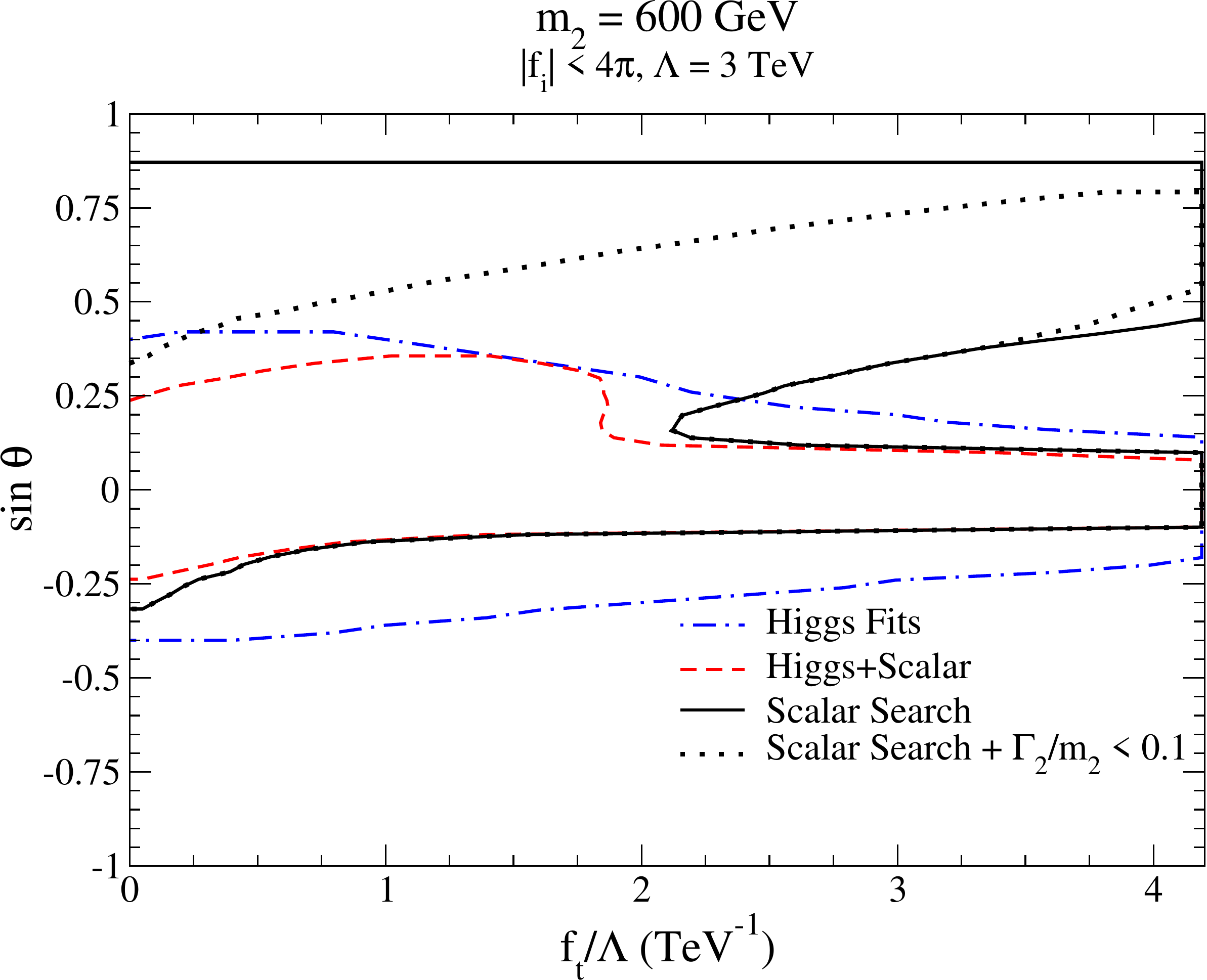}}\\\vspace{-0.25in}
\end{center}
\caption{\label{fig:Scalar2} 95\% CL allowed regions from  (black solid) heavy scalar searches, (black dotted) heavy scalar searches with $\Gamma_2<0.1\,m_2$, (blue dot-dashed) Higgs measurements, and (red dotted) combined heavy scalar searches and Higgs measurements.  The regions within the contours are allowed.  These are shown for (a,b,c) $f_{GG}$ and (d,e,f) $f_t$ vs $\sin\theta$ with all other parameters profiled over.  Three scalar masses are considered: (a,d) $m_2=200$~GeV, (b,e) $m_2=400$~GeV, and (c,f) $m_2=600$~GeV. The new physics scale is $\Lambda=3$~TeV.}
\end{figure}

Figure~\ref{fig:Scalar1} shows the two-dimensional 95\% CL allowed regions for $\lambda_{211}$ vs $\sin\theta$ with all other parameters profiled over.  Only two scalar masses are considered, $m_2=400$~GeV and $m_2=600$~GeV, since the decay $h_2\rightarrow h_1h_1$ is then kinematically possible.  The scalar searches do not put meaningful limits on $\lambda_{211}$: the upper bounds on $\lambda_{211}$ come from the theoretical requirement that the scalar potential be bounded and that the global minimum be the correct EWSB minimum as discussed in Sec.~\ref{sec:Model}.  As can be seen, the Higgs measurements mainly limit the values of $\sin\theta$.  

Many of the scalar searches that are included require that $h_2$ be a narrow resonance.  Hence, the results from requiring that the $h_2$ total width, $\Gamma_2$, be less than 10\% of the $h_2$ mass are also shown in Fig.~\ref{fig:Scalar1}.  For the scalar searches, the width constraint limits both $\sin\theta$ and $\lambda_{221}$.  When  $\sin\theta$ is large, the decays $h_2\rightarrow WW$ and $h_2\rightarrow ZZ$ are SM-like and large for large $m_2$~\cite{deFlorian:2016spz}.  Hence, $\Gamma_2/m_2<0.1$ places stronger constraints on $\sin\theta$ than just blindly applying the scalar searches.  Also, if $\lambda_{211}$ is too large the partial width of the decay $h_2\rightarrow h_1h_1$ becomes large.  As a result, requiring a narrow resonance puts strong constraints on $\lambda_{211}$.  Higgs measurements already strongly constrain $\sin\theta$, so the effect of requiring a narrow resonance is much less pronounced here.  When the Higgs measurements are combined with scalar searches, the narrow width requirement does not meaningfully constrain the parameter space.

The two-dimensional 95\% CL allowed regions for $f_{GG}$ vs $\sin\theta$ and $f_t$ vs $\sin\theta$ with all other parameters profiled over are shown in Fig.~\ref{fig:Scalar2}\footnote{The results for the remaining Wilson coefficients can be found in App.~\ref{app:limits}.}.  As with the $\lambda_{211}$ limits, requiring a narrow width in the scalar search results squeezes the allowed parameter region for $m_2=400$ and $m_2=600$~GeV.  The narrow width requirement does not significantly change the Higgs precision and combination constraints.  For the smaller $m_2$, the narrow width requirement makes no difference on any of the limits.

As can be seen in Fig.~\ref{fig:Scalar2}, Higgs measurements and scalar searches are complementary.  That is, the allowed regions for scalar searches and Higgs measurements do not fully overlap.  Indeed, the combined allowed region is smaller than the individual allowed regions.  This is particularly striking for $m_2=600$~GeV.

\begin{figure}[htb]
\begin{center}
\subfigure[]{\includegraphics[width=0.45\textwidth,clip]{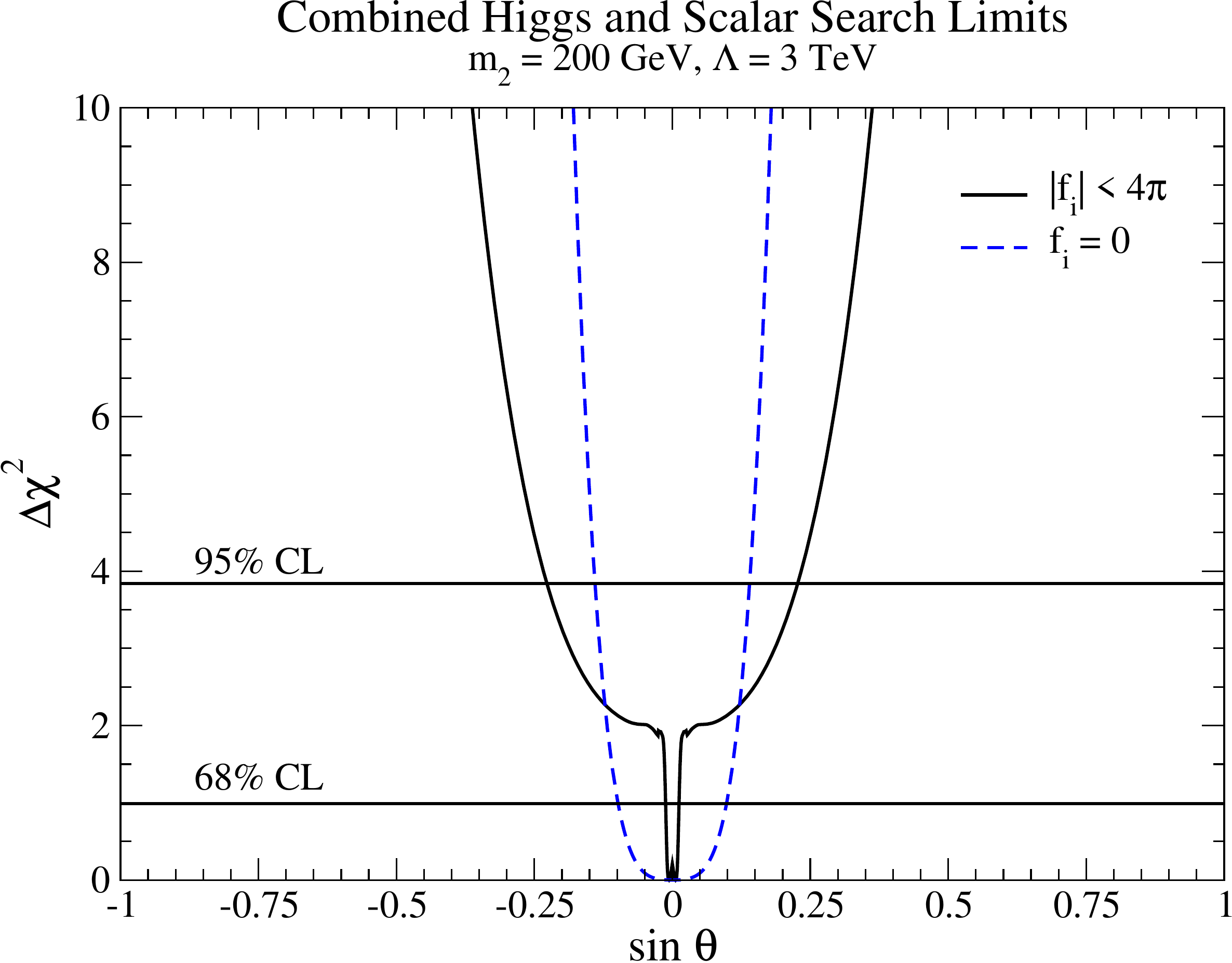}}
\subfigure[]{\includegraphics[width=0.45\textwidth,clip]{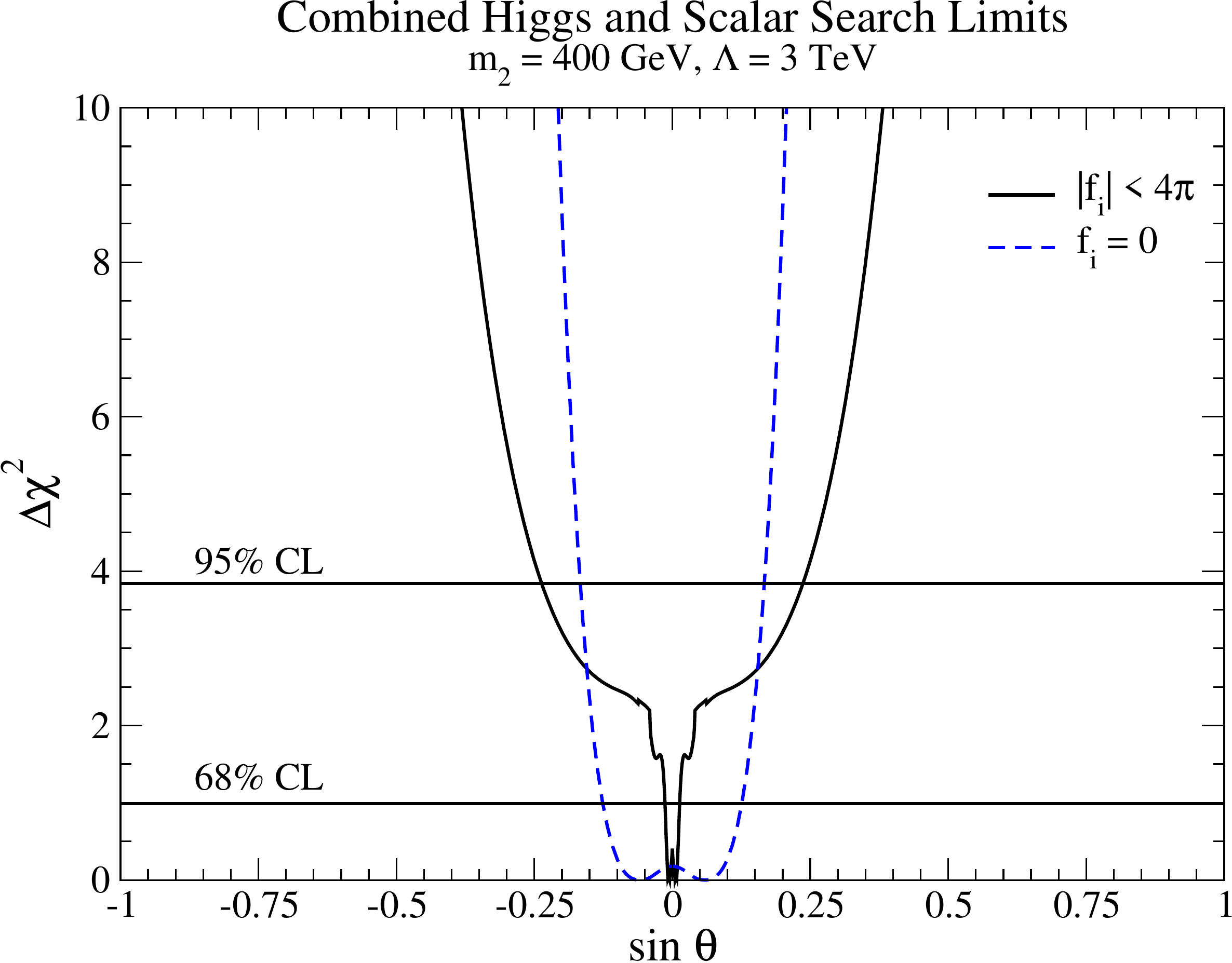}}
\subfigure[]{\includegraphics[width=0.45\textwidth,clip]{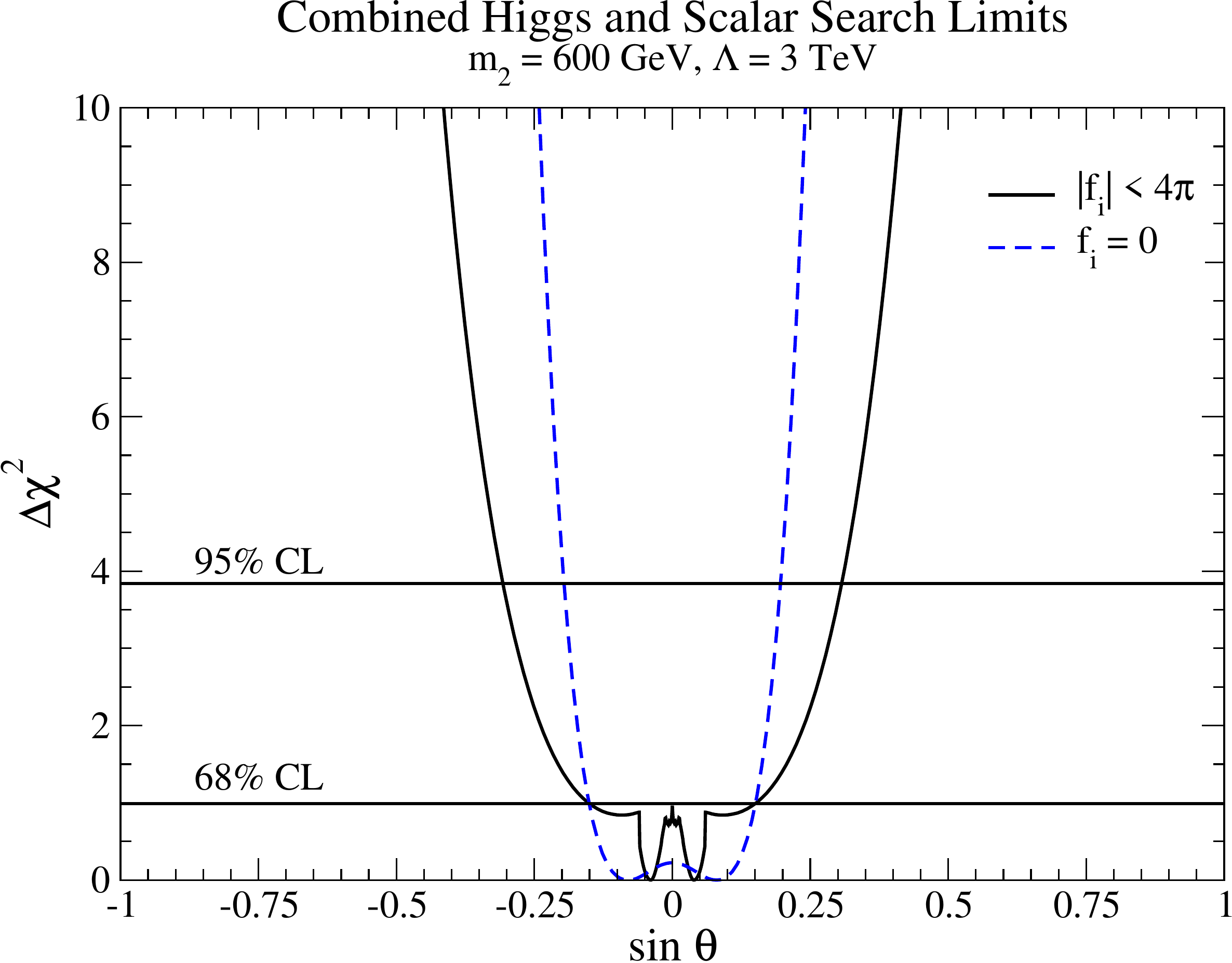}}
\end{center}
\caption{\label{fig:sthcomb} $\Delta\chi^2$ for combined Higgs measurements and scalar searches as a function of $\sin\theta$ with all other parameters profiled over.  Both (black solid) Wilson coefficients with $|f_i|<4\pi$ and (blue dashed) $f_i=0$, i.e. dimension-5 terms set to zero.  Three $h_2$ masses are considered: (a) $m_2=200$~GeV, (b) $m_2=400$~GeV, and (c) $m_2=600$~GeV.}
\end{figure}

Finally, in Fig.~\ref{fig:sthcomb}, we show the $\Delta\chi^2$ distributions as a function $\sin\theta$ for the BSM-EFT and renormalizable model with all other parameters profiled over.  In the BSM-EFT, the shape of $\Delta\chi^2$ changes dramatically.  The 95\% CL and 68\% CL allowed regions also change drastically and exactly how they change depends strongly on the $h_2$ mass.  It is clear that even 3 TeV new physics effects can make a significant impact on the interpretation of current measurements.

\section{Conclusion}
\label{sec:conc}
A common assumption of simplified models at the LHC is that there are a few new BSM particles that can be produced, while all other new particles are heavy and decoupled.  Under these assumptions, most studies of simplified models are renormalizable.  However, using EFT techniques, it is possible to test the basic assumption that all other new particles are indeed decoupled.  

In this paper, we studied a popular simplified model, the real singlet extended SM, and supplemented it with all possible dimension-5 operators involving the scalar singlet.  We studied the effects of the effective operators on the interpretation of Higgs signal strengths as well as searches for heavy new resonances.  As we showed, even if the new physics occurs at 3 TeV, the interpretation of these measurements and searches are changed drastically.  This study shows that even in the simplest of simplified model, the heavy new physics is not ``decoupled'' even when the BSM-EFT expansion is valid.  That is, it cannot be neglected and the BSM-EFT should generically be considered.

In addition to the numerical results, we also gave a comprehensive discussion of the counting in BSM-EFT for production and decay rates.  We showed that while in the linear SMEFT power counting is relatively straightforward, power counting in a BSM-EFT is strongly process and parameter space dependent.  We also developed a new proposal to consistently combine the limits from new resonance searches and precision measurements via an $\chi^2$.  This method allows for fluctuations in individual channels, while keeping the global $\chi^2$ within allowable limits.  This is unlike the usual cutoff method where all resonance cross sections are strictly cutoff at the observed limits~\cite{Bechtle:2011sb,Bechtle:2013wla,Bechtle:2020pkv}.

\section*{Acknowledgments}
We thank Jeong Han Kim, KC Kong, Tilman Plehn, Daniel Tapia Takaki, and Yajuan Zheng for helpful discussions.  Chris Rogan is thanked for reassuring IML that he is not crazy.  IML would like to thank the Institute for Theoretical Physics at Universit\"{a}t Heidelberg for their hospitality during the completion of this work.  This work was performed in part at the Aspen Center for Physics, which is supported by National Science Foundation grant PHY-1607611.  SA, IML, and MS are supported in part by United States Department of Energy grant number DE-SC0017988.  MS is supported in part by the United States Department of Energy under Grant Contract DE-SC0012704.  SA and MS are also supported in part by the State of Kansas EPSCoR grant program.  The data to reproduce the plots have been uploaded with the arXiv submission or is available upon request.

\appendix
\section{The Gluon Fusion Cross Section and Gluon Partial Width}
\label{app:WidthXSApprox}
In this appendix, we will inspect numbers for the gluon partial width and gluon fusion cross section and confirm the validity of the signal strength approximation in Eq.~(\ref{eq:ggF1}) in this EFT. The cross section results of Ref.~\cite{Deutschmann:2017qum} gives
\begin{eqnarray}
\frac{\sigma_{ggF}(pp\rightarrow h_1)}{\sigma_{ggF,SM}(pp\rightarrow h_1)}&=&\cos^2\theta+\frac{\cos\theta\,\sin\theta}{(\Lambda/{\rm TeV})}\left(0.517\,f_t+1.45\,f_{GG}\right)\nonumber\\
&&+\frac{\sin^2\theta}{(\Lambda/{\rm TeV})^2}\left(0.0626\,f_t^2+0.492\,f_{GG}^2+0.351\,f_t\,f_{GG}\right).
\end{eqnarray}
The corresponding gluon partial width calculation in the scalar EFT (dropping bottom quark EFT contributions for the purposes of comparison) gives
\begin{eqnarray}
\frac{\Gamma(h_1\rightarrow gg)}{\Gamma_{SM}(h_1\rightarrow gg)}&=&\cos^2\theta+\frac{\cos\theta\,\sin\theta}{(\Lambda/{\rm TeV})}\left(0.526\,f_t+1.38\,f_{GG}\right)\nonumber\\
&&+\frac{\sin^2\theta}{(\Lambda/{\rm TeV})^2}\left(0.0702\,f_t^2+0.481\,f_{GG}^2+0.367\,f_t\,f_{GG}\right).
\end{eqnarray}
Comparing the size of the linear terms, the $f_t$ terms differ by less than $2\%$ and the $f_{GG}$ terms differ by around $5\%$. For the quadratic terms, the $f_{GG}^2$ term and the $f_{GG} f_t$ term differ by around $2\%$ and $5\%$, respectively, while the $f_t^2$ term differs by around $12\%$. Naively, one might be worried that the large difference in the $f_t^2$ term might challenge the validity of the approximation in Eq.~(\ref{eq:ggF1}). However, as seen in Fig.~\ref{fig:Higgs} and the discussion that followed, the Higgs signal strengths constrained the mixing angle to be small, and so the linear terms dominated over the quadratic terms. We thus conclude that, for the parameter space allowed by observation, the approximation is good to within around $5\%$ or less.

\section{Feynman Rules}
\label{app:FeynRules}
\subsection{Trilinear Scalar Couplings}
The trilinear scalar couplings are defined as
\begin{eqnarray}
V(h_1,h_2)\supset \frac{1}{3!}\lambda_{111}h_1^3+\frac{1}{2}\lambda_{211}h_1^2h_2+\frac{1}{2}\lambda_{221} h_1h_2^2+\frac{1}{3!}\lambda_{222}h_2^3.
\end{eqnarray}
where $\lambda_{111}$ and $\lambda_{211}$ are given in Sec.~\ref{sec:Model}, and
\begin{eqnarray}
\lambda_{221}&=&\frac{m_1^2+2\,m_2^2}{v}\cos\theta\sin^2\theta+2\,b_3\cos^2\theta\sin\theta+a_2v\,\cos\theta\left(\cos^2\theta-2\sin^2\theta\right)\nonumber\\
&&+\frac{3\,a_3v^2}{2\Lambda}\cos^2\theta\sin\theta-\frac{a_4v^2}{\Lambda}\sin\theta\left(2\cos^2\theta-\sin^2\theta\right)\,,\label{eq:l221}\\
\lambda_{222}&=&-\frac{3\,m_2^2}{v}\sin^3\theta+2\,b_3\cos^3\theta-3\,a_2v\,\cos^2\theta\sin\theta+\frac{3\,a_3v^2}{2\,\Lambda}\cos^3\theta\\
&&+\frac{3\,a_4v^2}{\Lambda}\cos\theta\sin^2\theta\,.\nonumber
\end{eqnarray}

\subsection{$h_i-f-f$ and $h_i-V-V'$ couplings}
The vertex rules, with all momenta outgoing, are
\begin{eqnarray}
\mathcal{V}_{h_1 f f} &=& -i \frac{m_f}{v}(\cos\theta + \frac{f_f}{\Lambda}v \sin\theta)\,,\\
\mathcal{V}_{h_1 g_\mu(p_1) g_\nu(p_2)} &=& -i \frac{g_s^2}{4 \pi^2}\frac{f_{GG}}{\Lambda}\sin\theta\, (\eta_{\mu\nu} p_1 \cdot p_2 - p_{1\nu} p_{2\mu})\,,\\
\mathcal{V}_{h_1 \gamma_\mu(p_1) \gamma_\nu(p_2)} &=& -i \frac{e^2}{4 \pi^2}\frac{f_{BB}+f_{WW}}{\Lambda}\sin\theta\, (\eta_{\mu\nu} p_1 \cdot p_2 - p_{1\nu} p_{2\mu})\,,\\
\mathcal{V}_{h_1 W^{+}_\mu(p_1) W^{-}_\nu(p_2)} &=& 2i \frac{M_W^2}{v}\eta_{\mu\nu}\cos\theta-i \frac{g^2}{4 \pi^2}\frac{f_{WW}}{\Lambda}\sin\theta\, (\eta_{\mu\nu} p_1 \cdot p_2 - p_{1\nu} p_{2\mu})\,,\\
\mathcal{V}_{h_1 Z_\mu(p_1) Z_\nu(p_2)} &=& 2i \frac{M_Z^2}{v}\eta_{\mu\nu}\cos\theta-i \frac{g_Z^2}{4 \pi^2}\frac{f_{ZZ}}{\Lambda}\sin\theta (\eta_{\mu\nu} p_1 \cdot p_2 - p_{1\nu} p_{2\mu})\,,\\
\mathcal{V}_{h_1 Z_\mu(p_1) \gamma_\nu(p_2)} &=& - i \frac{g_Z e}{2 \pi^2}\frac{f_{Z\gamma}}{\Lambda}\sin\theta\, (\eta_{\mu\nu} p_1 \cdot p_2 - p_{1\nu} p_{2\mu})\,,\\
\mathcal{V}_{h_2 f f} &=& -i \frac{m_f}{v}(-\sin\theta\, + \frac{f_f}{\Lambda}v \cos\theta)\,,\\
\mathcal{V}_{h_2 g_\mu(p_1) g_\nu(p_2)} &=& -i \frac{g_s^2}{4 \pi^2}\frac{f_{GG}}{\Lambda}\cos\theta\, (\eta_{\mu\nu} p_1 \cdot p_2 - p_{1\nu} p_{2\mu})\,,\\
\mathcal{V}_{h_2 \gamma_\mu(p_1) \gamma_\nu(p_2)} &=& -i \frac{e^2}{4 \pi^2}\frac{f_{BB}+f_{WW}}{\Lambda}\cos\theta\, (\eta_{\mu\nu} p_1 \cdot p_2 - p_{1\nu} p_{2\mu})\,,\\
\mathcal{V}_{h_2 W^{+}_\mu(p_1) W^{-}_\nu(p_2)} &=& -2i \frac{M_W^2}{v}\eta_{\mu\nu}\sin\theta-i \frac{g^2}{4 \pi^2}\frac{f_{WW}}{\Lambda}\cos\theta\, (\eta_{\mu\nu} p_1 \cdot p_2 - p_{1\nu} p_{2\mu})\,,\\
\mathcal{V}_{h_2 Z_\mu(p_1) Z_\nu(p_2)} &=& -2i \frac{M_Z^2}{v}\eta_{\mu\nu}\sin\theta-i \frac{g_Z^2}{4 \pi^2}\frac{f_{ZZ}}{\Lambda}\cos\theta\, (\eta_{\mu\nu} p_1 \cdot p_2 - p_{1\nu} p_{2\mu})\,,\\
\mathcal{V}_{h_2 Z_\mu(p_1) \gamma_\nu(p_2)} &=& - i \frac{g_Z e}{2 \pi^2}\frac{f_{Z\gamma}}{\Lambda}\cos\theta\, (\eta_{\mu\nu} p_1 \cdot p_2 - p_{1\nu} p_{2\mu})\,.
\end{eqnarray}

with $g_Z=\frac{g}{\cos\theta_W}$, $f_{ZZ}=f_{BB} \sin^4\theta_w+f_{WW} \cos^4\theta_W$, $f_{Z\gamma}=f_{BB} \sin^2\theta_w+f_{WW} \cos^4\theta_W$, and $\theta_W$ is the weak mixing angle.

\section{Signal Strengths/Bounds}
\label{app:data}
\subsection{Higgs Signal Strengths}
We now give the signal strengths used in our fits.  These are chosen to be the measured signal strengths with the most integrated luminosity in a given channel.  All results are from Run 2, with up to 139~fb$^{-1}$ of accumulated data.  The guide to nonobvious abbreviations: ggF = gluon fusion, VBF = vector boson fusion, $Wh_1$ = Higgs associated production with a $W$, $Zh_1$ = Higgs associated production with a $Z$, $Vh_1$ = combination of $Wh_1$ and $Zh_1$, and $VV^*$ = combination of $ZZ^*$ and $WW^*$.

 \begin{table}[H]
 \begin{center}
\resizebox{\textwidth}{!}{
\begin{tabular}{|c||c|c|c|c|c|c|c|c|}
\hline\hline
  & $\gamma\gamma$ & $ZZ^*$  & $WW^*$ & $VV^*$ & $\tau\tau$  & $bb$ & $\mu\mu$& $Z\gamma$ \\\hline
ggF & $1.03^{+0.11}_{-0.11}$~\cite{ATLAS:2020qdt} & 0.94$^{+0.11}_{-0.10}$~\cite{ATLAS:2020qdt}&1.08$^{+0.19}_{-0.18}$~\cite{ATLAS:2020qdt}&*&$1.02^{+0.60}_{-0.55}$~\cite{ATLAS:2020qdt}& *&$0.61^{+0.75}_{-0.75}$~\cite{Aad:2020xfq}&*\\\hline
VBF & $1.31^{+0.26}_{-0.23}$~\cite{ATLAS:2020qdt} & $1.25^{+0.50}_{-0.41}$~\cite{ATLAS:2020qdt} & $0.60^{+0.36}_{-0.34}$~\cite{ATLAS:2020qdt}&*& $1.15^{+0.57}_{-0.53}$~\cite{ATLAS:2020qdt}&$0.99^{+0.35}_{-0.34}$~\cite{Aad:2020ago}&$1.8^{+1.0}_{-1.0}$~\cite{Aad:2020xfq}&*\\\hline
$Vh_1$ & $1.32^{+0.33}_{-0.30}$~\cite{ATLAS:2020qdt} & $1.53^{+1.13}_{-0.92}$~\cite{ATLAS:2020qdt}&*&*&*&$1.02^{+0.18}_{-0.17}$~\cite{ATLAS:2020qdt}&*&*\\\hline
$Wh_1$ & * & * & $2.3^{+1.2}_{-1.0}$~\cite{Aad:2019lpq} & * & * & * &*&* \\\hline
$Zh_1$ & * & * & $2.9^{+1.9}_{-1.3}$~\cite{Aad:2019lpq} & * & * & * &*&* \\\hline
$t{\overline{t}}h_1+th_1$ & $0.90^{+0.27}_{-0.24}$~\cite{ATLAS:2020qdt} & * & * & $1.72^{+0.56}_{-0.53}$~\cite{ATLAS:2020qdt} & $1.20^{+1.07}_{-0.93}$~\cite{ATLAS:2020qdt} & *&* &*\\\hline
$t{\overline{t}}h_1$ & * & * & * & * & * & $0.43^{+0.36}_{-0.33}$~\cite{ATLAS-CONF-2020-058} &*&*\\\hline
$Vh_1+t{\overline{t}}h_1$ & * & *&*&*&*&*&$5.0^{+3.5}_{-3.5}$~\cite{Aad:2020xfq}&*\\\hline
ggF+VBF+$t{\overline{t}}h_1$ & * & * & * & * & *&* &*&$2.0^{+1.0}_{-0.9}$~\cite{Aad:2020plj}\\\hline\hline
  \end{tabular}}
 \caption{\label{tab:ATLAS}ATLAS signal strengths at 13 TeV.  The asterisk indicates those signal strengths are not used in our fits.  The production modes are listed in the rows and the decay modes are listed in the columns.}
 \end{center}
 \end{table}
 
 \begin{table}[H]
\begin{center}
\resizebox{\textwidth}{!}{
\begin{tabular}{|c||c|c|c|c|c|c|}
\hline\hline
  & $\gamma\gamma$ & $ZZ^*$  & $WW^*$ &  $\tau\tau$  & $bb$ &$\mu\mu$ \\\hline
ggF & $0.98^{+0.13}_{-0.10}$~\cite{CMS:2020omd} & $0.98^{+0.12}_{-0.11}$~\cite{CMS:2020gsy}&$1.28^{+0.20}_{-0.19}$~\cite{CMS:2020gsy}&$0.39^{+0.38}_{-0.39}$~\cite{CMS:2020gsy}& $2.45^{+2.53}_{-2.35}$~\cite{CMS:2020gsy}&$0.63^{+0.65}_{-0.64}$~\cite{CMS:2020eni}\\\hline
VBF & $1.15^{+0.36}_{-0.31}$~\cite{CMS:2020omd} & $0.57^{+0.46}_{-0.36}$~\cite{CMS:2020gsy}& $0.63^{+0.65}_{-0.61}$~\cite{CMS:2020gsy}& $1.05^{+0.30}_{-0.29}$~\cite{CMS:2020gsy}&*&$1.36^{+0.69}_{-0.61}$~\cite{CMS:2020eni}\\\hline
$Vh_1$ & $0.71^{+0.31}_{-0.28}$~\cite{CMS:2020omd} & $1.10^{+0.96}_{-0.74}$~\cite{CMS:2020gsy}&*& *&*&$5.48^{+3.10}_{-2.83}$~\cite{CMS:2020eni}\\\hline
$Wh_1$ & * & *& $2.85^{+2.11}_{-1.87}$~\cite{CMS:2020gsy}&$3.01^{+1.65}_{-1.51}$~\cite{CMS:2020gsy}&$1.27^{+0.42}_{-0.40}$~\cite{CMS:2020gsy}&*\\\hline
$Zh_1$ & * & *& $0.90^{+1.77}_{-1.43}$~\cite{CMS:2020gsy}&$1.53^{+1.60}_{-1.37}$~\cite{CMS:2020gsy}&$0.93^{+0.33}_{-0.31}$~\cite{CMS:2020gsy}&*\\\hline
$t{\overline{t}}h_1+th_1$ &$1.40^{+0.33}_{-0.27}$~\cite{CMS:2020omd} & $0.13^{+0.93}_{-0.13}$~\cite{CMS-PAS-HIG-19-001} &  *& *&*&*\\\hline
$t{\overline{t}}h_1$ & *
& *& $0.93^{+0.48}_{-0.45}$~\cite{CMS:2020gsy}& $0.92^{+0.26}_{-0.23}$~\cite{CMS:2020iwy}&$1.13^{+0.33}_{-0.30}$~\cite{CMS:2020gsy}&$2.32^{+2.27}_{-1.95}$~\cite{CMS:2020eni}\\\hline
$th_1$ & * & *& * & $5.7^{+4.1}_{-4.0}$~\cite{CMS:2020iwy}& * &* \\\hline
  \end{tabular}}
 \caption{\label{tab:CMS}CMS signal strengths at 13 TeV.  The asterisk indicates those signal strengths are not used in our fits.  The production modes are listed in the rows and the decay modes are listed in the columns.}
 \end{center}
 \end{table}

\subsection{Scalar Search Bounds}
Now we give the relevant observed and expected scalar cross section upper bounds from ATLAS in Tab.~\ref{tab:ATLsearch} and CMS in Tab.~\ref{tab:CMSsearch}.  For searches for $pp\rightarrow h_2\rightarrow h_1h_1$, several of the experimental papers~\cite{Aaboud:2018ftw,Aaboud:2018zhh,Aaboud:2018ksn,CMS-PAS-HIG-18-013} report bounds on the production cross section $\sigma(pp\rightarrow h_2\rightarrow h_1h_1)$, not including the $h_1$ decays.  To do this, the CMS and ATLAS collaborations assume that $h_1$ decays are SM-like.  However, in our model we also change the branching ratios of $h_1$ and we need bounds on the cross section $\sigma(pp\rightarrow h_2\rightarrow h_1h_1\rightarrow 2X\,2Y)$ including $h_1$ decays, where $X$ and $Y$ are $h_1$ decay products.   Hence, when the experimental searches are reported as bounds on $\sigma(pp\rightarrow h_2\rightarrow h_1h_1)$ we multiply the bounds by the relevant SM $h_1$ branching ratios as provided by the LHC Higgs Cross Section Working Group~\cite{deFlorian:2016spz}.  This step eliminates the assumption that $h_1$ decays are SM-like and provides the relevant cross section bounds for our model.
\begin{table}[H]
  \begin{center}
    \label{tab:table1}
\resizebox{\textwidth}{!}{
    \begin{tabular}{|c||c|c||c|c||c|c|} 
    \hline
      $\sigma$(pb) & \multicolumn{2}{|c||}{$m_2=200$ GeV} & \multicolumn{2}{|c||}{$m_2=400$ GeV} & \multicolumn{2}{|c|}{$m_2=600$ GeV}\\
      \hline
      & Obs. & Exp. & Obs. &Exp.&Obs. & Exp. \\\hline
      $pp\rightarrow h_2 \rightarrow Z\gamma$ & * & * & 0.027 pb ~\cite{Aaboud:2017uhw} & 0.027 pb ~\cite{Aaboud:2017uhw} & 0.0081 pb ~\cite{Aaboud:2017uhw} & 0.0125 pb ~\cite{Aaboud:2017uhw} \\\hline
      $pp\rightarrow h_2 \rightarrow \gamma\gamma$ & 0.0053 pb ~\cite{ATLAS:2020tws} & 0.0041 pb ~\cite{ATLAS:2020tws} & 0.0013 pb ~\cite{ATLAS:2020tws} & 0.0010 pb ~\cite{ATLAS:2020tws} & 0.00057 pb ~\cite{ATLAS:2020tws} & 0.00046 pb ~\cite{ATLAS:2020tws} \\\hline
      $pp\rightarrow h_2 \rightarrow ZZ$ & 0.11 pb ~\cite{Aad:2020fpj} & 0.12 pb ~\cite{Aad:2020fpj} & 0.050 pb ~\cite{Aad:2020fpj} & 0.039 pb ~\cite{Aad:2020fpj} & 0.023 pb ~\cite{Aad:2020fpj} & 0.017 pb ~\cite{Aad:2020fpj} \\\hline
      $pp\rightarrow h_2 \rightarrow WW$ & 6.4 pb ~\cite{Aaboud:2017gsl} & 5.6 pb ~\cite{Aaboud:2017gsl} & 1.29 pb ~\cite{Aaboud:2017gsl}& 1.22 pb ~\cite{Aaboud:2017gsl} & 0.30 pb ~\cite{Aaboud:2017gsl}  & 0.43 pb ~\cite{Aaboud:2017gsl} \\\hline
      $pp\rightarrow h_2 \rightarrow h_1h_1 \rightarrow b\bar{b}b\bar{b}$ & * & * & 0.222 pb ~\cite{Aaboud:2018knk} & 0.210 pb ~\cite{Aaboud:2018knk} & 0.0273 pb ~\cite{Aaboud:2018knk} & 0.0391 pb ~\cite{Aaboud:2018knk}\\\hline
      $pp\rightarrow h_2 \rightarrow h_1h_1 \rightarrow b\bar{b}\gamma\gamma$ & * & * & 0.00209 pb ~\cite{Aaboud:2018ftw} & 0.00176 pb ~\cite{Aaboud:2018ftw} & $7.11\times10^{-4}$ pb ~\cite{Aaboud:2018ftw} & $8.33\times10^{-4}$ pb ~\cite{Aaboud:2018ftw}\\\hline
      $pp\rightarrow h_2 \rightarrow h_1h_1 \rightarrow b\bar{b}\tau^{+}\tau^{-}$ & * & * & 0.0397 pb ~\cite{Aaboud:2018sfw} & 0.0445 pb ~\cite{Aaboud:2018sfw} & 0.00907 pb ~\cite{Aaboud:2018sfw} & 0.0117 pb ~\cite{Aaboud:2018sfw}\\\hline
      $pp\rightarrow h_2 \rightarrow h_1h_1 \rightarrow b\bar{b}WW$& * & * & * & * & 0.301 pb ~\cite{Aaboud:2018zhh} & 0.319 pb ~\cite{Aaboud:2018zhh}\\\hline
      $pp\rightarrow h_2 \rightarrow h_1h_1 \rightarrow \gamma\gamma WW$ & * & * & 0.0090 pb ~\cite{Aaboud:2018ewm} & 0.00630 pb ~\cite{Aaboud:2018ewm} & * & *\\\hline
     $pp\rightarrow h_2 \rightarrow h_1h_1 \rightarrow WWWW$ & * & * & 0.242 pb ~\cite{Aaboud:2018ksn} & 0.309 pb ~\cite{Aaboud:2018ksn} & * & *\\\hline
      $pp\rightarrow h_2 \rightarrow \tau^{+}\tau^{-}$ & 0.23 pb ~\cite{Aad:2020zxo} & 0.424 pb ~\cite{Aad:2020zxo} & 0.0836 pb ~\cite{Aad:2020zxo} & 0.0408 pb ~\cite{Aad:2020zxo} & 0.015 pb ~\cite{Aad:2020zxo} & 0.0128 pb ~\cite{Aad:2020zxo} \\\hline
      $pp\rightarrow h_2 \rightarrow \mu^{+}\mu^{-}$ & 0.0442 pb ~\cite{Aaboud:2019sgt}  & 0.0258 pb ~\cite{Aaboud:2019sgt} & 0.0074 pb ~\cite{Aaboud:2019sgt} & 0.0079 pb ~\cite{Aaboud:2019sgt} & 0.006 pb ~\cite{Aaboud:2019sgt} & 0.0041 pb ~\cite{Aaboud:2019sgt} \\\hline
      $pp\rightarrow h_2 \rightarrow jj$ & *  & * & *& * & 9.95 pb ~\cite{Aaboud:2018fzt} & 10.5 pb ~\cite{Aaboud:2018fzt} \\\hline
    \end{tabular}}
    \caption{\label{tab:ATLsearch}Observed (Obs.) and expected (Exp) 95 $\%$ CL ATLAS upper limit on cross section times branching ratio for heavy resonances at center of mass energy 13 TeV.  The asterisk indicates that either there were no relevant results or they were not included in the fit.}
  \end{center}
\end{table}

\begin{table}[H]
  \begin{center}
    \label{tab:table2}
\resizebox{\textwidth}{!}{
    \begin{tabular}{|l||c|c||c|c||c|c|} 
    \hline
      $\sigma$(pb) & \multicolumn{2}{|c||}{$m_2=200$ GeV} & \multicolumn{2}{|c||}{$m_2=400$ GeV} & \multicolumn{2}{|c|}{$m_2=600$ GeV}\\
      \hline
      & Obs. & Exp. & Obs. & Exp. &Obs. & Exp. \\\hline
      $pp\rightarrow h_2 \rightarrow Z\gamma$ & *  & * & 0.022 pb ~\cite{Sirunyan:2017hsb} & 0.027 pb ~\cite{Sirunyan:2017hsb} & 0.016 pb ~\cite{Sirunyan:2017hsb} & 0.013 pb ~\cite{Sirunyan:2017hsb}\\\hline
      $pp\rightarrow h_2 \rightarrow \gamma\gamma$ & * & * & * & * & 0.0014 ~\cite{Sirunyan:2018wnk} pb & 0.0016 pb ~\cite{Sirunyan:2018wnk}\\\hline
      $pp\rightarrow h_2 \rightarrow ZZ$ & 0.24 pb ~\cite{Sirunyan:2018qlb} & 0.33 pb ~\cite{Sirunyan:2018qlb} & 0.067 pb ~\cite{Sirunyan:2018qlb} & 0.083 pb ~\cite{Sirunyan:2018qlb} & 0.025 pb ~\cite{Sirunyan:2018qlb} & 0.032 pb ~\cite{Sirunyan:2018qlb} \\\hline
      $pp\rightarrow h_2 \rightarrow WW$ & 6.8 pb ~\cite{CMS:2019kjn} & 5.9 pb ~\cite{CMS:2019kjn} & 1.0 pb ~\cite{CMS:2019kjn} & 1.21 pb ~\cite{CMS:2019kjn} & 0.43 pb ~\cite{CMS:2019kjn} & 0.33 pb ~\cite{CMS:2019kjn} \\\hline
      $pp\rightarrow h_2 \rightarrow h_1h_1 \rightarrow b\bar{b}b\bar{b}$ & * & * & 0.163 pb ~\cite{Sirunyan:2018zkk} & 0.179 pb ~\cite{Sirunyan:2018zkk} & 0.0355 pb ~\cite{Sirunyan:2018zkk} & 0.0396 pb ~\cite{Sirunyan:2018zkk} \\\hline
      $pp\rightarrow h_2 \rightarrow h_1h_1 \rightarrow b\bar{b}\gamma\gamma$ & * & * & 0.0012 pb ~\cite{Sirunyan:2018iwt} & 0.0012 pb ~\cite{Sirunyan:2018iwt} & $4.4\times10^{-4}$ pb ~\cite{Sirunyan:2018iwt} & $4.4\times10^{-4}$ pb ~\cite{Sirunyan:2018iwt} \\\hline
     $pp\rightarrow h_2 \rightarrow h_1h_1 \rightarrow b\bar{b}\tau^{+}\tau^{-}$ & * & * & 0.0860 pb ~\cite{Sirunyan:2017djm} & 0.106 pb ~\cite{Sirunyan:2017djm} & 0.0314 pb ~\cite{Sirunyan:2017djm} & 0.0200 pb ~\cite{Sirunyan:2017djm} \\\hline
 $pp\rightarrow h_2 \rightarrow h_1h_1 \rightarrow b\bar{b}ZZ$ & * & * & 0.659 pb ~\cite{Sirunyan:2020qcq} & 0.92 pb ~\cite{Sirunyan:2020qcq} & 0.16 pb ~\cite{Sirunyan:2020qcq} & 0.305 pb ~\cite{Sirunyan:2020qcq} \\\hline
      $pp\rightarrow h_2 \rightarrow \tau^{+}\tau^{-}$ & 0.823 pb ~\cite{Sirunyan:2018zut} & 0.887 pb ~\cite{Sirunyan:2018zut} & 0.0743 pb ~\cite{Sirunyan:2018zut} & 0.0851 pb ~\cite{Sirunyan:2018zut} & 0.0495 pb ~\cite{Sirunyan:2018zut} & 0.0294 pb ~\cite{Sirunyan:2018zut} \\\hline
      $pp\rightarrow h_2 \rightarrow \mu^{+}\mu^{-}$ & 0.0175 pb ~\cite{Sirunyan:2019tkw}  & 0.0112 pb ~\cite{Sirunyan:2019tkw} & 0.0028 pb ~\cite{Sirunyan:2019tkw} & 0.003 pb ~\cite{Sirunyan:2019tkw} & 0.00175 pb ~\cite{Sirunyan:2019tkw} & 0.00175 pb ~\cite{Sirunyan:2019tkw}\\\hline
      $pp\rightarrow h_2 \rightarrow dijets(gg)$ & * & * & * & * & 71.8 pb ~\cite{Sirunyan:2018xlo} & 38.4 pb ~\cite{Sirunyan:2018xlo}\\\hline
    \end{tabular}}
    \caption{\label{tab:CMSsearch}Observed (Obs.) and expected (Exp) 95 $\%$ CL CMS upper limit on cross section times branching ratio for heavy resonances at center of mass energy 13 TeV.  The asterisk indicates that either there were no relevant results or they were not included in the fit.}
  \end{center}
\end{table}

\section{95\% C.L. Limits}
\begin{figure}[hb!]
\begin{center}\vspace{-0.1in}
\subfigure[]{\includegraphics[width=0.39\textwidth,clip]{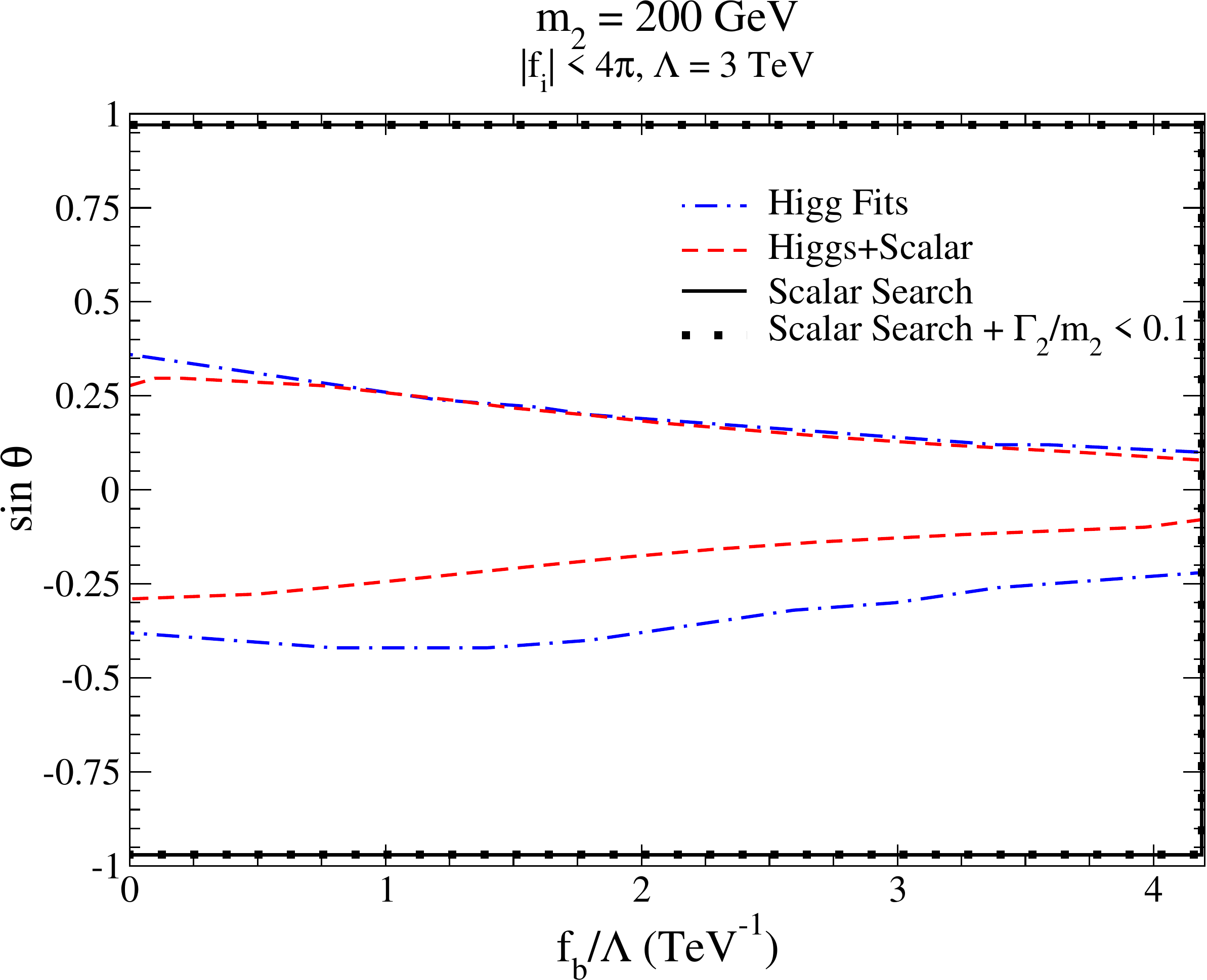}}
\subfigure[]{\includegraphics[width=0.39\textwidth,clip]{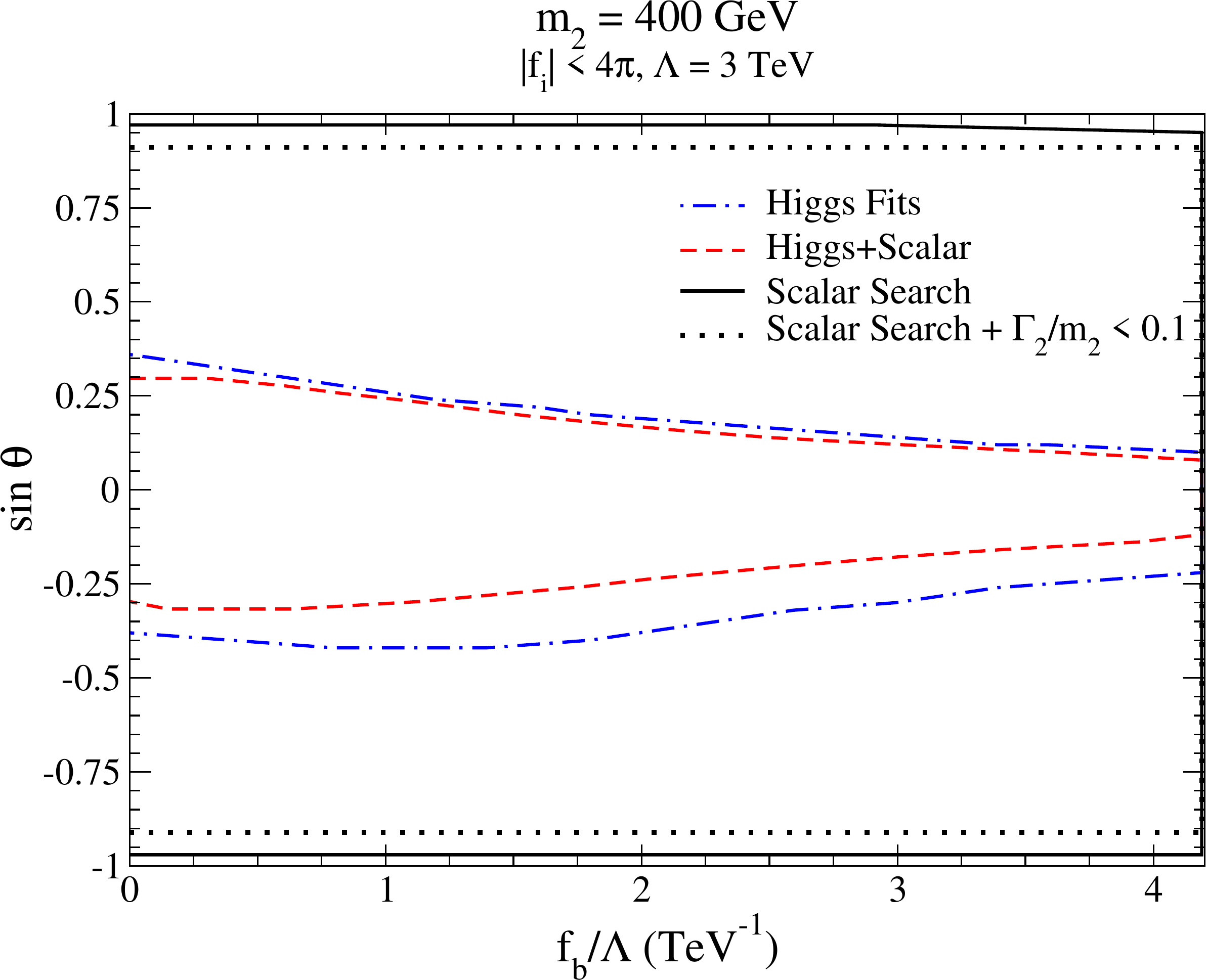}}\\\vspace{-0.1in}
\subfigure[]{\includegraphics[width=0.39\textwidth,clip]{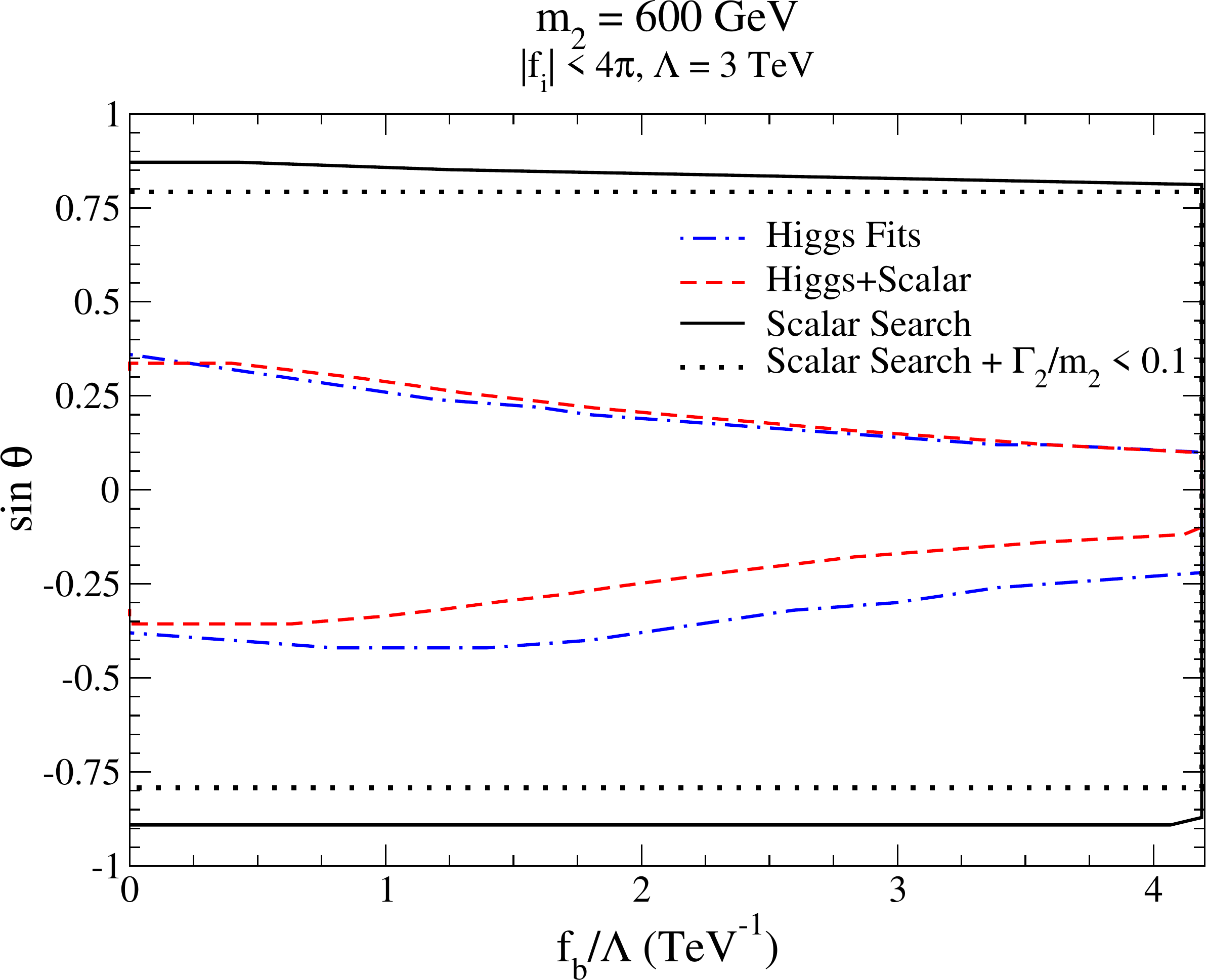}}
\subfigure[]{\includegraphics[width=0.39\textwidth,clip]{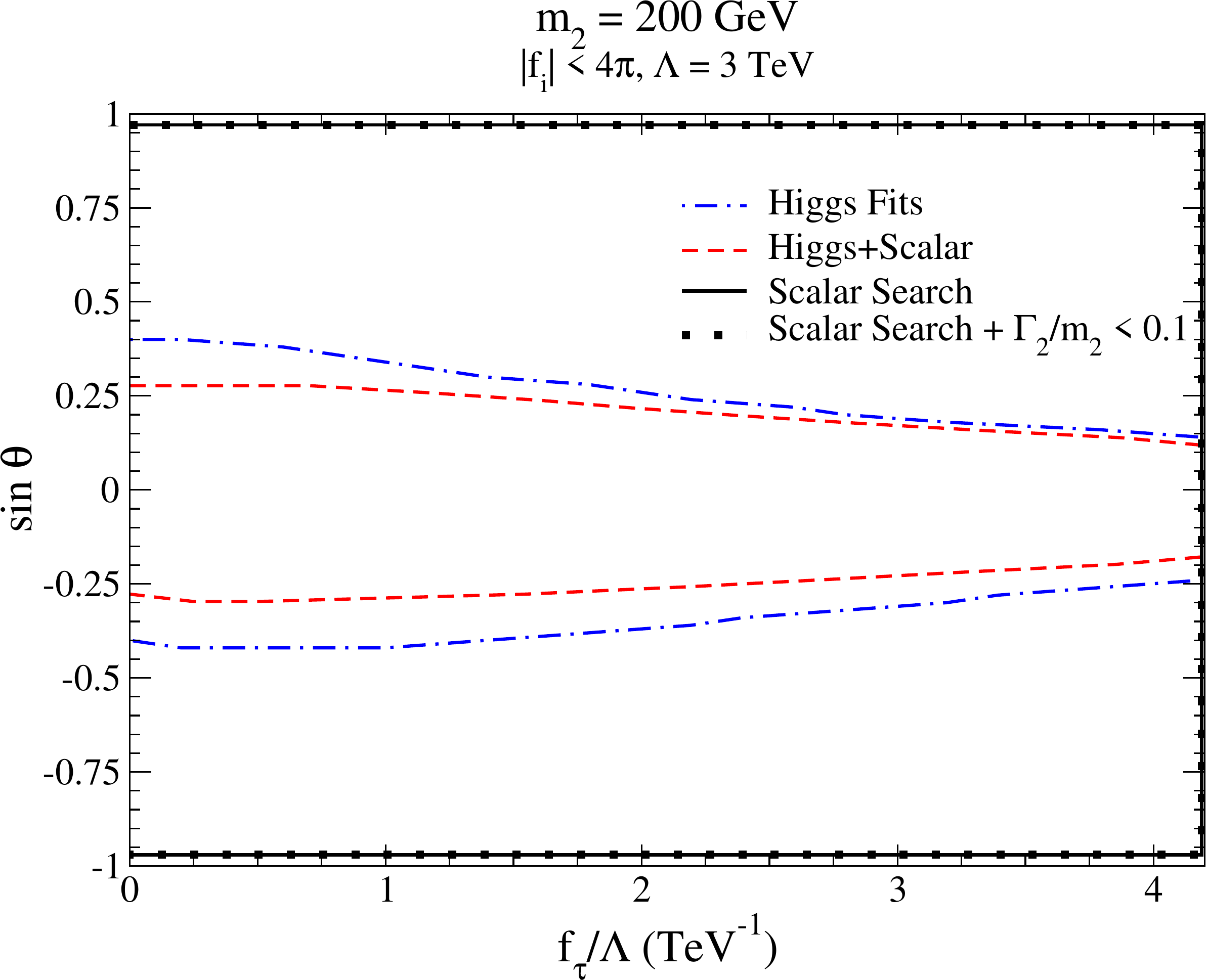}}\\\vspace{-0.1in}
\subfigure[]{\includegraphics[width=0.39\textwidth,clip]{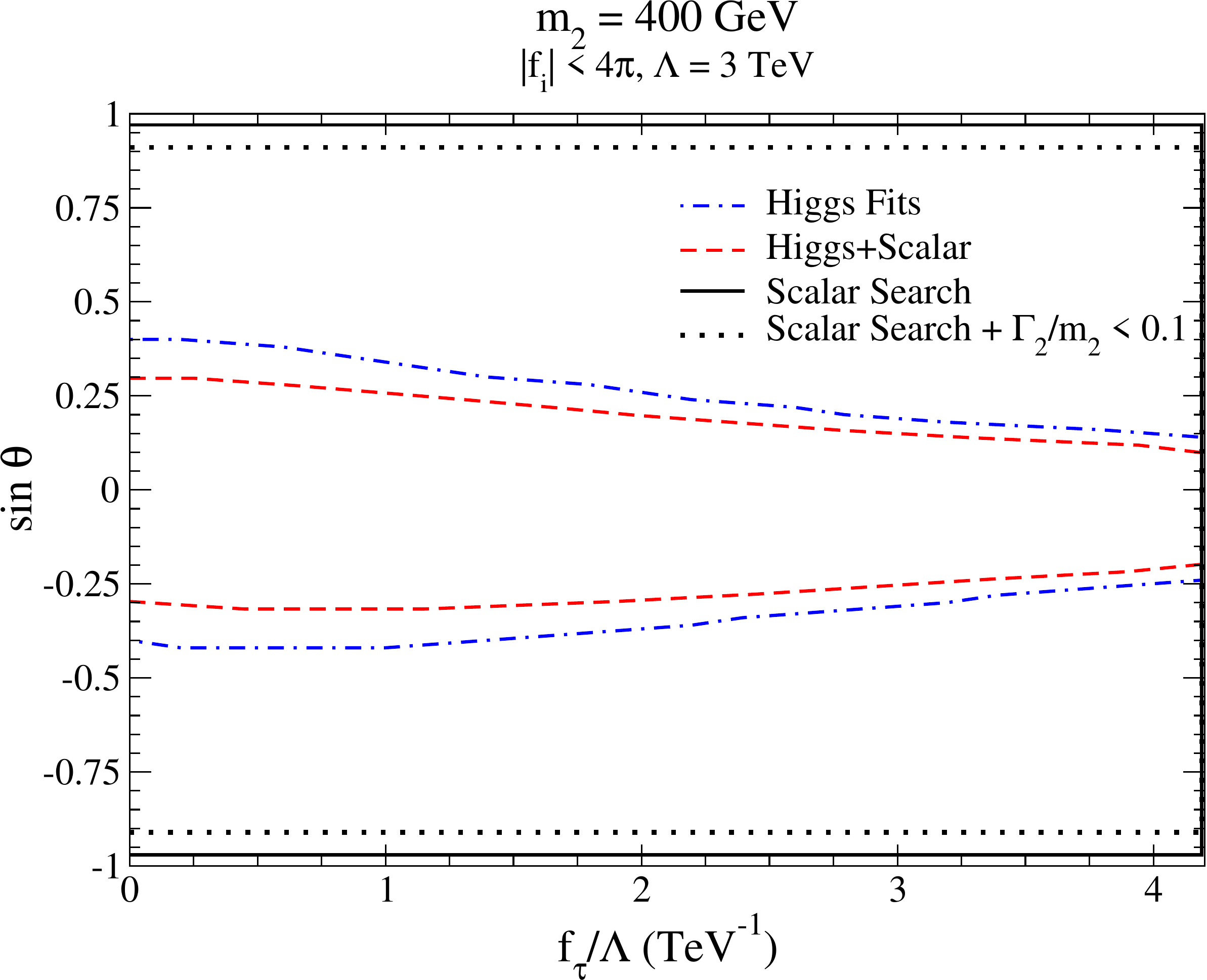}}
\subfigure[]{\includegraphics[width=0.39\textwidth,clip]{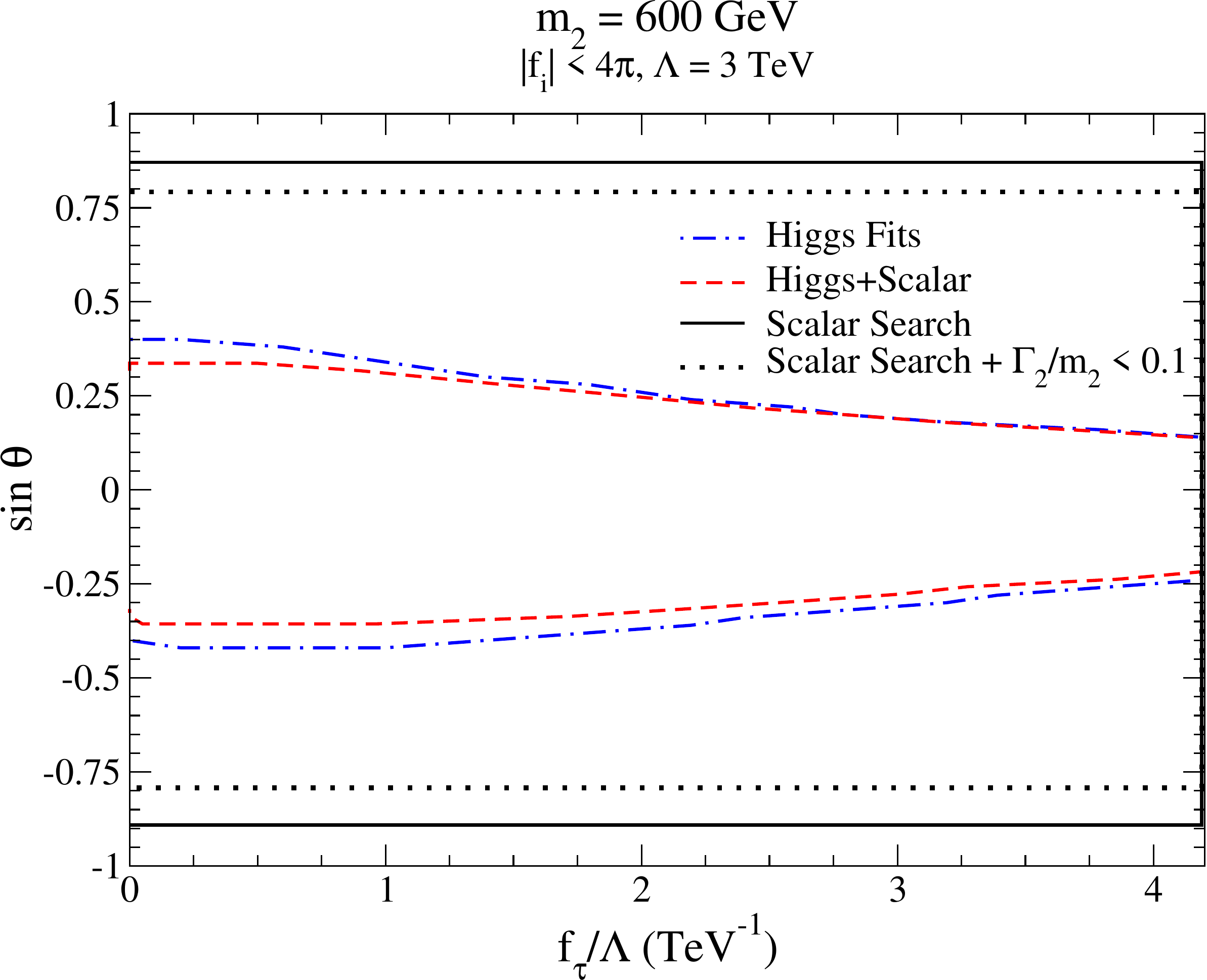}}\\\vspace{-0.25in}
\end{center}
\caption{Same as Fig.~\ref{fig:Scalar2} for (a,b,c) $f_{b}$and (d,e,f) $f_\tau$ vs $\sin\theta$ with all other parameters profiled over.  Three scalar masses are considered: (a,d) $m_2=200$~GeV, (b,e) $m_2=400$~GeV, and (c,f) $m_2=600$~GeV. The new physics scale is $\Lambda=3$~TeV.\label{fig8}}\vspace{-0.3in}
\end{figure}

\begin{figure}[H]
\begin{center}
\subfigure[]{\includegraphics[width=0.42\textwidth,clip]{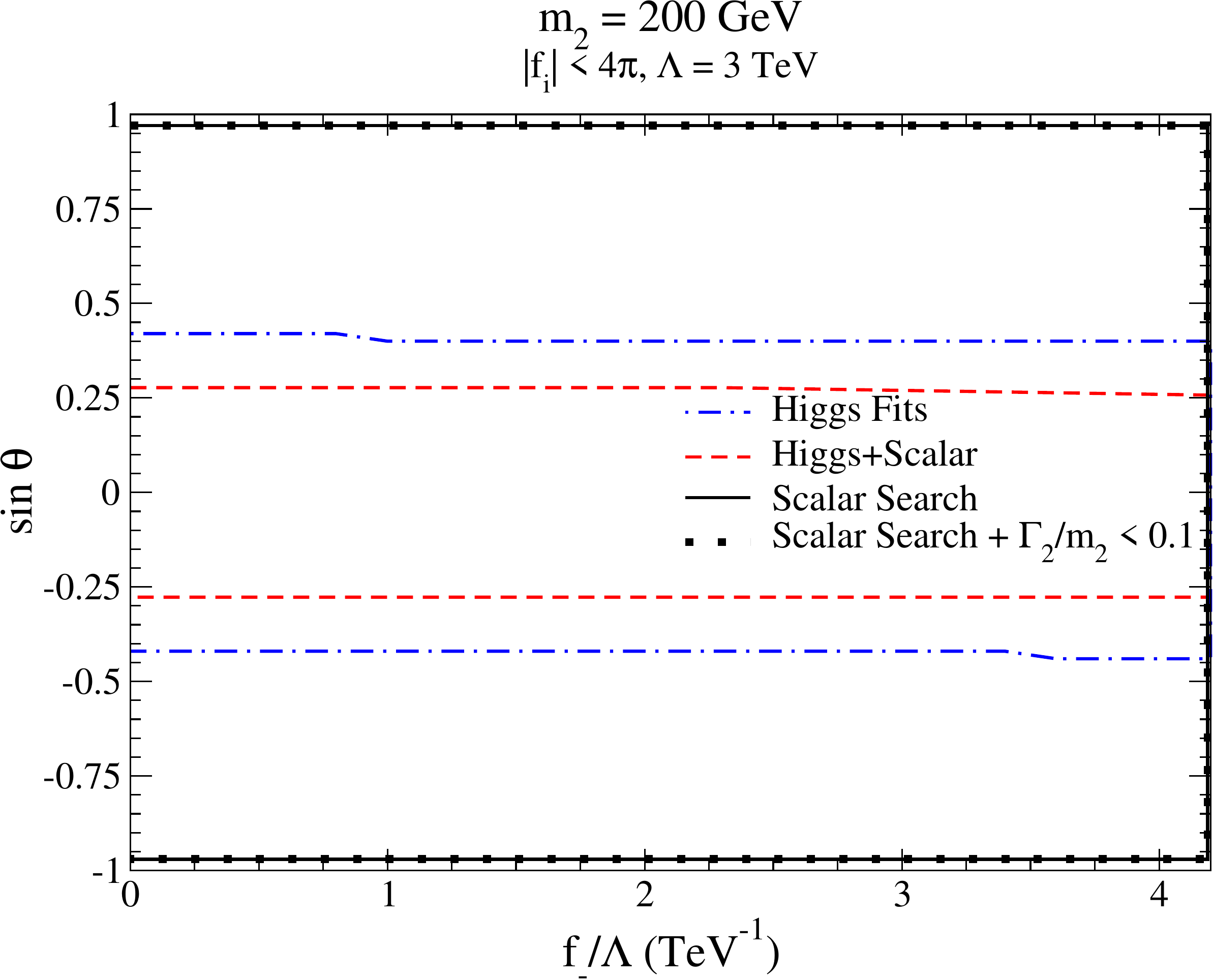}}
\subfigure[]{\includegraphics[width=0.42\textwidth,clip]{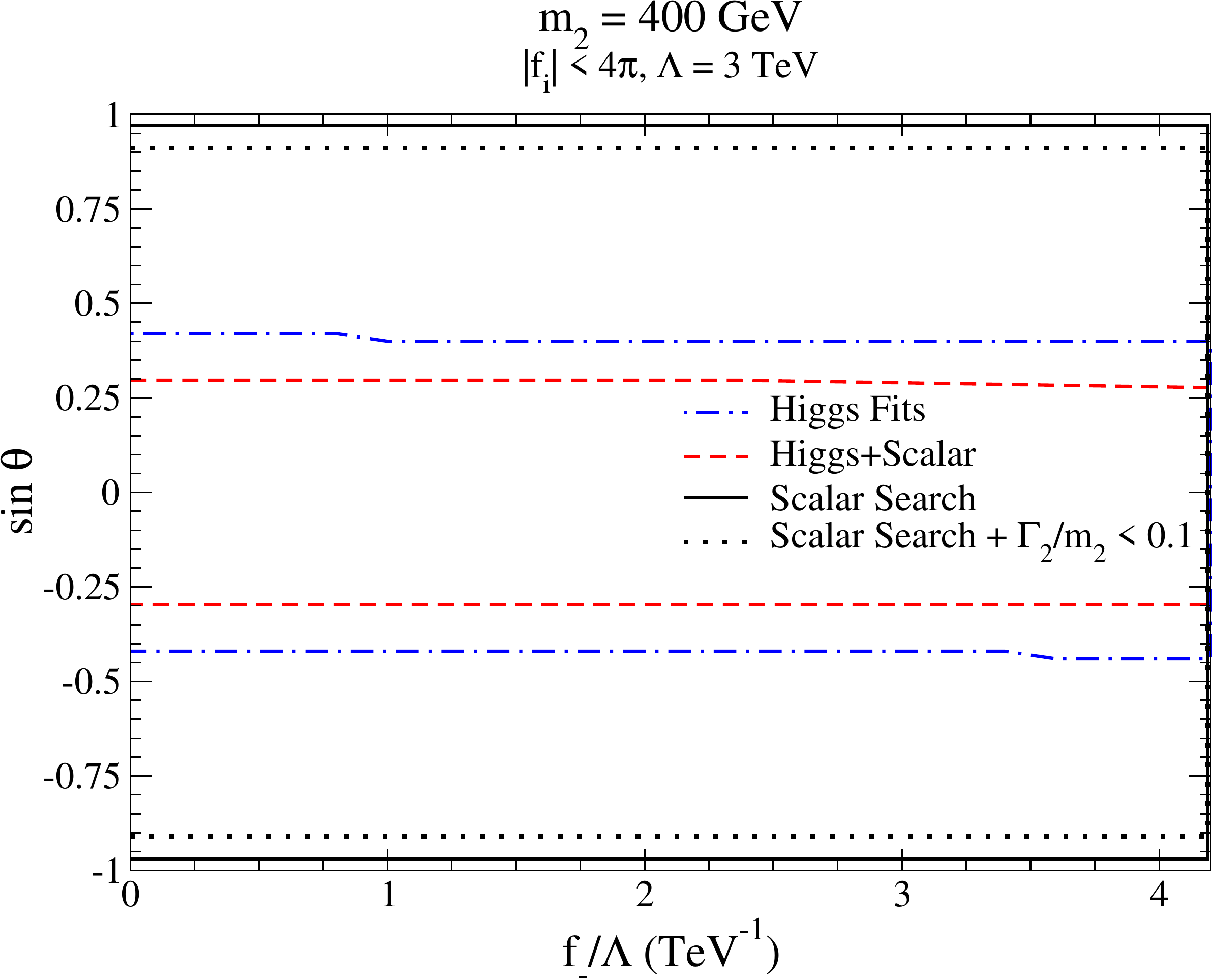}}\\\vspace{-0.1in}
\subfigure[]{\includegraphics[width=0.42\textwidth,clip]{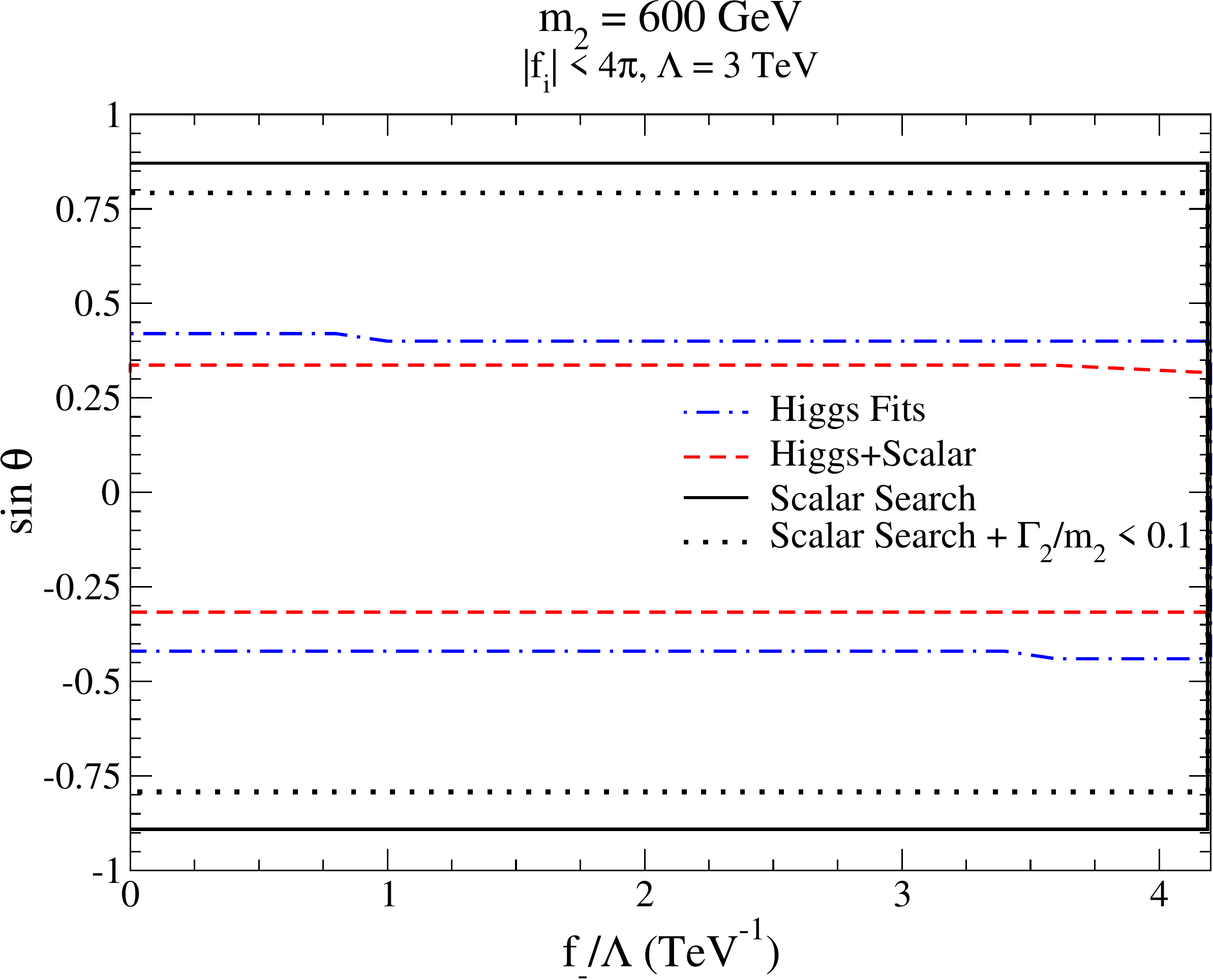}}
\subfigure[]{\includegraphics[width=0.42\textwidth,clip]{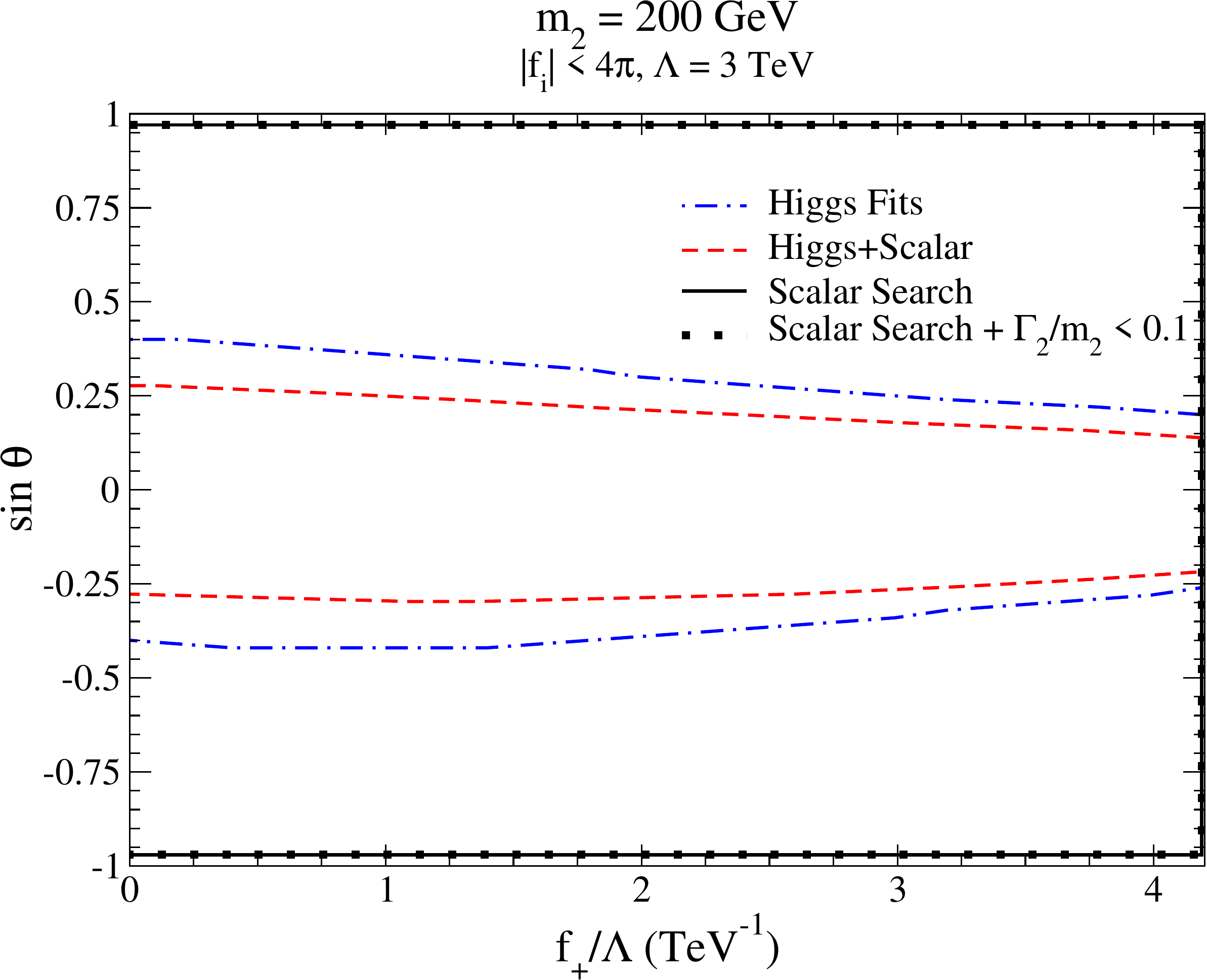}}\\\vspace{-0.1in}
\subfigure[]{\includegraphics[width=0.42\textwidth,clip]{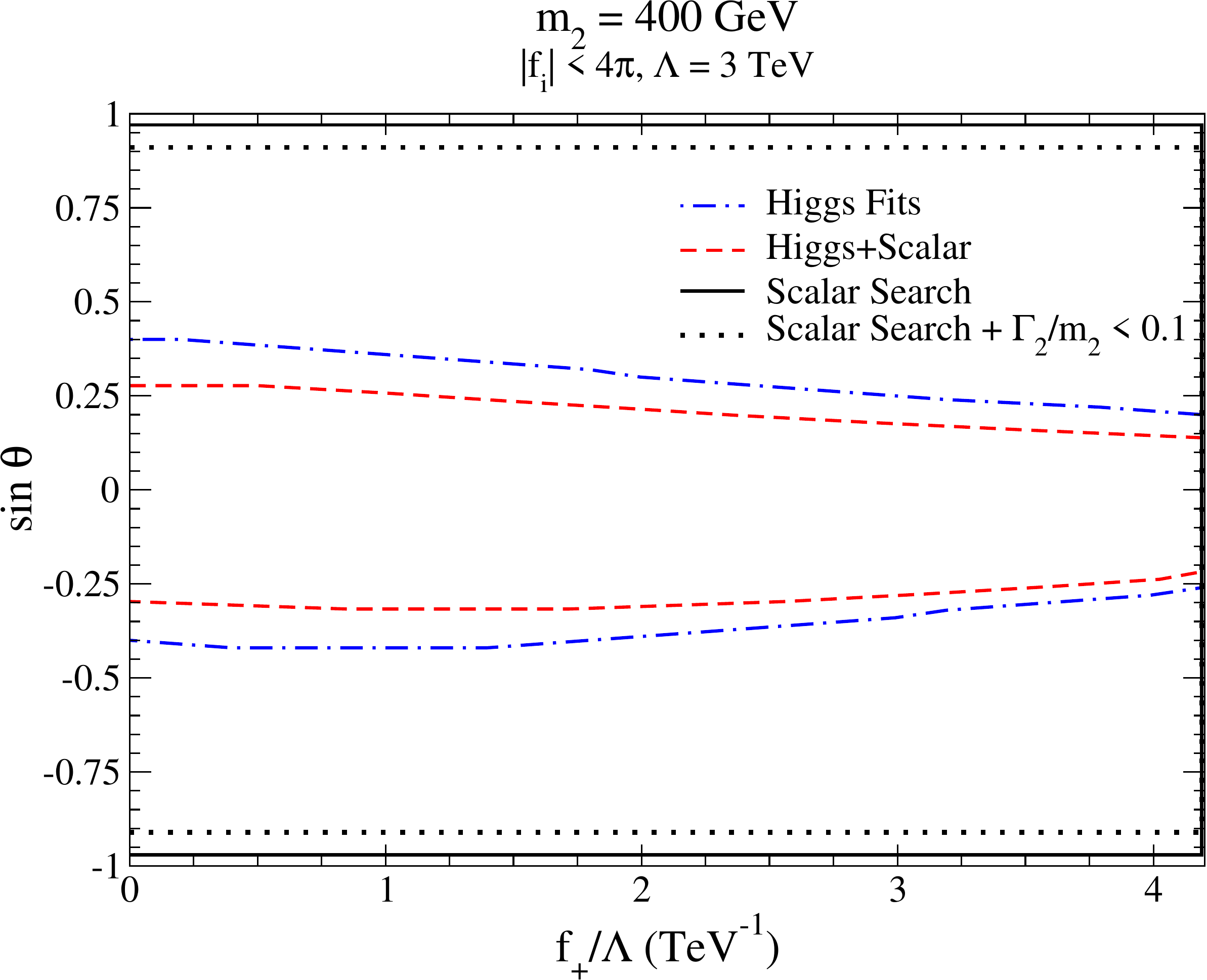}}
\subfigure[]{\includegraphics[width=0.42\textwidth,clip]{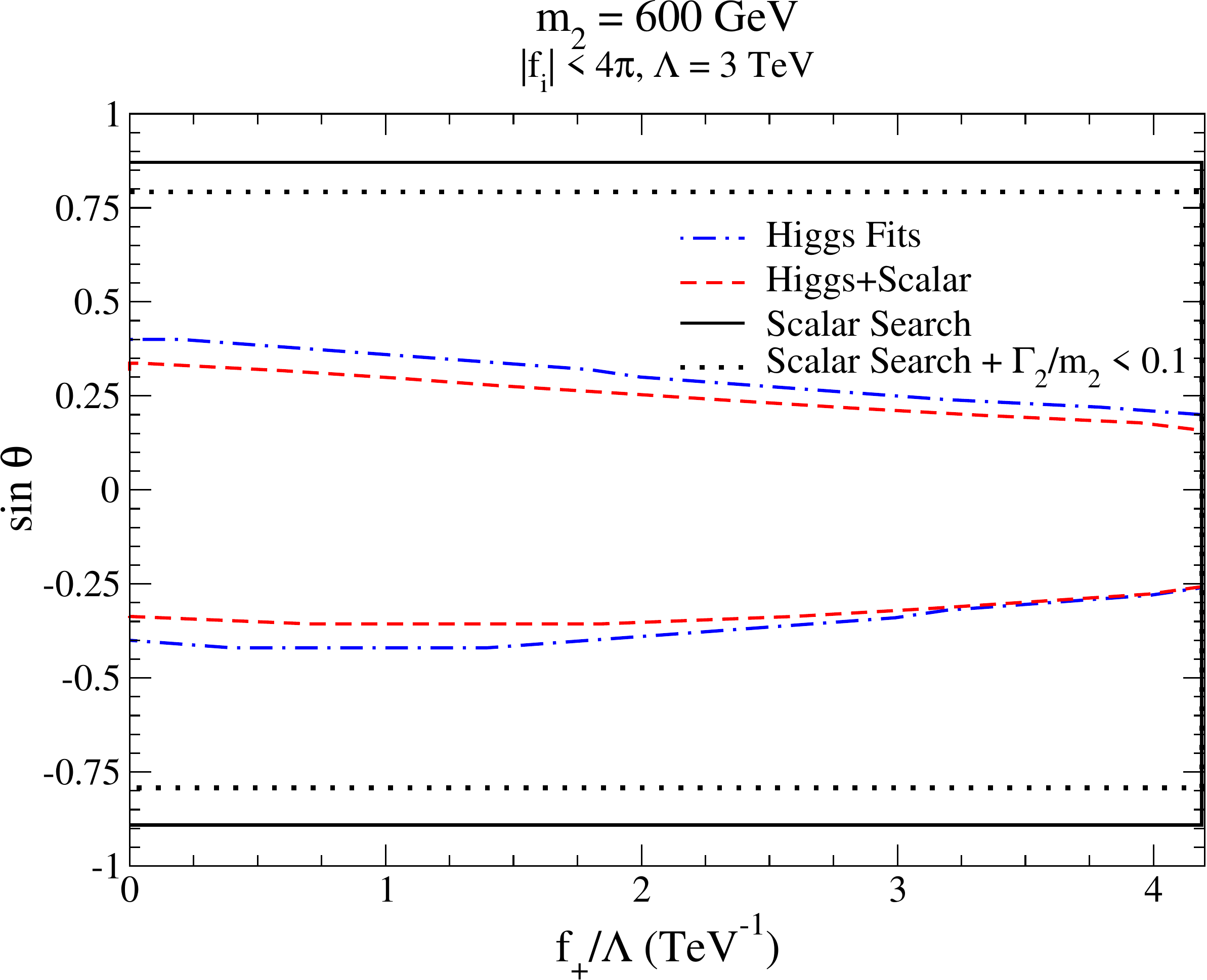}}\\\vspace{-0.25in}
\end{center}
\caption{Same as Fig.~\ref{fig:Scalar2} for (a,b,c) $f_-$and (d,e,f) $f_+$ vs $\sin\theta$ with all other parameters profiled over.  Three scalar masses are considered: (a,d) $m_2=200$~GeV, (b,e) $m_2=400$~GeV, and (c,f) $m_2=600$~GeV. The new physics scale is $\Lambda=3$~TeV.\label{fig9}}
\end{figure}

\begin{figure}[H]
\begin{center}
\subfigure[]{\includegraphics[width=0.42\textwidth,clip]{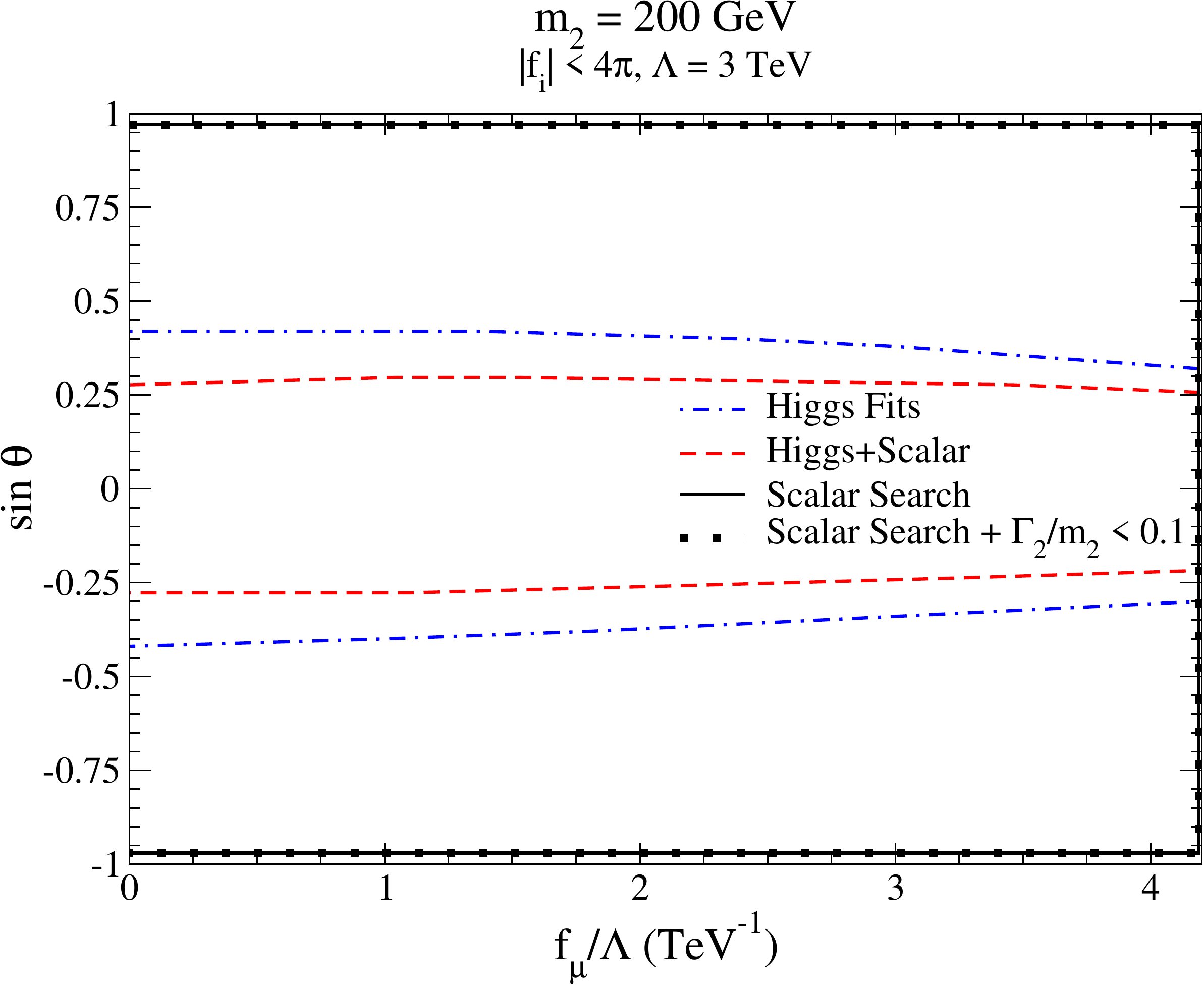}}
\subfigure[]{\includegraphics[width=0.42\textwidth,clip]{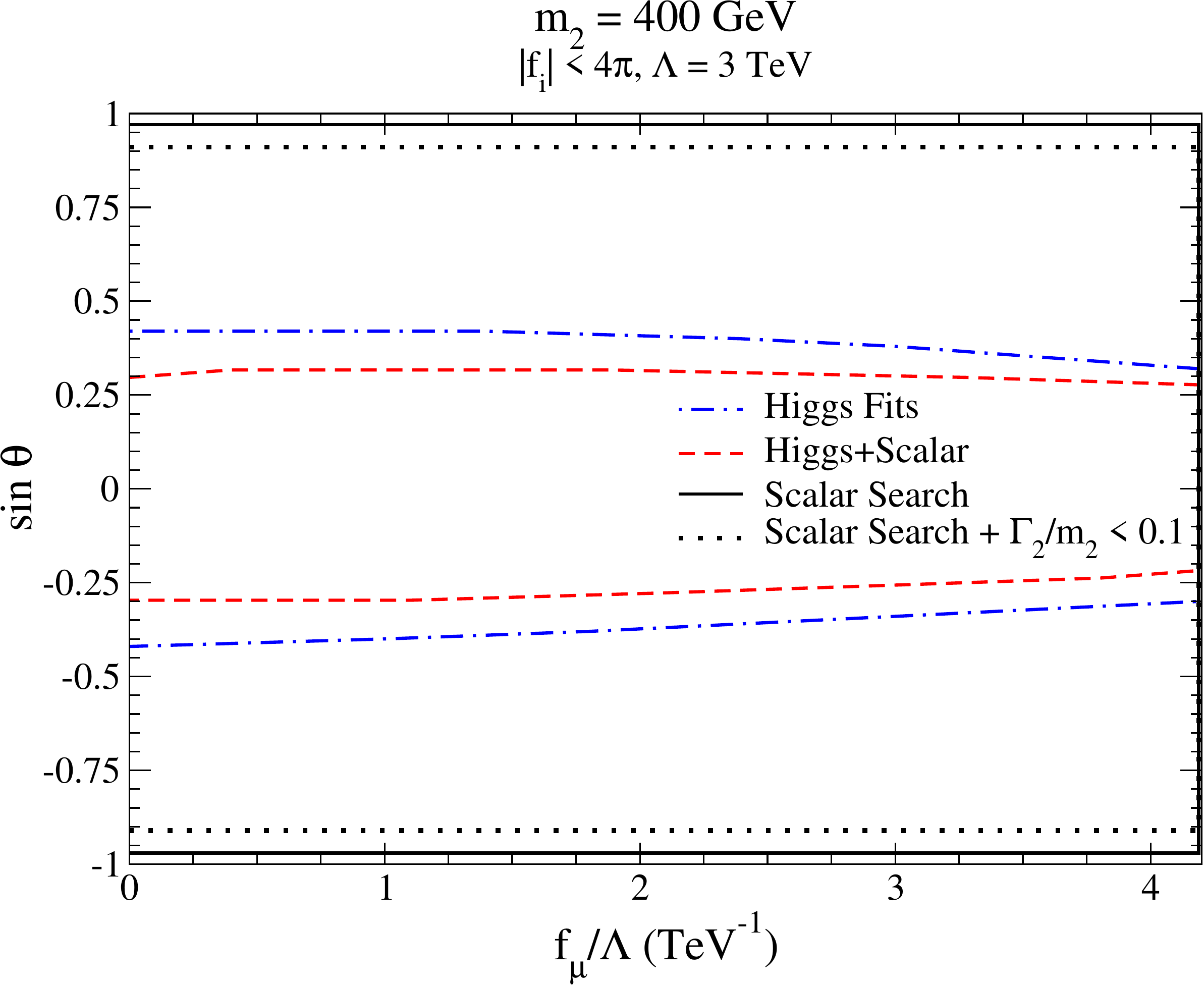}}
\subfigure[]{\includegraphics[width=0.42\textwidth,clip]{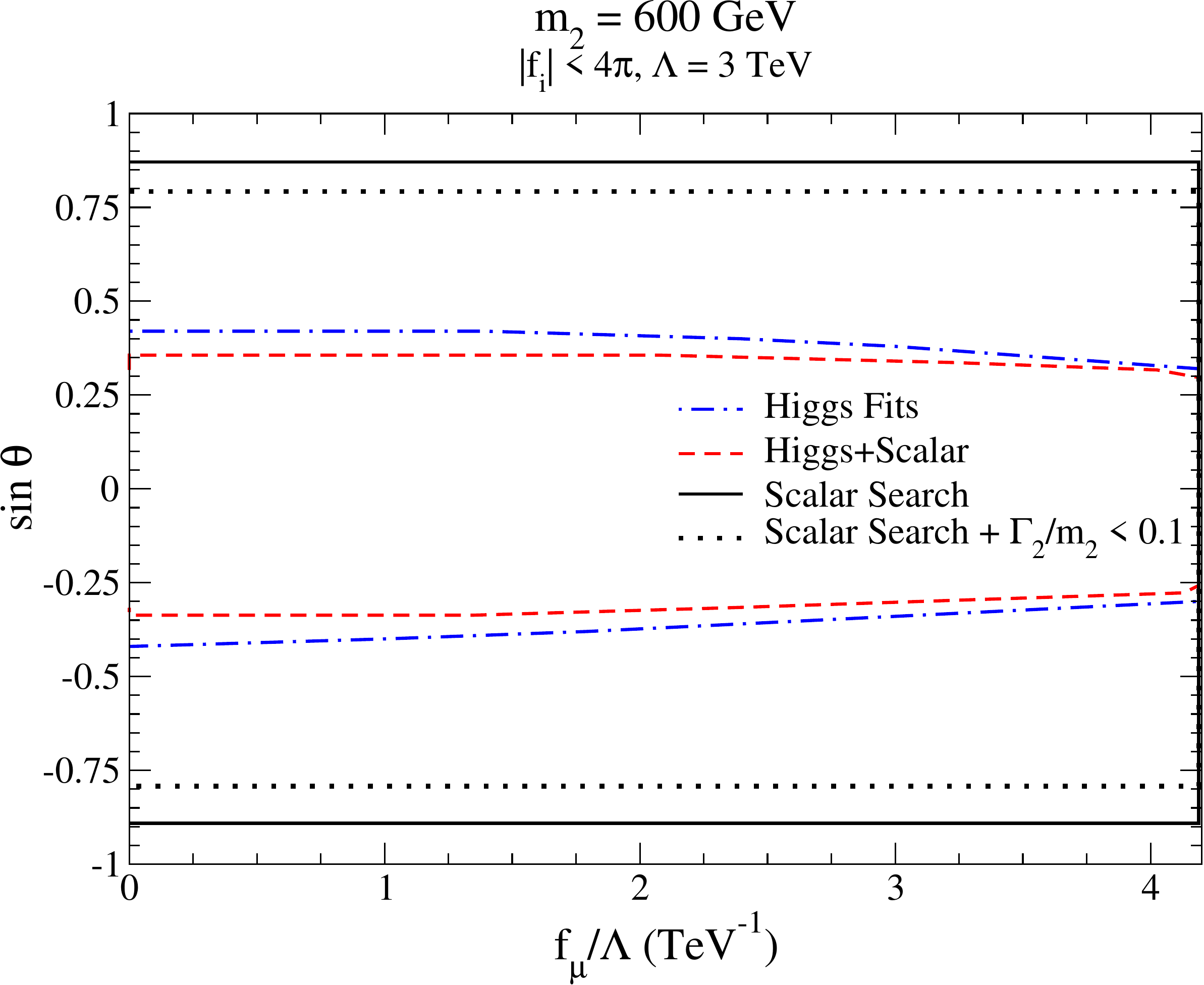}}
\end{center}
\caption{Same as Fig.~\ref{fig:Scalar2} for $f_{\mu}$ vs $\sin\theta$ with all other parameters profiled over.  Three scalar masses are considered: (a,d) $m_2=200$~GeV, (b,e) $m_2=400$~GeV, and (c,f) $m_2=600$~GeV. The new physics scale is $\Lambda=3$~TeV.\label{fig10}}
\end{figure}

\vspace{-0.15in}
\label{app:limits}
Here we show all other Wilson coefficient 95\% CL allowed regions not shown in the main text.  Fig.~\ref{fig8} shows the limits on $f_{b}$ and $f_{\tau}$, Fig.~\ref{fig9} shows the limits on $f_{-}$ and $f_{+}$, and Fig.~\ref{fig10} shows the limits on $f_{\mu}$.\vspace{-0.1in}

\clearpage
\bibliographystyle{myutphys}
\bibliography{singlet}

\end{document}